\documentclass[twocolumn,ap,pre,superscriptaddress,nofootinbib,showpacs]{revtex4-1}
\usepackage{graphicx}
\usepackage{times}
\usepackage{bm}
\usepackage[linktocpage=true]{hyperref}

\begin{document}

\title{Majorana Fermions in Semiconductor Nanowires:  Fundamentals, Modeling, and Experiment}
\author{Tudor D. Stanescu}
\affiliation{Department of Physics, West Virginia University, Morgantown, WV 26506.}
\author{Sumanta Tewari}
\affiliation{Department of Physics and Astronomy, Clemson University, Clemson, SC 29634, USA}

\begin{abstract}
After a recent series of rapid and exciting developments, the long search for the Majorana fermion - the elusive quantum entity at the border between particles and antiparticles  - has produced the first positive experimental results, but is not over yet. Originally proposed by E. Majorana in the context of particle physics, Majorana fermions  have a condensed matter analog in the zero--energy bound states emerging in topological superconductors.  A promising route to engineering topological superconductors capable of hosting Majorana zero modes consists of proximity coupling semiconductor thin films or nanowires with strong spin--orbit interaction to conventional s--wave superconductors in the presence of an external Zeeman field. The Majorana zero mode is predicted to emerge above a certain critical Zeeman field as a zero--energy state localized near the order parameter defects, viz., vortices for thin films and wire-ends for the nanowire. These Majorana bound states are expected to manifest non--Abelian quantum statistics, which makes them ideal building blocks for fault--tolerant topological quantum computation.
This review provides an update on current status of the search for Majorana fermions in semiconductor nanowires by focusing on the recent developments, in particular the period following the first reports of experimental signatures consistent with the realization of Majorana bound states in semiconductor nanowire--superconductor hybrid structures. We start with a discussion of the fundamental aspects of the subject, followed by considerations on the realistic modeling which is a critical bridge between theoretical predictions based on idealized conditions and the real world, as probed experimentally. The last part is dedicated to a few intriguing issues  that were brought to the fore by the recent encouraging experimental advances.
\end{abstract}

\maketitle

\tableofcontents

\section{Introduction}

In relativistic quantum mechanics  spin--1/2 fermions are described by the solutions of the Dirac equation~\cite{Dirac1928},
\begin{equation}
( i\gamma^{\mu}\partial_{\mu} -m )\psi = 0.
\label{eq:DE}
\end{equation}
Here, $\gamma^{\mu}$ ($\mu = 0,1,2,3$) is a set of $4\times4$ matrices satisfying the anti--commutation relations
$\{\gamma^{\mu},\gamma^{\nu}\}=2g^{\mu\nu}$, where  $g^{\mu\nu}$ represents  the Minkowski metric and $\gamma_0\gamma_{\mu}\gamma_0 = \gamma_{\mu}^{\dagger}$~\cite{Peskin1995}. In general, the matrices $\gamma^{\mu}$ have complex elements, making Eq.~(\ref{eq:DE}) a set of coupled differential equations with complex coefficients.Thus, the general solution $\psi(x)$ of Eq.~(\ref{eq:DE}) representing the fermion field is a complex bi--spinor that is not an eigenstate of the charge conjugation (CC) operator. Since charge conjugation, $\psi\rightarrow\psi^*$, maps a particle into its anti--particle~\cite{GellMann1955},  a complex solution of Eq.~(\ref{eq:DE}) represents a fermion that has a distinct anti--fermion, with the same mass and spin but opposite charge and magnetic moment, as its counterpart. Therefore, the field of a relativistic fermion that coincides with its own antiparticle, should it exist, is necessarily an eigenstate of CC that must be described by a real solution $\chi(x)$ of Eq.~(\ref{eq:DE}). Real solutions of the Dirac equation are possible, provided one can find a suitable representation of the matrices $\gamma^{\mu}$characterized by purely imaginary non--zero matrix elements. Such a representation of the $\gamma^{\mu}$ matrices that renders Eq.~(\ref{eq:DE}) purely real was found by E. Majorana in 1937 \cite{Majorana1937}. The real solutions $\chi(x)$ correspond to charge--neutral fermions, known as Majorana fermions (MFs), representing particles that are their own antiparticles. Charge neutrality and the identification of the particle with its own anti--particle are not uncommon features among bosons - photons and $\pi^0$--mesons being two standard examples - but they are quite special among fermions. The neutron, for example, which is a charge--neutral fermion, has an anti--particle (the anti--neutron) that is distinguished from the neutron by the sign of its magnetic moment. Also, neutrinos produced in beta-decay are thought to be charge--neutral and have a small but non--zero rest mass (so that they cannot be Weyl fermions \cite{Weyl1929}, which are massless), but whether they are Dirac or Majorana fermions is still an unsettled question in particle physics.

While Majorana fermions have remained undetected in high energy physics for almost 70 years, in the last decade they have made an emphatic entrance into the realm of condensed matter physics \cite{Nayak2008,Wilczek2009,Beenakker2011,Alicea2012,Leijnse2012}. In this context, the term Majorana fermion does not refer to an elementary particle, but rather to \textit{quasiparticles} corresponding to collective excitations of an underlying complicated many--body ground state.  As shown by N. Read and D. Green~\cite{Read2000}, in the so--called weak--pairing phase of a two--dimensional (2D) spinless $p_x+ip_y$ superconductor (or superfluid),  the  quasiparticles satisfy an equation - the Bogoliubov--de Gennes (BdG) equation - that is similar to the Majorana form of the Dirac equation. These quasiparticles represent the solid state analog of the Majorana fermions from high--energy physics and, in general, are characterized by a finite mass and have finite energy. However, in the context of the recent developments, the term ``Majorana fermion'' has a slightly different meaning.  In a superconductor, the spinless fermionic Bogoliubov quasiparticles satisfy particle--hole symmetry, $\gamma_E^{\dagger}=\gamma_{-E}$, and, consequently, the zero--energy quasiparticles can be viewed as fermions that are identical with their own anti--quasiparticles,  $\gamma_0^{\dagger}=\gamma_{0}$. These BdG quasiparticles with zero excitation energy emerge as  \textit{localized} states bound to defects in the superconductor, where the order parameter amplitude vanishes, e.g., vortices and sample edges~\cite{Read2000}.  Furthermore, in two dimensions the quantum quasiparticles associated with these zero--energy bound states were shown to obey a form of quantum statistics known as non-Abelian statistics \cite{Moore1991,Nayak1996,Read2000,Ivanov2001,Stern2004}. Exchanging  two particles that obey non-Abelian statistics represents a non--commutative operation. Recently, the interest in the properties and the possible realization of non-Abelian zero--energy quasiparticles has increased dramatically, after they have been proposed~\cite{Kitaev2003} as possible building blocks for fault tolerant topological quantum computation (TQC)~\cite{Nayak2008}.

In this review,  the terms ``{\em Majorana fermion}", ``{\em zero--energy Majorana state}'', or ``{\em Majorana bound state}'' refer to localized, charge--neutral, zero--energy states that occur at defects and boundaries in topological superconductors. The creation operator for such a zero energy state is a hermitian second quantized operator $\gamma^{\dagger}=\gamma$ that anti--commutes with other fermion operators and satisfy the relation $\gamma^2=1$ . In a superconductor, a non--degenerate localized zero--energy eigenstate enjoys a form of \textit{topological protection} that makes it immune to any local perturbation that does not close the superconducting gap. Such perturbations cannot move the state away from zero energy because of the particle--hole symmetry and the non--degeneracy condition.
 Since small perturbations to the BdG differential equation are not expected to change the total number of solutions,  it necessarily follows that local perturbations (i.e., perturbations that do not couple pairs of MFs) leave the non--degenerate zero energy eigenvalues unperturbed.
 This argument also implies that the zero--energy solutions are characterized by vanishing expectation values for \textit{any} local physical observable, such as charge, mass, and spin. Otherwise, in the presence of a non--zero average, a local field that couples to the corresponding operator would shift the energy of the state. Thus, our use of the term ``Majorana fermion" is  more restrictive than that in the particle physics literature: the MFs in this review are charge--less, mass--less, spin--less, i.e.,  free of any internal quantum number, and obey non--Abelian statistics, while the MFs in particle physics satisfy standard Fermi statistics and, although charge--neutral, since they are eigenstates of the charge--conjugation operator, can be massive and spin-full . Finally, as zero energy MFs in solid state systems are topologically protected, they can be removed from zero energy only by driving the system through a topological quantum phase transition (TQPT)~\cite{Volovik1988} characterized by the closing of the energy gap at the topological critical point, where the MFs become entangled with other gapless states~\cite{Read2000}.

Majorana fermions have recently been proposed in low temperature solid--state systems in the context of fractional quantum Hall (FQH) effect \cite{Nayak1996,Read2000,Wen1993,Moore1991,Read1992,DSarma2005,Stern2006,Bonderson2006},  spinless chiral $p$-wave superconductors/superfluids \cite{Read2000,Kitaev2001,Ivanov2001,Stern2006}, in particular in strontium ruthenate, via the realization of the so--called half--quantum vortices \cite{DSarma2006,Chung2007,Jang2011}, heterostructures of topological insulators (TI) and superconductors~\cite{Fu2008,Fu2009,Fu2009a,Akhmerov2009,Linder2010,Cook2011}, metallic surface states~\cite{Potter2012}, metallic ferromagnet--cuprate high-T$_c$ superconductor~\cite{Takei2013}, non--centro--symmetric superconductors~\cite{Sato2009a,Ghosh2010,Tanaka2010}, superconductors with odd--frequency pairing~\cite{Tanaka2012,Asano2012}, helical magnets~\cite{Martin2012}, carbon nanotubes and graphene~\cite{Sau2011,Klinovaja2012,Klinovaja2012a,Klinovaja2013}, spin--orbit--coupled ferromagnetic Josephson junctions~\cite{Black2011a}, chains of magnetic atoms on a superconductor~\cite{Nadj2013}, and cold fermion systems with p--wave Feshbach resonance~\cite{Regal2003,Ticknor2004,Schunck2005,Tewari2007}, or with s--wave Feshbach resonance plus artificially generated spin--orbit coupling and Zeeman splitting~\cite{Zhang2008,Sato2009}. It has also been shown that MFs can exist as quasiparticles localized in the topological defects and at the boundaries of a spin--orbit (SO) coupled electron--doped semiconductor 2D thin film \cite{Sau2010b,Sau2010a} or 1D nanowire \cite{Sau2010a,Lutchyn2010a,Oreg2010} with proximity induced s--wave superconductivity and an externally induced Zeeman splitting. More recently, it was shown that similar mechanisms can lead to the emergence of MFs in hole--doped semiconductor structures~\cite{Mao2011,Mao2012}. Although the physics responsible for the emergence of the zero energy MFs is identical in both 2D and 1D systems (in 3D a similar mechanism produces a so--called Weyl superconductor/superfluid \cite{Gong2011,Sau2012a} characterized by  gapless, topologically protected, Weyl fermion nodes in the bulk), the nanowire setting has some experimental advantages, such as the option of generating the Zeeman splitting by using a parallel magnetic field \cite{Lutchyn2010a,Oreg2010} and a significantly enhanced gap (the so--called mini--gap) that protects the end--localized MFs from thermal effects \cite{Tewari2012}. In addition, the nanowires can be arranged in networks that allow the maniputation of MFs for TQC~\cite{Alicea2011,Halperin2012}. 

The 1D version of the semiconductor--superconductor heterostructure~\cite{Sau2010a,Lutchyn2010a,Oreg2010} -- the so--called semiconductor Majorana nanowire -- is a realization of the 1D topological superconductor (TS) model first proposed by Kitaev \cite{Kitaev2001} in the context of TQC.  In the presence of a small Zeeman splitting $\Gamma$, the Majorana nanowire with proximity induced $s$-wave superconductivity is in a conventional superconducting state with no MFs, while for $\Gamma$ larger than a critical value $\Gamma^c$, localized MFs exist at the ends of the quantum wire. 
We emphasize that, in the proposals for the realization of Majorana fermions using semiconductor--superconductor hybrid structures, the  Zeeman splitting plays the key role of lifting the degeneracy associated with the fermion doubling problem~\cite{Nielssen1983,Santos2010}. In practice, the required Zeeman splitting can be obtained either by applying an external magnetic field, or by proximity to a ferromagnetic insulator. We note that the realization of Majorana fermions in a 3D topological insulator--superconductor heterostructure does not require a Zeeman field, since the fermion doubling problem is avoided by the spatial separation between the surface states localized on opposite surfaces. Furthermore, the Zeeman splitting does not represent the only solution to the  fermion doubling problem~\cite{Santos2010}. For example, in the case of a topological insulator nanowire proximity coupled to a superconductor~\cite{Cook2011,Cook2012}  the cure of the fermion doubling problem is ensured by the orbital effect of a magnetic field applied along the wire, while the corresponding Zeeman splitting is negligible. More generally, each proposal for realizing MFs in solid state systems has to contain a specific solution to the  fermion doubling problem. 
Finally, we note that the MFs emerging in a Majorana nanowire can be detected by measuring the zero--bias conductance peak (ZBCP) associated with tunneling into the end of the wire~\cite{Sengupta2001,Tewari2008,Sau2010a,Law2009,Flensberg2010,Wimmer2011}, or by detecting the predicted characteristic fractional AC Josephson effect \cite{Kitaev2001,Kwon2004,Fu2009,Lutchyn2010a,Oreg2010,Jiang2011}. While these techniques have already been  implemented experimentally, measurements involving the direct observation of non--Abelian Majorana interference will ultimately be required to  validate the existence of the zero--energy Majorana bound states. 
In the last year, the semiconductor Majorana wire, which is the focus of this review, has attracted considerable attention as a result  of the recently reported experimental evidence for both the ZBCP~\cite{Mourik2012,Deng2012,Das2012,Churchill2013,Finck2013} and the fractional AC Josephson effect in the form of doubled Shapiro steps~\cite{Rokhinson2012}.

\section{Theoretical background}

\subsection{Spinless $(p_x+ip_y)$ superconductor/superfluid}

The spinless $(p_x+ip_y)$ superconductor (superfluid) is the canonical system that supports zero energy MFs localized at the defects of the order parameter,  such as vortices and sample edges~\cite{Read2000}. In 2D the mean field Hamiltonian for such a system is given by,
\begin{equation}
H_{2D}^{p} = \sum_{p}\xi_p c_p^{\dagger}c_p + \Delta_0 \sum_p\left[(p_x + ip_y)c_p^{\dagger}c_{-p}^{\dagger} + h.c.\right],
\label{eq:Hp}
\end{equation}
where $\xi_p=p^2/2m-\epsilon_F$, with $\epsilon_F$ the Fermi energy, and spin indices are omitted because the system is considered spinless (or spin-polarized). The lowest--energy solution of the BdG equations $H_{2D}^p\Psi(r)=E\Psi(r)$ near a vortex or near the sample edges (where the order parameter $\Delta_0$ vanishes), calculated under appropriate conditions, is non--degenerate and has zero energy ($E=0$), while the  corresponding second--quantized operator (the creation operator for the Bogoliubov state) is hermitian, $\gamma^{\dagger}=\gamma$. In 1D the zero energy MF solutions occur near the ends of the wire and, as argued by Kitaev \cite{Kitaev2001}, should be observable in a fractional AC Josephson effect--type experiment. Although a spinless p--wave superconductor does not exist in nature,  this example makes it clear that the three main ingredients that are important for realizing zero--energy MFs in a condensed matter system are superconductivity,  chirality (i.e., the $p_x+ip_y$ orbital form of the order parameter), and the spinless or spin--polarized nature of the system (which ensures that the zero energy solution localized at a defect of the order parameter is non--degenerate).

Despite the possibilities opened by superconducting strontium ruthenate~\cite{Mackenzie2003} and cold fermion systems in the presence of a p--wave Feshbach resonance~\cite{Regal2003,Ticknor2004,Schunck2005,Tewari2007}  of realizing the physical conditions necessary for the emergence of MFs, actually realizing and observing the MFs in these systems are challenging tasks. In strontium ruthenate, even if the required half--quantum vortices can be realized, the mini--gap $\sim \Delta^2/\epsilon_F \sim 0.1$mK (with $\epsilon_F$ the Fermi energy and $\Delta$ the magnitude of the p--wave superconducting order parameter) that separates the zero energy MF bound states from the higher energy regular BdG excitations also localized at the vortex cores is unrealistically small. On the other hand, in cold fermion systems with p--wave Feshbach resonance in the unitary limit, even if the mini--gap $\sim \Delta^2/\epsilon_F \sim \epsilon_F$ can be relatively large, the short lifetimes of the p--wave pairs and molecules represents a major experimental challenge.

The first major attempt to reproduce the essential physics of spinless p--wave superconductors using only s--wave  pairing was made in Ref.~[\onlinecite{Fu2008}], by using a heterostructure consisting of a 3D strong topological insulator and an s--wave superconductor. In the cold fermion context, Ref.~[\onlinecite{Zhang2008}] showed that (also see Ref.~[\onlinecite{Sato2009}]) an artificial, laser--induced, Rashba spin--orbit coupling and a Zeeman field, in conjunction with s--wave pairing interactions, can reproduce the topological superfluidity and MFs of spinless chiral p--wave superfluids using an s--wave Feshbach resonance. Later, the same approach that was used in the cold fermion context was applied to a solid state heterostructure where the spin--orbit coupling, Zeeman field, and s--wave superconductivity are provided by a semiconductor, a magnetic field or the proximity to a magnetic insulator, and the proximity to an s--wave superconductor, respectively. The proposed setups are shown in Fig. \ref{Fig2_1}. As discussed below, in the semiconductor--superconductor heterostructure the two main ingredients (in addition to superconductivity) that are essential for realizing MFs, i.e., chirality and spinlessness , are provided by a Rashba--type spin--orbit coupling and by restricting the low--energy physics of the semiconductor to a single (or an odd number of) relevant Fermi surfaces.

\subsection{Semiconductor--superconductor heterostructure}

The Rashba SO--coupled semiconductor (e.g., InSb, InAs) with proximity induced s--wave superconductivity and  Zeeman splitting
is mathematically described by the following Bardeen-Cooper-Schrieffer (BCS)-type
Hamiltonian:
\begin{equation}  \label{eq:polar_bulk_H}
H=(\eta \bm k^2-\mu)\tau_z + \Gamma \hat{\bm{n}}\cdot \bm\sigma+\frac{\alpha}{2}(\bm k\times \bm\sigma)\cdot\hat{\bm z} \tau_z+\Delta\tau_x,
\end{equation}
where $\eta = 1/m^*$, with $m^*$ being the effective mass of the charge--carriers, $\mu$ is the chemical potential measured from the bottom of a pair of spin sub--bands (in this review by ``band", we always mean a pair of a spin--split sub--bands, the band themselves being separated by the energy gaps due to lateral confinement), $\alpha$ is the Rashba spin--orbit coupling constant, and $\Delta$ is the s--wave superconducting pair--potential in the semiconductor, which is assumed to be proximity induced from an adjacent superconductor. In addition,
 $\hat{\bm{n}}$ is a suitably chosen direction of the applied Zeeman spin splitting given by $\Gamma = \frac{1}{2} g\mu_B B$ with $g$ the effective
Land$\acute{e}$ g-factor, $B$ the applied magnetic field and $\mu_B$ the Bohr magneton. Note that $H$ is written in terms of the $4$--component Nambu spinor $(u_\uparrow(\bm r),u_\downarrow(\bm r),v_\downarrow(\bm r),-v_\uparrow(\bm r))$, and that the Pauli matrices  $\sigma_{x,y,z},\tau_{x,y,z}$ act on the spin and particle-hole spaces, respectively. For a Zeeman splitting $\Gamma$ larger that a critical value, $H$ describes a 2D
topological superconductor when the direction of the Zeeman field is perpendicular to the plane containing the vector $\bm k=(k_x, k_y)$, i.e.,  for $\hat{\bm{n}}=\bm{\hat{z}}$.  A 1D topological superconductor with  $\bm k=k_x$ obtains for either $\hat{\bm{n}}=\bm{\hat{x}}$ or $\hat{\bm{n}}=\bm{\hat{z}}$. In general, a finite topological superconducting gap develops only if the component of the Zeeman field perpendicular to the effective k--dependent Rashba spin--orbit field  is larger than a critical value.

\begin{figure}[tbp]
\begin{center}
\includegraphics[width=0.4\textwidth]{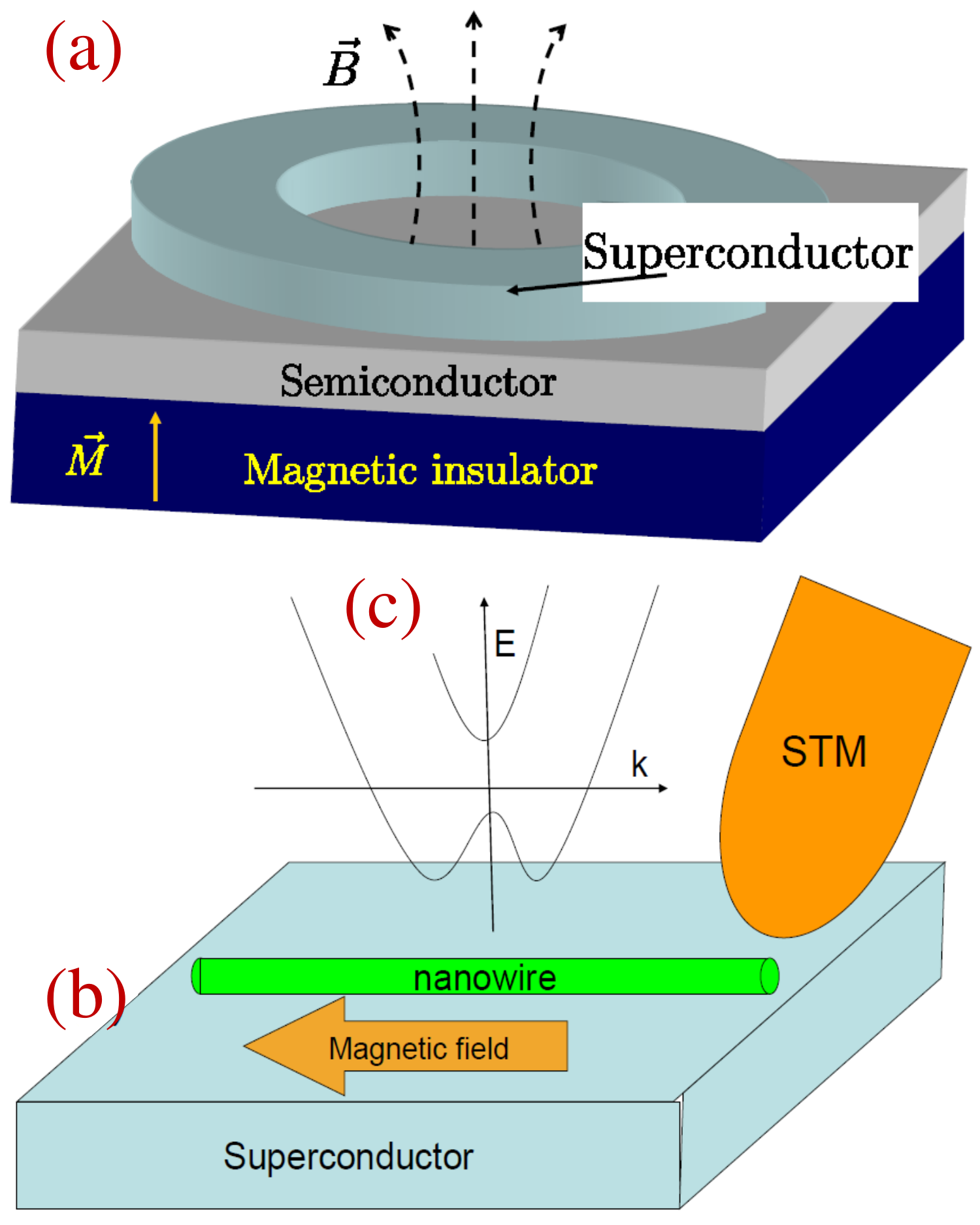}
\vspace{-3mm}
\end{center}
\caption{(Color online) Proposed setups for realizing MFs in semiconductors. (a) Two--dimensional semiconductor with Rashba SO--coupling  proximity coupled to a superconductor and a magnetic insulator.  The MFs are predicted to occur at the order parameter defects such as vortices. The required Zeeman field  oriented  perpendicular to the semiconductor ($\hat{\bm{n}}=\bm{\hat{z}}$ in Eq.~(\ref{eq:polar_bulk_H})) it proximity--induced by coupling the semiconductor to  a magnetic insulator. If the semiconductor has both Rashba and Dresselhaus SO couplings, the required Zeeman splitting can be parallel to the surface and can be induced by a parallel magnetic field~\cite{Alicea2010}.
(b) One dimensional setup with Zeeman splitting parallel to the nanowire ($\hat{\bm{n}}=\bm{\hat{x}}$ in Eq.~(\ref{eq:polar_bulk_H})) generated by an external magnetic field. The end--of--wire MFs can be probed as zero bias conductance peaks in a local tunneling measurement.  (c) The spectrum of the spin--split sub--bands (or the two top--most occupied sub--bands in a quasi--1D system) in the presence of Zeeman splitting. Figure adapted from Ref. [\onlinecite{Sau2010b}] (panel a) and Ref. [\onlinecite{Sau2010a}] (panels b and c).}
\vspace{-2mm}
\label{Fig2_1}
\end{figure}

The Hamiltonian in Eq.~(\ref{eq:polar_bulk_H}) has recently been studied extensively in the context of 2D, 1D and quasi--1D (multichannel) semiconductors \cite{Zhang2008,Sato2009,Sau2010b,Sau2010,Tewari2010,Alicea2010,Lutchyn2010a,Oreg2010,Sau2010a,Potter2010,Lutchyn2011a,Stanescu2011,Potter2011}. Two setups proposed for the experimental realization of the Majorana physics described by this Hamiltonian are shown in Fig. \ref{Fig2_1}.  The system is characterized by a topological quantum critical point (TQCP) that corresponds to the critical Zeeman field $\Gamma_c = \sqrt{\Delta^2 +\mu^2}$,  where the quantity $C_0=( \Delta^2 + \mu^2-\Gamma^2 )$ changes sign. For $C_0 > 0$, the (low--$\Gamma$) state is an ordinary, non--topological superconductor that includes perturbative effects from the Zeeman field and tha spin--orbit coupling but does not host MFs. However, for $C_0 < 0$, the (high--$\Gamma$) state has non--perturbative effects from $\alpha$ and can support zero--energy MF states localized at the defects of the pair--potential $\Delta$.
 Interestingly, the proximity--induced pair--potential $\Delta$ itself remains non--zero and continuous across the TQCP \cite{Sau2010a,Tewari2012b}, and, consequently,  the two superconducting states  break exactly the same symmetries, namely the gauge and time--reversal symmetries. In the absence of topological defects and boundaries, the single particle spectrum of the high--$\Gamma$ phase is similar to that of the low--$\Gamma$ phase, as both are completely gapped in the bulk. However, the high--$\Gamma$ topological state can be distinguished from the non--topological superconductor at $\Gamma < \Gamma_c$  by probing the topological defects and the boundaries, as they host MFs only for $\Gamma > \Gamma_c$.

The topological quantum critical point $\Gamma_c$ is marked by the vanishing of the single--particle minimum excitation gap $E_0$. By diagonalizing the Hamiltonian in Eq.~(\ref{eq:polar_bulk_H}) one  obtains the lower--branch of the quasiparticle excitation spectrum,
\begin{equation}
E_k^2=\Delta^2+\tilde{\epsilon}^2+r_{k}^{2}-2 \sqrt{\Gamma^2\Delta^2+\tilde{\epsilon}^2r_{k}^{2}},  \label{eq:spectrum}
\end{equation}
where $\tilde{\epsilon}=\eta k^2-\mu$ and $r_{k}^{2}=\Gamma^2+\alpha^2 k^2$. For $\Gamma$ near $\Gamma_c$, the
minimum of $E_k$ is at $k=0$, which corresponds to the minimum quasiparticle gap $E_0$,
\begin{equation}
E_0=|\Gamma-\sqrt{\Delta^2+\mu^2}|.
\label{eq:E0}
\end{equation}
Note that, $E_0$ vanishes exactly at $C_0=0$, marking the location of the TQCP as a function of $\Gamma$, $\Delta$, or $\mu$. Furthermore,
it has been shown that the quantity $C_0$ is the Pfaffian of the BdG Hamiltonian at $k=0$, $(C_0=Pf(H(k=0)\sigma_y\tau_y))$ \cite{Ghosh2010}. The sign of $C_0$, which determines whether the system has $0$ or $1$ MFs, is related to the Pfaffian topological invariant $Q$ (see Eq.~(\ref{eq:Q})) given by the product of the Pfaffians of the BdG Hamiltonian at $k=0$ and $k=\pi$ \cite{Kitaev2001}. Since the sign of the Pfaffian of the BdG Hamiltonian at $k=\pi$ is generically positive, the sign of $C_0$ by itself determines whether the state of the semiconductor hosts $0$ or $1$ MFs (or, in general, an even or odd number of Majorana bound states). The topological invariant of the semiconductor-superconductor heterostructure is discussed in more detail in section \ref{TopInv}.

\subsection{Topological class}

Recent work \cite{Schnyder2008,Kitaev2009,Ryu2010} has established that the quadratic Hamiltonians describing gapped topological insulators and superconductors can be classified into 10 distinct topological classes that can be characterized by certain topological invariants. The 2D spinless chiral $(p_x+ip_y)$ superconductor in the weak pairing  phase ($\mu>0$) belongs to the topological class D, whish is characterized by an integer $Z$ toplogical invariant. This implies that, under appropriate conditions, the system can support $Z$ number of chiral Majorana edge modes that remain protected against small perturbations. The analogous system of a 2D Rashba--coupled semiconductor with $\Gamma > \Gamma_c$ is also characterized by a $Z$ invariant and supports $Z$ gapless chiral Majorana edge modes with dispersion $E_{\rm edge}(k)=(\Delta/k_F) k$. Since the 2D system is in class D, does it automatically imply that the corresponding 1D or quasi-1D system is also in class D? The topological class of the 1D nanowire can be guessed from a heuristic dimensional reduction argument. The MF mode in a 1D wire (say, along the $x$-axis) can be viewed as the dimensionally--reduced version of the chiral Majorana edge mode of a 2D system on an edge parallel to the $y$ axis. From the mathematical equivalence, $H_{BdG}^{2D}(k_y = 0)=H_{BdG}^{1D}$, where $H_{BdG}^{1D}$ is the BdG Hamiltonian near an end of the wire, and the property $E_{\rm edge}(0)=0$,  it follows that  the end of the wire supports a zero energy eigenstate. Since in 2D there are  $Z$ allowed chiral edge modes, it follows that in 1D there must be, under appropriate conditions, $Z$ zero energy MFs. Thus, the nanowire should also be characterized by a $Z$ (not $Z_2$) invariant, and, consequently, the purely 1D Majorana nanowire should belong to the BDI  topological class.

To reveal the appropriate BDI classification for the nanowire and the associated $Z$ invariant, it is necessary to identify a hidden chirality symmetry  of the 1D system\cite{Tewari2012a,Tewari2012}. In 1D, the Hamiltonian in Eq.~(\ref{eq:polar_bulk_H}) anticommutes with a unitary operator ${\cal{S}}=\tau_x$,  where the chirality symmetry operator ${\cal{S}}$ can be written as the product of an artificial `time reversal' operator ${\cal{K}}$ and a particle--hole transformation operator $\Lambda=\tau_x\cdot{\cal{K}}$. Here, ${\cal{K}}$ is the complex conjugation operator.
The existence of all three symmetries - `time reversal', particle-hole, and chirality - ensures that the Hamiltonian is in the BDI symmetry class \cite{Schnyder2008,Kitaev2009,Ryu2010} characterized by an integer topological invariant. Under invariance of this symmetry, the strictly 1D nanowire can support an arbitrary integer number $N$ ($=Z$) of protected zero energy MFs at each end. The situation is similar to that of a 1D spinless p--wave superconductor that was shown~\cite{Niu2012} to support any integer number of MFs at \textit{each end}. However, for the semiconductor Majorana wire, the experimentally realistic case of a \textit{quasi}--1D system retains only an \textit{approximate} chirality symmetry (the chirality symmetry is weakly broken by the inter--band Rashba coupling, which no longer commutes with ${\cal{S}}$~\cite{Tewari2012,Lim2012}) and allows multiple \textit{near zero energy} modes at each end for values of the Zeeman coupling close to the confinement energy gap and above. The existence of the near zero modes suppresses the gap protecting the MF end modes in certain parameter regimes characterized by $\Gamma$ larger than the confinement energy ~\cite{Tewari2012,Lim2012}. Further consequences of the chirality symmetry for the electrical conductance are discussed in Ref. \cite{Diez2012}.

\subsection{Topological invariant}\label{TopInv}

 The topological class and topological invariant of the semiconductor-superconductor heterostructure is analogous to that of a spinless $p_x+ip_y$ superconductor. For the topological invariant of the 2D spinless $p_x+ip_y$ superconductor (which is in class D), we rewrite the Hamiltonian in Eq.~(\ref{eq:Hp})in the particle--hole basis $(c_{\mathbf{k}}^{\dagger},c_{-\mathbf{k}})$ as,
 \begin{equation}
 H_{2D}^{p}=\xi_{\mathbf{k}}\tau_z + \Delta_xk_x\tau_x -\Delta_yk_y\tau_y,
 \label{eq:Hp1}
 \end{equation}
where we have allowed for different pair potentials $\Delta_x,\Delta_y$ along the $x,y$ directions. Writing the Hamiltonian in terms of the Anderson
pseudo--spin vector~\cite{Anderson1958} $\vec{d}({\mathbf{k}})$ as  $H_{2D}^{p} (\mathbf{k})=\vec{d}(\mathbf{k}).\vec{\tau}$, one can observe that in $D=2$ all three components of $\vec{d}$ are non-zero. The topological invariant is an integer, as $Z$ is the relevant homotopy group $\pi_2(S^2)$ of the mapping from the 2D $k$ space to the 2--sphere of the 3--component unit vector $\hat{d}=\vec{d}/|\vec{d}|$ \cite{Volovik1988,Read2000}. On the other hand, in $D=1$, the corresponding Hamiltonian can be made purely real (e.g.,  $\Delta_y$ drops out from Eq.~(\ref{eq:Hp1}) if the system is along the $x$--axis) and the vector $\vec{d}$ has only two components. Now since the $k$-space is also one--dimensional, the topological invariant must again be in $Z$ (class BDI) since $\pi_1(S^1)=Z$. This invariant is simply the winding number,
\begin{equation}
N=\frac{1}{2\pi}\int_0^{2\pi} d\theta(k),
\label{eq:winding1}
\end{equation}
where $\theta(k)$ is the angle the unit vector $\hat{d}$ makes with, say, the $z$-axis in the $x-z$ plane. The winding number counts the number of times the 2--component vector $\hat{d}$ makes a complete circle as $k$ varies in the 1D Brillouin zone.

The topological invariant for the semiconductor-superconductor heterostructure can be defined in a similar way to the spinless chiral p--wave superconductor. One needs, however, to account for the higher dimensionality of the Hamiltonian matrix in Eq.~(\ref{eq:polar_bulk_H}) and the consequent generalization of the winding number invariant. It is clear that in $D=2$ the Hamiltonian in Eq.~(\ref{eq:polar_bulk_H}) cannot be made real because of the complex Rashba term. In contrast, in $D=1$ $H$ can be made purely real,
 but the components of the $\vec{d}$-vector in the $4\times4$ Hamiltonian are themselves $2\times2$ matrices.  More generally, the BdG Hamiltonian of a TS system in $D=1$, despite being real (thus preserving the chiral symmetry given by the operator ${\cal S}=\tau_x$), can be a large $2N\times2N$ square matrix. Using the general framework for chiral symmetric systems \cite{Ryu2002,Zak1989}, the integer topological invariant for this system can still be defined \cite{Tewari2012a,Tewari2012} by generalizing the concept of the $\vec{d}$-vector winding number for arbitrary dimensional matrices. Since, as discussed in the last subsection, $H$ anticommutes with the chirality operator ${\cal{S}}={\cal{K}}\cdot\Lambda=\tau_x$, $H$ can be expressed as a block off--diagonal matrix  in a basis that diagonalizes the unitary operator $\cal{S}$:
\begin{equation}
U \widetilde{H}(k) U^\dagger = \left(\begin{array}{cc}0&A(k)\\ A^T(-k) & 0\end{array}\right).
\label{eq:Off-Diag}
\end{equation}
Following Ref.~[\onlinecite{Tewari2012a,Tewari2012}] one can now define the variable,
\begin{equation}
z(k)=\exp(i\theta(k))=Det(A(k))/|Det(A(k))|,
\label{eq:Z}
\end{equation}
 and calculate the topological invariant,
 \begin{equation}
W=\frac{-i}{\pi}\int_{k=0}^{k=\pi} \frac{d z(k)}{z(k)},\label{eq:W}
\end{equation}
which is an integer,  including zero ($W\in Z$).

If the chirality symmetry is broken (it is weakly broken for quasi--1D multi--band wires with transverse Rashba coupling $\alpha_y \neq 0$), the number of exact zero energy MF at each end of the wire goes back to 0 or 1. The corresponding $Z_2$ topological invariant is
\begin{equation}
Q={\rm sgn}\left\{Pf[H(k=0)\sigma_y\tau_y]\times Pf[H(k=\pi)\sigma_y\tau_y]\right\},
\label{eq:Q}
\end{equation}
and the topological class for the quasi-1D system reduces from BDI to D. $Q = +1 (-1)$ represents the topologically trivial (non-trivial) state with 0 (1) MF at each end of the wire. In the presence of the chirality invariance in 1D, it can be shown that $Q$ gives the parity of the integer topological invariant $W$, in analogy to the 2D case where $Q$ for the semiconductor--superconductor heterostructure gives the parity~\cite{Ghosh2010} of the first Chern number topological invariant, which is an integer.

 \subsection{Minigap}
The minigap for topological superconductors is defined as the energy of the first excited regular fermion state (above the zero energy MF state) in either the bulk or the edge of the system. This gap is responsible for the thermal protection of the MFs and, consequently,  for the non--trivial physics associated with them, such as non--Abelian statistics, by protecting the Majorana bound state from mixing with regular fermion states above the minigap. In naturally occurring quasi--2D chiral--p wave superconductors, such as strontium ruthenate, even if the MFs are realized in the cores of the half--quantum vortices, the minigap $\delta\sim\Delta^2/\epsilon_F \sim 0.1$mK is unrealistically small. In chiral $p$-wave superfluids, potentially realizable using cold fermions with p--wave Feshbach resonance, this is less of a problem because in this case $\Delta \sim \epsilon_F$ and the minigap $\Delta^2/\epsilon_F$ is of order $\Delta$. In solid state hybrid structures, the minigap has to be large enough so that the experiments can access the MF physics at realistically achievable temperatures. This is one important experimental aspect in which the semiconductor nanowire heterostructure has significant advantage over the corresponding 2D systems.  In the 1D system, the minigap naturally scales with the induced superconducting gap (see equation (5)), while  this is generically not the case for the 2D systems (see, however, Ref.~[~\onlinecite{Sau2010}]).

The reason why the minigap in the nanowires is greatly enhanced over the 2D heterostructures can be understood from the dimensional reduction arguments. In going from the 2D plane to the 1D wire, as the width $L_y$ in the $y$-direction is reduced, the energies of the quantized chiral edge modes scale as $1/L_y$ (recall that $E(k_y)=(\Delta/k_F)k_y$), i.e. as the inverse of the confinement size of the wire in the transverse direction. Since $1/L_y$ diverges in the strict 1D limit $L_y \rightarrow 0$, it implies that the minigap becomes arbitrarily large, in fact of the order of the bulk quasiparticle gap $E_{0}$. In principle, the end--state MFs protected by such a large minigap can be probed in local tunneling experiments without having to worry about other low energy states with energies comparable to experimental temperatures.

In practice, for clean quasi--1D nanowires ($L_x \gg L_y > L_z$) with hard--wall confinement, the lowest excited state above the MF end state (for $\Gamma >\Gamma_c$, but less than the confinement energy gap) has an energy that is a significant fraction of $E_0$~\cite{Stanescu2011}. In quasi--1D wires with the Zeeman splitting comparable to the confinement energy gap, the minigap can be drastically suppressed due to the existence of multiple near zero energy modes at the same end, which is a manifestation of the approximate hidden chirality symmetry~\cite{Tewari2012}.

 \subsection{Phase diagram of semiconductor Majorana wire, sweet spots, and approximate chirality symmetry}

In a strictly 1D wire with proximity--induced superconductivity the Majorana--supporting topological phase emerges at Zeeman fields higher than the critical value~\cite{Sau2010a,Sato2009}
\begin{equation}
\Gamma_c = \sqrt{\Delta^2 + \mu^2}.  \label{Gammac}
\end{equation}
Consequently, for $\Gamma<\Gamma_c$ the superconducting phase is topologically trivial, while for $\Gamma>\Gamma_c$  the topological phase is realized.  How is this simple picture modified in a quasi--1D system with multiple occupied confinement bands? If the applied Zeeman field is much smaller than the spacings between the confinement bands, nothing much is changed, except that $\mu$ in Eq.~(\ref{Gammac}) has to be understood as the chemical potential measured from the bottom of the top--most occupied band (also called the Majorana band). However, the phase diagram is non--trivially modified  when the Zeeman field is comparable with or higher than the spacings between the confinement bands.

\begin{figure}[tbp]
\begin{center}
\includegraphics[width=0.48\textwidth]{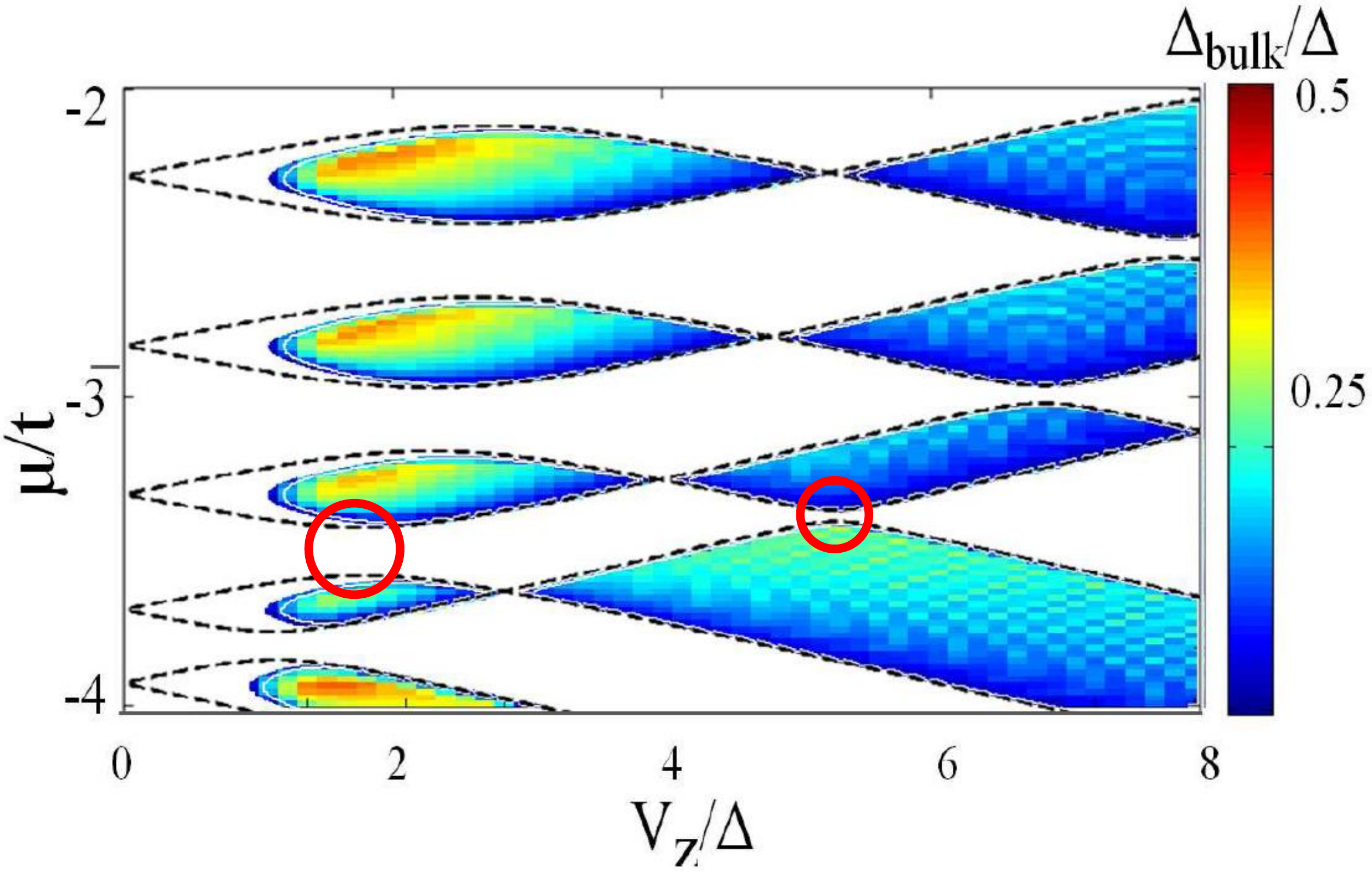}
\vspace{-7mm}
\end{center}
\caption{(Color online) Phase diagram as a function of chemical potential and Zeeman field oriented perpendicular to the SM--SC interface.   White regions correspond to the topologically trivial phase, while colored regions indicate topologically non--trivial SC phases. The dashed lines show the location of the normal state SM sub--bands. Note the characteristic anti--crossings between the sub--bands of the transverse $n_y$ bands with opposite parity, e.g., between $n_y$ and $n_y+1$. For a Zeeman field oriented along the wire (see Fig. \ref{Fig4_3}), these bands cross at the ``sweet spots'' (for example, the regions marked by red circles).
Figure adapted from Ref.~\cite{Potter2011}.}
\vspace{-6mm}
\label{Fig4_2}
\end{figure}

\begin{figure}[tbp]
\begin{center}
\includegraphics[width=0.48\textwidth]{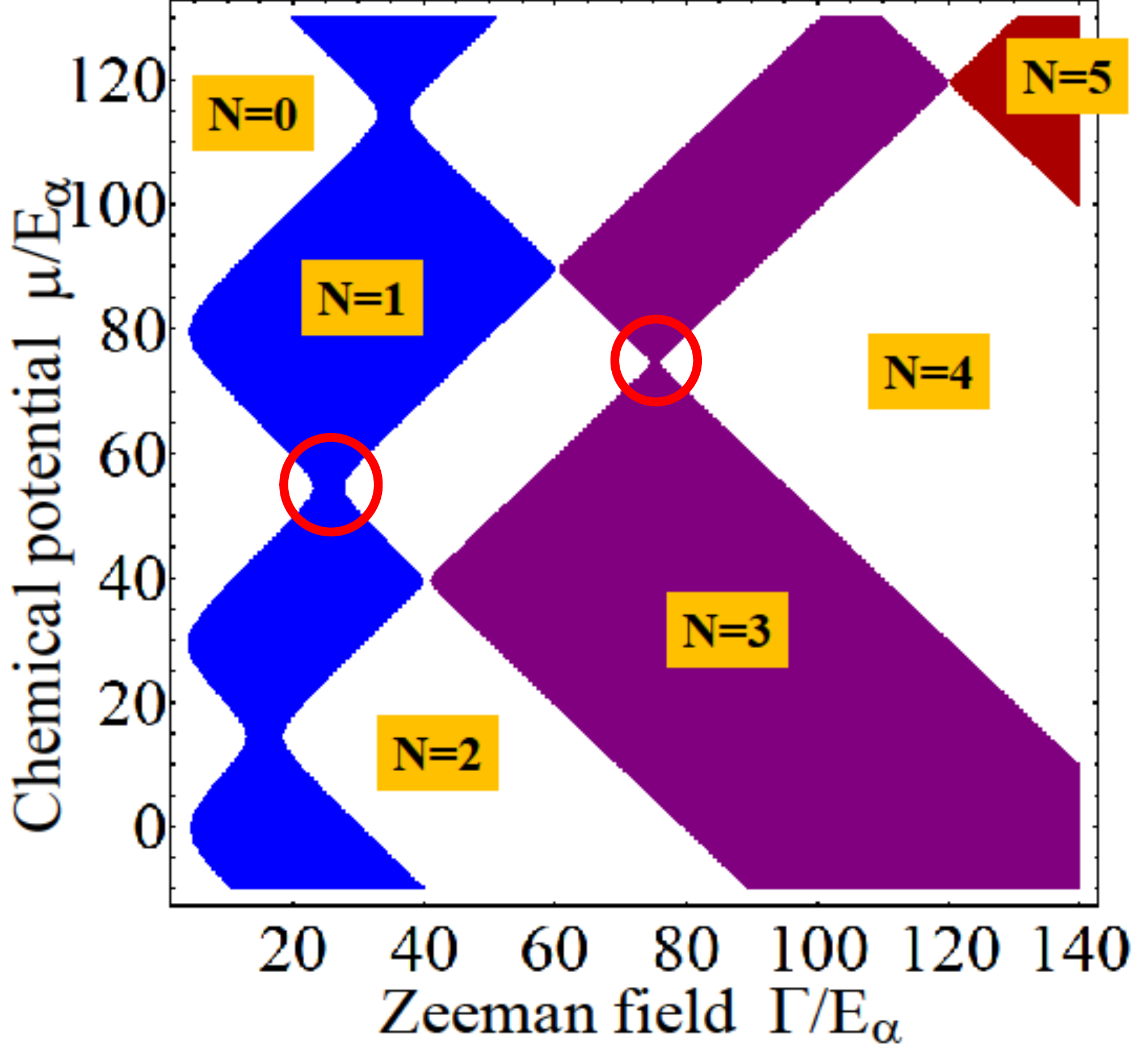}
\vspace{-7mm}
\end{center}
\caption{(Color online)
Phase diagram of the multiband nanowire with the Zeeman field oriented along the wire. Superconducting phases characterized by an odd (even) number $N$ of nearly--zero energy modes are topologically nontrivial (trivial). The phase boundaries correspond to topological quantum phase transitions characterized by the vanishing of the bulk quasiparticle gap. The energy unit is given by the spin--orbit characteristic energy $E_{\alpha}=m_{\rm eff} \alpha_R^2$.  Nonuniform SM-SC coupling induces inter--band pairing and stabilizes the topological phase near the sweet spots. Figure adapted from Ref.~\cite{Stanescu2011}.}
\vspace{-6mm}
\label{Fig4_3}
\end{figure}

Consider a finite--width wire oriented along the $x$ direction and proximity--coupled to an s--wave superconductor, with the interface perpendicular to the $z$ direction. The effective spin--orbit field generated by Rashba coupling is oriented along the $y$ direction. Two phase diagrams of this quasi--1D Majorana wire are shown in figures \ref{Fig4_2} and \ref{Fig4_3}.
The main difference between the two diagrams is due to the different orientations of the Zeeman field: in Fig. \ref{Fig4_2} the field is perpendicular to the semiconductor-superconductor interface, while in Fig.  \ref{Fig4_3} it is oriented along the wire.  In both cases the Zeeman field is perpendicular to the effective spin--orbit field generated by the Rashba coupling. Note that, when the magnetic field is parallel to the $z$ axis, the normal state spectrum is characterized by anti--crossings of the sub--bands corresponding to transverse $n_y$ modes with opposite parity (see Fig. \ref{Fig4_2}, red circles). By contrast,  when the magnetic field is parallel to the wire, these bands cross at the ``sweet spots''~\cite{Lutchyn2011a}, critical points where (in the absence of inter--band pairing) two topologically trivial and two topologically nontrivial phases meet.  We emphasize that inter--band pairing stabilizes the topological phase in the vicinity of the sweet spots (see  Fig. \ref{Fig4_3}; more details about sweet spot physics are provided in Appendix \ref{App0}). By contrast, rotating the field away from the $x$ direction favors the topologically trivial SC phase.

The phase diagram in Fig. \ref{Fig4_3} reveals that, upon increasing the Zeeman field, the system undergoes a series of topological quantum phase transitions as it traverses successive topologically trivial and nontrivial phases. These phases are characterized by a certain number $N$ of low--energy modes localized at each end of the wire. In the thin wire limit, $L_y\rightarrow 0$, these modes become zero--energy Majorana bound states. In a finite width wire, for $N$ even, the low--energy modes acquire a small non--zero gap, while for $N$ odd, $N-1$ modes become gaped and one mode (the Majorana mode) remains gapless. The opening of the small low--energy gap is due to the inter--band Rashba coupling~\cite{Lim2012,Tewari2012}, (see, for example,  the inter--band Rashba term $q_{n_y n_y^\prime}$ given by  Eq. (\ref{qnn} in Appendix \ref{App1}), that beaks the hidden chirality symmetry of the purely 1D model~\cite{Tewari2012a,Tewari2012}.

Finally, we note that in the limit of weak semiconductor--superconductor (SM--SC) coupling, the location of the phase boundaries is closely related to the non--superconducting SM spectrum~\cite{Gibertini2012}. However, the locations of the phase boundaries (in particular the boundary between the $N=0$ and $N=1$ phases) depends strongly on the coupling strength~\cite{Stanescu2011}. Specifically, from Eq. (\ref{Gammac}) the minimal Zeeman field necessary for reaching the topological phase in the single band case is $\Gamma_{c0}=\Delta$. This critical field corresponds to the transition between the $N=0$ and $N=1$ phases at low values of the chemical potential.  For a multi--band system, we have to take into account the band dependence of the induced gap $\Delta_{n n^\prime}$ and the proximity--induced energy renormalization $\widetilde{Z}_{nn^\prime}$ (see section \ref{Sec3_3} for details). Within the decoupled--band static approximation (valid when the effective SM--SC coupling is much weaker than the inter--band spacing and the SC bulk gap), we find that the critical Zeeman field required for entering the topological phase with $N=1$ is characterized by multiple minima $\Gamma_{cn}=\gamma_{nn}$ corresponding to values of the chemical potential at the bottom of each band $n$. Here, $\gamma_{nn}$ is the effective SM--SC coupling corresponding to band $n$. Consequently, increasing the coupling strength will move that phase boundary toward higher Zeeman fields. In addition, the presence of proximity--induced inter--band pairing widens the topological phase near the sweet spots and further reduces the dependence of $\Gamma_{c}$ on $\mu$~\cite{Stanescu2011}. Experimentally, this may results in an apparent insensitivity of the critical magnetic field associated with the emergence of the Majorana bound state on the chemical potential. In addition, the inter--band pairing arising from a non uniform coupling with the bulk superconductor results in a phase diagram characterized by a non-simply connected structure~\cite{Stanescu2011}, which allows  topological adiabatic pumping~\cite{Gibertini2013}.

\subsection{Probing Majorana fermions I: zero bias conductance peak and fractional AC Josephson effect}

The simplest way to establish the possible presence of a Majorana bound state at the end of a spin--orbit coupled semiconductor wire with  proximity-- induced superconductivity is by performing a tunneling spectroscopy measurement. Since the MF state is a zero energy bound state
localized at the end of the wire,  it is expected to produce a zero bias conductance peak in the tunneling
conductance $dI/dV$, similar to other zero energy boundary states in superconductors \cite{Ekin1997,Wagenknecht2008}. Here, $I$ is the current from the lead to the Majorana wire and $V$ is the applied voltage difference.
The zero bias conductance peak associated with tunneling into MFs is a result of resonant \textit{local} Andreev reflection of the lead electrons
 at the lead--nanowire interface \cite{Law2009}. In a local Andreev process, the MF localized at the other end of the wire plays no role, provided the wire is long enough so that the two MF wavefunctions do not overlap. If there is a significant overlap between the MF wavefunctions,  two other processes can also contribute to the transport current. One process corresponds to the direct transport of electrons via the conventional (Dirac) fermion state formed by the two overlapping MFs \cite{Semenoff2006,Bolech2007,Tewari2008}. The second process is the crossed Andreev reflection, in which an incident electron from a lead coupled to one end of the wire is followed by a hole ejected into the lead coupled to the other end \cite{Nilsson2008}. These two processes contribute to the non--local conductance or transconductance through the Majorana wire that we discuss in the next subsection. We note that the local Andreev reflection process typically gives rise to a non--zero sub--gap conductance even in the absence of MFs at the wire ends \cite{Blonder1982}. To suppress this background conductance, a gate--induced barrier potential may be applied at the lead--nanowire interface.

\begin{figure}[tbp]
\begin{center}
\includegraphics[width=0.48\textwidth]{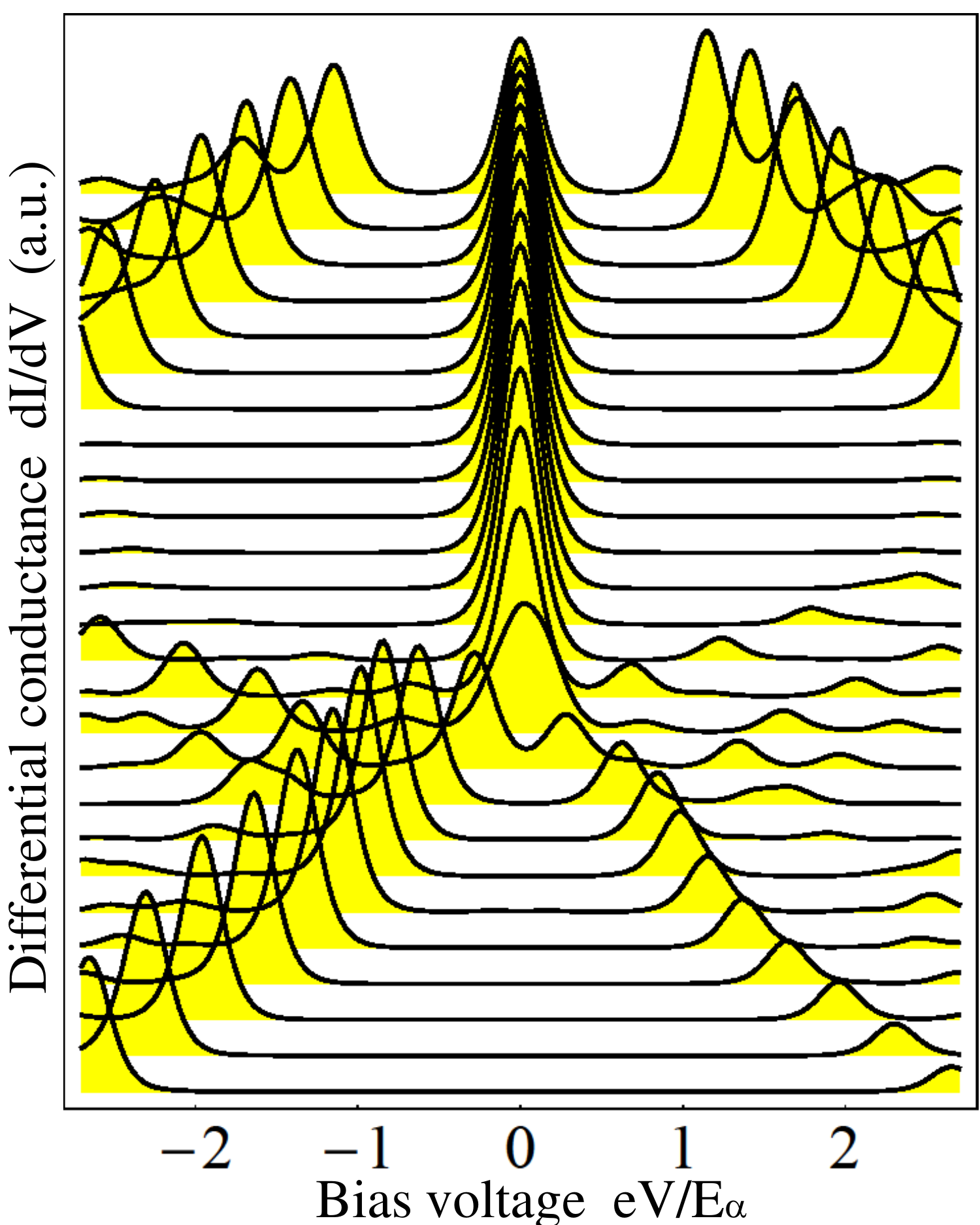}
\vspace{-7mm}
\end{center}
\caption{(Color online) Predicted differential conductance  for tunneling into the end of a semiconductor nanowire coupled to a superconductor. The lines represent cuts (shifted for clarity) corresponding to different values of the Zeeman field. As the field is increased, the superconducting quasiparticle gap closes at the critical Zeeman field $\Gamma_c$ that marks the topological quantum phase transition. By further increasing the Zeeman field, the presence of a zero energy MF at the wire end is revealed as a pronounced zero bias conductance peak. In this calculation, the chemical potential is set near the middle of an inter--band gap, close to a sweet spot (see Fig. \ref{Fig4_3}). For values of the chemical potential close to the bottom of a band, the signature associated with the gap closing at the TQPT is strongly suppressed, although the presence of a zero bias peak still reveals the MF in the topological phase (for details see Sec. \ref{Sec5_2}, in particular Fig. \ref{Fig5_3}) . Figure adapted from Ref.~\cite{Stanescu2011}.}
\vspace{-5mm}
\label{Fig2_2}
\end{figure}

The presence of a zero bias conductance peak represents a necessary, but not sufficient, condition for the existence of MFs that was theoretically proposed \cite{Sengupta2001,Tewari2008,Sau2010a,Law2009,Flensberg2010,Wimmer2011,Stanescu2011,Qu2011,Fidkowski2012,Sau2011b} (see Fig. \ref{Fig2_2}) and experimentally observed in recent measurements on quasi--1D semiconductor--superconductor hybrid structures \cite{Mourik2012,Deng2012,Das2012}. The main problem concerning the unambiguous identification of such a zero bias peak with end--localized MFs is that other sources of zero-- or low--energy states localized at the wire--boundary \cite{Kells2012,Liu2012} may produce similar zero--bias peaks. A  more conclusive signature of MFs in a charge transport measurement is the quantized value of the $T=0$ zero bias conductance \cite{Law2009}. This value should be $2e^2/h$ for tunneling into a non--degnerate MF at the wire end, zero for the wire in the topologically trivial phase with no low energy end state, and $4e^2/h$ in the presence of a conventional (non--Majorana) zero energy state at the wire end. Such a conductance quantization, which can be taken as a smoking gun signature of MFs in semiconductor Majorana wires, was not yet observed and may be difficult to obtain under realistic experimental conditions~\cite{Lin2012,Lutchyn2013}.

Another proposal for detecting  the existence of MFs in semiconductor wires involves the Josephson effect. 
In a Josephson junction (JJ) between two topologically trivial s--wave superconductors maintained at a phase difference $\phi$, the Josephson supercurrent $I$ is related to $\phi$ as $I=I_c \sin\phi$, where $I_c$ is the Josephson critical current.  In a Josephson junction between topological superconducting wires with end--state MFs, this relation changes to $I=I_c\sin\phi/2$~\cite{Kitaev2001,Kwon2004,Fu2009}. Consequently, if the phase difference $\phi$ is controlled by a magnetic flux threaded through a loop containing the Josephson junction, the periodicity of the Josephson current  changes from $2\pi$ to $4\pi$ as the semiconductor Majorana wire is tuned through the TQPT~\cite{Lutchyn2010a,Oreg2010}. In practice, the doubling of the period of the Josephson effect can be measured by applying a small voltage $V$ across the junction and measuring the AC Josephson effect. The frequency of the AC Josephson effect in the TS phase, $eV/\hbar$, is half the conventional Josephson frequency, $2eV/\hbar$, obtained in the topologically trivial superconducting  phase. Note that the use of the AC Josephson effect is necessary to avoid the inelastic relaxation between the two branches of the local Andreev bound states involved in the Josephson effect\cite{Kwon2004}. 

There are several ways to understand the change in the periodicity of the AC Josephson effect in the presence of MFs. Mathematically, it can be understood as a change in the dependence of the energies of the Andreev bound states at the junction on the phase difference $\phi$. While in the absence of the MFs the energy of the Andreev bound state goes as $E\propto \cos \phi$, with MFs this formula changes to $E\propto \cos \phi/2$. Since the Josephson supercurrent $I$ is related to the Andreev bound state energy as $I=(2e/\hbar)dE/d\phi$, it follows that the period of the supercurrent, $I \propto \sin \phi/2$, is doubled in the presence of the MFs. In order for this period doubling to be observable, it is necessary for the two branches of the Andreev bound states, $\pm E$, to maintain their occupation number as $\phi$ is varied adiabatically. For DC Josephson effect this is impossible at thermodynamic equilibrium due to quasiparticle poisoning effect (the higher (lower) energy state is unoccupied (occupied)) \cite{Kwon2004}. In AC Josephson effect, provided the period of the Josephson oscillations is shorter than the inelastic relaxation between the two branches $\pm E(\phi)$, the Josephson period doubling may be observable via the fractional frequency $eV/\hbar$. 

The doubling of the Josephson period in $\phi$ can also be understood from purely topological considerations. The fermion parity of the ground state of the topological superconductor (which is in the form of a ring with a weak link in the Josephson set up) flips as  the threaded flux changes by $2\pi$. With a $2\pi$ flux change the superconductor is thus in an excited state with an electron from the ground state ejected to the junction and held between the two MF states. Only with a total flux change of $4\pi$ the ground state Fermion parity flips back and the electron can leave the state formed by the MFs. This makes the junction return to the original state and consequently doubles the period of the Josephson effect from $2\pi$ to $4\pi$ \cite{Read2000,Kitaev2001}.

Even though the fractional Josephson effect is a robust signature of MFs, such an effect cannot be ruled out in ballistic S--N--S junctions and in JJs made of 1D p--wave superconducting wires such as the quasi--1D organic superconductors \cite{Kwon2004}. However, in fermion parity protected superconductors, the MF mediated fractional Josephson effect is topologically protected while there is no such robustness  (for instance, to disorder in the ballistic S--N--S junctions, or to magnetic fields in p--wave wires) in the other two systems.  Nevertheless, the fractional Josephson effect in semiconductor Majorana wires is susceptible to the quasiparticle poisoning effect \cite{Kwon2004}, as well as to the non--adiabaticity effects~\cite{Sau2012b}. The latter effect, in particular, shows that the fractional Josephson effect cannot be taken as an unambiguous signature of MFs because half or other fractional frequencies are in principle possible even in junctions between conventional superconductors. For recent work on the fractional Josephson effect in the limits of strong tunneling and long junctions, see Refs. \cite{Nogueira2012} and  \cite{Beenakker2013}.
Experimental evidence of fractional AC Josephson effect in the form of doubled Shapiro steps  has been reported in semiconductor Majorana wires~\cite{Rokhinson2012}. In this review we will not further discuss this effect, but will focus instead on  experiments  targeting the zero bias conductance peak.

\subsection{Probing Majorana fermions II: quantum non-locality, transconductance, and interference}

Neither the non--quantized zero bias conductance peak, nor the fractional AC Josephson effect could constitute a sufficient proof for the existence
of MFs in semiconductor--superconductor heterostructures, as in principle they can arise even in the absence of MFs. However, a conclusive
experimental proof could be obtained by making use of the intrinsic non--local properties of these states. In MFs, non--locality stems from the absence of an occupation number associated with them individually. To define the electron occupation number in terms of  spatially separated MFs, one must consider a pair of MFs, $\gamma_a, \gamma_b$, and define the second quantized electron creation operator as $d^\dagger=\gamma_a+i\gamma_b$.
The quantum state of the system is then determined by the eigenvalues of the electron occupation number operator $n_d=d^\dagger d= 0,1$.  Since $n_d$ is related to the MF operators by
\begin{equation}
n_d=\frac{1+i\gamma_a\gamma_b}{2},
\end{equation}
it follows that the state of the whole system is determined by non--local correlations between the  spatially separated MFs $\gamma_a$ and $\gamma_b$. An idea to probe this non-locality involves  injecting an electron into one end of the Majorana wire and retrieving it at the opposite end. By connecting  leads to the left and the right ends of a wire that hosts MF states localized near the two ends, one could imagine that an electron injected into the end $a$ flips the occupation number $n_d$ from $n_d=0$ to $n_d=1$. The injected electron can then escape from the end  $b$, flipping the occupation number of the nanowire state back to $n_d=0$. Such a process, where an electron can enter from one end and exit at the other end \textit{as an electron}, can be viewed as  Majorana--assisted electron transfer. As has been discussed before \cite{Semenoff2006,Bolech2007,Tewari2008}, such a transfer should not violate the
locality and causality principles. In other words, in the absence of an overlap between the two MF wavefunctions, the probability of an electron appearing at end $b$ becomes completely independent of the lead--wire voltage difference applied at the opposite and~\cite{Tewari2008}. Hence, there is no violation of causality.  
The non--local conductance (transconductance), given by $dI_b/dV_a$, can only be non--zero when there is a finite overlap between the two MF wavefunctions. However, in addition to the Majorana--assisted electron transfer, a second source \cite{Nilsson2008} for non--zero transconductance is represented by the crossed Andreev reflection (CAR), also known as Cooper pair splitting. In this process an electron injection into a Majorana bound state at one end is followed by the emission of a hole from a second Majorana state at the other end. The net result is the injection of a Cooper pair in the topological superconductor nanowire. The CAR contribution to transconductance, which is opposite in sign to the contribution arising from Majorana assisted electron transfer, is also non--zero only when the wavefunctions of the two MF states overlap \cite{Nilsson2008}. It has been shown that \cite{Sau2012c}, in the case of symmetric tunneling between the two leads and the MFs ($t_a=t_b=t$), the transconductance in the semiconductor Majorana wire is given by,
\begin{equation}
\frac{dI_b}{d V_a}=\delta \frac{32 V_a}{16 \Gamma^2+(\delta^2-V_a^2)^2+8\Gamma^2(\delta^2+V_a^2)}.
\label{Paraconductance}
\end{equation}
Here, $V_a$ is the voltage at end $a$, $I_b$ is the current at end $b$, $\Gamma\propto t^2$ is the lead--induced broadening
of the MF level, and $\delta$ is the overlap integral between the two MF wavefunctions. Note that the transconductance vanishes in the limit $\delta \rightarrow 0$, which is consistent with  earlier results~\cite{Bolech2007,Tewari2008,Nilsson2008}. Even when the direct wavefunction overlap of the MFs is vanishingly small, an effective coupling between the MFs, and consequently a non-zero Majorana assisted electron transfer amplitude, may be present if the topological superconductor has an appreciable charging energy \cite{Fu2010}. In this case, both the Majorana assisted electron transfer and CAR can result in a non--zero transconductance between the leads. Recently, the shot noise and the current--current correlations due to CAR have been proposed as possible experimental signatures of end state MFs~\cite{Liu2012a}. 

Even the MF induced transconductance, while interesting and nontrivial, cannot be considered as a definitive signature of MF modes because conventional near--zero energy states (such as those produced by localized impurities) trapped near the contacts with the leads can also produce such non--local signatures in the presence of superconductivity \cite{Sau2012c}. On the other hand, a non--local tunneling spectroscopy \textit{interference} experiment, similar to earlier interference based proposals in topological insulators and superconductors \cite{Akhmerov2009,Fu2009a,Sau2011a,Fu2010}, has recently been suggested as capable to provide a direct verification of the non--local physics of 1D wires containing end MFs \cite{Sau2012c}. The effect requires non--local fermion parity~\cite{Turner2011}, which is unique to topological systems and cannot be emulated by conventional near--zero--energy Andreev states or any other local excitations near the wire ends. The proposed scheme, where the tunneling
amplitude is measured as a flux dependence of an energy level and the fermion--parity is fixed by a superconducting single--electron transistor configuration, is suitable for unambiguous experimental testing of the presence of MFs in semiconductor Majorana wires. We note that other methods for detecting MFs have also been proposed~\cite{Bermudez2009,Bermudez2010,Sticlet2012,Chevallier2012,Mei2012,Appelbaum2013,Nadj2013,Pachos2013}.

\section{Realistic modeling of semiconductor Majorana wires}\label{Sec3}

 The construction of the effective low--energy model for a Majorana hybrid structure involves three main steps: i) developing a tight--binding model for the component that provides spin--orbit coupling (e.g., semiconductor, topological insulator, etc.) and projecting onto a reduced low--energy subspace, ii) incorporating the superconducting proximity effect, and iii) defining an effective Hamiltonian based on a linear approximation for the frequency--dependent proximity--induced self--energy. Below, we provide the relevant details and point out the main approximations involved in this construction. In addition, we discus several aspects of Majorana--supporting hybrid nanostructures that are not captured by simple
models of ideal 1D systems, but represent critical components of the experimental realizations of Majorana nanowires, such as the presence of disorder and smooth confining potentials.

The existence of Majorana fermions in an ideal topological SC system does not depend on the details of the Hamiltonian. However, under realistic laboratory conditions, the stability of the Majorana mode and the low--energy phenomenology of the heterostructure depend critically on a large number of specific  parameters, including details of the  electronic band structure,  nanostructure size and geometry, coupling at the interface, type and strength of disorder, and applied gate potentials. To account for the complex phenomenology of a Majorana--supporting structure and make connection with experimental observations, it is key to incorporate these factors in the theoretical model. The general form of a Hamiltonian that describes a semiconductor--superconductor (SM--SC) hybrid structure is
\begin{equation}
H_{\rm tot} = H_{\rm SM} + H_{\rm int} + H_{\rm Z} + H_{\rm V} + H_{\rm SC} + H_{\rm SM-SC}, \label{Htot}
\end{equation}
where $H_{\rm SM}$ is a non--interacting model for the semiconductor component (or, in general, other spin--orbit coupled material, e.g., topological insulator), $H_{\rm int}$ contains many--body electronic interactions, $H_{\rm Z}$ describes the applied Zeeman field, $H_{\rm V}$ contains terms that account for disorder and gate potentials, $H_{\rm SC}$ is the Hamiltonian for the superconductor, and $H_{\rm SM-SC}$ describes the semiconductor--superconductor coupling. Here, we will not address the problem of electron--electron interaction and will assume that $H_{\rm int}=0$. However, we emphasize that Coulomb interaction plays an important role in low--dimensional systems, as it generates a density--dependent renormalization of  the electrostatic potential. Consequently, the Zeeman field $\Gamma$ and the chemical potential $\mu$ of a Majorana wire are not independent variables\cite{DSarma2012} and, for a given configuration of external gate potentials, the system is characterized by a specific function $\mu=\mu(\Gamma)$.  This has important experimentally observable consequences\cite{DSarma2012}, as we will discuss in Section \ref{Sec5}.  At the Hartree level, the effects of Coulomb interaction can be incorporated using a self--consistent scheme similar to that described in Ref. \cite{Galanakis2012}. 
Below, we derive the low--energy effective model for the heterostructure described by Eq. (\ref{Htot}) under the assumption $H_{\rm int}=0$.

\subsection{Tight--binding models for semiconductor nanowires}\label{Sec3_1}

The first element in the development of a low--energy effective theory for a Majorana--supporting hybrid structure is a tight--binding model for the spin--orbit coupled component of the system (e.g., electron--doped SM, hole--doped SM, topological insulator, etc.) corresponding to $H_{\rm SM}$ in Eq. (\ref{Htot}). In general, when choosing the model, one has to strike a balance between accuracy and simplicity and one has to consider two key aspects: i) the dimensional reduction from three dimensions (3D) to quasi--2D or quasi--1D, and ii) the projection onto a reduced low--energy subspace. To illustrate some of the possible issues, we consider the case of electron--doped and hole--doped semiconductors.

The simplest tight--binding model for an electron--doped wire (or a thin film) is a two--band model with nearest neighbor hopping. The natural starting point for constructing such a model is a 3D Hamiltonian for conduction electrons in the effective mass approximation,
\begin{equation}
{\cal H}_0({\bm k}) = \sum_{{\bm k}, \sigma} \left(\frac{\hbar^2 k^2}{2 m^*}-\mu\right)c_{{\bm k}\sigma}^\dagger c_{{\bm k}\sigma},  \label{H0c}
\end{equation}
where $c_{{\bm k}\sigma}$ is the annihilation operator for a particle with wave vector  ${\bm k}$ and spin $\sigma$,  $m^*$ the effective mass of the conduction band, and $\mu$ is the chemical potential. In a thin film, the motion in the transverse direction is quantized and the transverse modes can be obtained by solving the quantum problem involving the single--particle Hamiltonian corresponding to Eq. (\ref{H0c}) with the substitution $k_z\rightarrow - i \partial_z$ and the confining potential $V(z)$. In the presence of a transverse field that breaks inversion symmetry, the spin and orbital degrees of freedom are coupled and the effective Rashba--type spin--orbit interaction (SOI) is described by
\begin{equation}
{\cal H}_{\rm SOI}({\bm k}) =\alpha_R \sum_{\bm k} c_{{\bm k}}^\dagger(k_y \sigma_x - k_x\sigma_y)c_{{\bm k}},  \label{HSOIc}
\end{equation}
where ${\bm k} = (k_x, k_y)$, $\alpha_R$ is the Rashpa coefficient, $\sigma_i$ are Pauli matrices, we have used the spinor notation $c_{{\bm k}}^\dagger = (c_{{\bm k}\uparrow}^\dagger, c_{{\bm k}\downarrow}^\dagger)$, and we have assumed that only one transverse mode is relevant to the low--energy physics. The physics described in the long wavelength limit by the Hamiltonian ${\cal H}_0 + {\cal H}_{\rm SOI} + {\cal H}_{\rm V}$, where  ${\cal H}_{\rm V}$ represents the contribution from the confining potential, can be also determined using a tight--binding model defined on a lattice. For example, considering a simple cubic lattice with lattice constant $a$ and nearest--neighbor hopping, the semiconductor Hamiltonian reads
\begin{eqnarray}
H_{\rm SM} &=& H_0 +H_{\rm SOI} = -t_0\sum_{{\bm i}, {\bm \delta}, \sigma}c_{{\bm i}\!+\!{\bm \delta}\sigma}^{\dagger}c_{{\bm j}\sigma} -\mu \sum_{{\bm i}, \sigma} c_{{\bm i}\sigma}^{\dagger}c_{{\bm i}\sigma}  \nonumber \\
&+& \frac{i \alpha}{2}\sum_{{\bm i},{\bm \delta}}\left[ c_{{\bm i}+{\bm \delta}_x}^{\dagger}{\sigma}_y c_{{\bm i}} -  c_{{\bm i}+{\bm \delta}_y}^{\dagger}{\sigma}_x c_{{\bm i}} + {\rm h.c.} \right],  \label{HSM}
\end{eqnarray}
 where $H_0$ includes the first two terms and describes nearest--neighbor hopping on a simple cubic lattice and the last term represents the Rashba spin-orbit interaction. Here, ${\bm i}=(i_x, i_y,i_z)$ labels the lattice sites, ${\bm \delta}\in\{{\bm \delta}_x, {\bm \delta}_y, {\bm \delta}_z\}$ are nearest--neighbor position vectors. The confining potential involves the additional local term $H_{\rm V} = \sum_{{\bm i}, \sigma} V({\bm i})c_{{\bm i}\sigma}^{\dagger}c_{{\bm i}\sigma}$. The parameters of the tight--binding Hamiltonian (\ref{HSM}) are determined by the condition that $H_{\rm SM}$ and ${\cal H}_0 + {\cal H}_{\rm SOI} $ describe the same physics in the long--wavelength limit ${\bm k}\rightarrow 0$. Consequently, the hopping parameter in  (\ref{HSM}) is  $t_0 = \hbar^2a^{-2}/2 m^*$ and the Rashba coupling is $\alpha=\alpha_R/a$.

For hole--doped SMs, an effective tight--binding Hamiltonian can be obtained using a similar approach and starting from the Luttinger 4--band model~\cite{Luttinger1955,Luttinger1956},  ${\cal H}_0({\bm k}) = \sum_{\bm k}c_{\bm k}^\dagger[h_L({\bm k}) -\mu] c_{\bm k}$, where $c_{\bm k}$ and $c_{\bm k}^\dagger$ are four--component spinors. The single--particle Luttinger Hamiltonian is
\begin{eqnarray}
 h_L({\bm k}) &=& \frac{-\hbar^2}{m_0}\left[\frac{2\gamma_1+5\gamma_2}{4}{k}^2\right. \label{hL} \\
 &-&\left.\gamma_2 \sum_ik_i^2 {J}_i^2 - \gamma_3\sum_{i\neq j} k_i k_j {J}_i {J}_j\right],   \nonumber
\end{eqnarray}
where $m_0$ is the free electron mass, ${\bm k} = (k_x, k_y, k_z)$ is the wave--vector, ${\bm J} = ({J}_x, {J}_y, {J}_z)$ is a set of $4\times 4$ matrices representing  angular momentum $3/2$, and $\gamma_1$, $\gamma_2$, and $\gamma_3$ are the Luttinger parameters. We assume that the dominant contribution to the valence band comes from p--like orbitals, $|X\rangle$,  $|Y\rangle$, and  $|Z\rangle$, and that in the presence of SOI the top valence band corresponds to the eigenstates of the total angular momentum momentum ${\bm J}$ with $j=3/2$,
\begin{eqnarray}
\Phi_{\pm\frac{1}{2}} &=& \frac{1}{\sqrt{6}}(\mp |X\rangle - i |Y\rangle)\chi_\mp + \sqrt{\frac23}|Z\rangle\chi_\pm,  \nonumber \\
\Phi_{\pm\frac{3}{2}}  &=& \frac{1}{\sqrt{2}}(\mp|X\rangle - i |Y\rangle)\chi_\pm,  \label{Phim}
\end{eqnarray}
where $m=\pm 1/2, \pm 3/2$ are the eigenvalues of $J_z$ and $\chi_\pm$ are spin eigenstates with $S_z=\pm1/2$. The effective tight--binding model is constructed using the eigenstates (\ref{Phim}) as a  basis. More specifically, we consider an $fcc$ lattice and nearest--neighbor hopping between the states $\Phi_m({\bm i})$ and $\Phi_{m^\prime}({\bm j})$, where ${\bm i}$ and ${\bm j}$ are nearest neighbor sites, with  hopping matrix element $t_{\bm i j}^{m m^\prime}$. The tight--binding Hamiltonian has the form
\begin{equation}
H_0 = \sum_{m, m^\prime}\sum_{\bm i, j}t_{\bm i j}^{m m^\prime}c_{{\bm i} m}^\dagger c_{{\bm j} m^\prime}-\mu \sum_{{\bm i}, m} c_{{\bm i} m}^\dagger c_{{\bm i} m}, \label{H0L}
\end{equation}
where $c_{{\bm i} m}^\dagger$ is the creation operator for the state  $\Phi_m({\bm i})$.
If we considering the $xy$ plane perpendicular to the $(0,0,1)$ crystal axis,  each site has four in--plane and eight out--of--plane nearest--neighbors. The corresponding hopping  matrix elements can be expressed in terms of three independent parameters, $t_1$, $t_2$, and $t_3$.  For example,  the in--plane diagonal components are $t_{{\rm in}}^{\pm 3/2 \pm 3/2} = t_1/2+2t_2$ and   $t_{{\rm in}}^{\pm 1/2 \pm 1/2} = t_1/2-2t_2$, while the diagonal out--of--plane hoppings can be written as $t_{{\rm out}}^{\pm 3/2 \pm 3/2} = t_1/2-t_2$, and $t_{{\rm out}}^{\pm 1/2 \pm 1/2} = t_1/2+t_2$. Finally, the values of the independent parameters are determined by the condition that the low--energy, long--wavelength spectrum of the lattice Hamiltonian (\ref{H0L}) be identical with the spectrum of the Luttinger model.  Explicitly, we have  $t_i= \gamma_i \hbar^2/m_0 a^2$, $i=1, 2,3$, where $a$ is the lattice constant.
\begin{figure}[tbp]
\begin{center}
\includegraphics[width=0.48\textwidth]{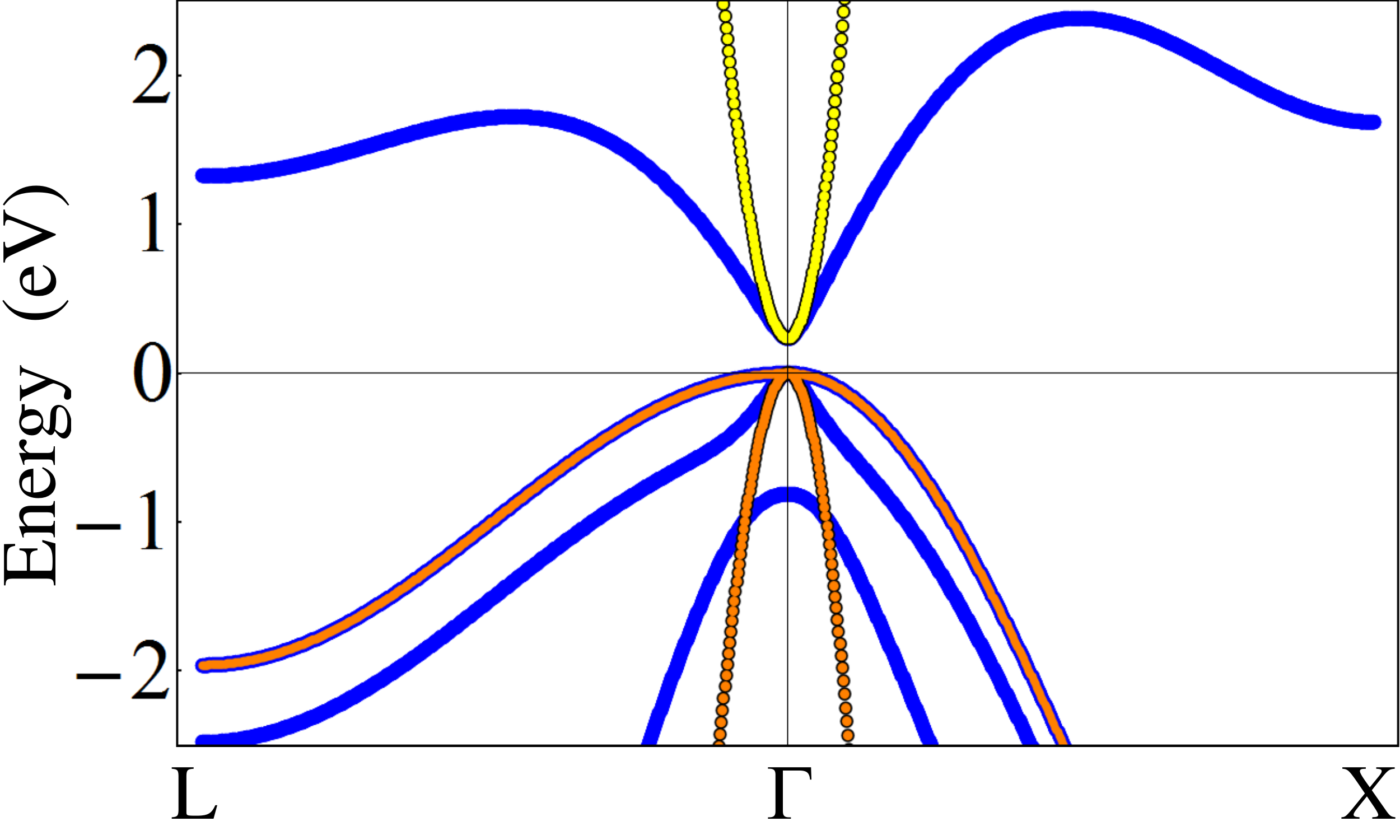}
\vspace{-7mm}
\end{center}
\caption{(Color online) Comparison of 3D spectra corresponding to the 8--band Kane--type model (blue lines), the 4--band Luttinger--type model given by Eq. (\ref{H0L}) (orange), and the 2--band model described by $H_0$ in Eq. (\ref{HSM}) (yellow). Each band is double spin--degenerate. Note that the top valence band (heavy hole band) has the same dispersion in the 8--band and 4--bad models, while quantitative agreement between the light--hole bands and between the conduction bands is present only in the vicinity of the $\Gamma$ point, i.e., for ${\bm k}\rightarrow 0$. The numerical values of model parameters correspond to $InSb$.}
\vspace{-2mm}
\label{Fig3_1}
\end{figure}
For a hole--doped thin film with structural inversion asymmetry, the Rashba--type spin--orbit coupling is modeled phenomenologically by adding a term similar to $H_{\rm SOI}$ in Eq. (\ref{HSM}), but for spin $3/2$, i.e., with ${\bm \sigma}\rightarrow{\bm J}$. This SOI term induces a splitting of the top valence band (the heavy hole band) that is proportional to $k^3$ in the limit $k\rightarrow 0$. This is contrast with the linear splitting that characterizes the Rashba splitting of the conduction band modeled by the Hamiltonian $H_{\rm SM}$ from Eq. (\ref{HSM}). Note that for both the 2--band model (\ref{HSM}) and the 4--band Luttinger--type model, the effective Rashba coefficient $\alpha$ is an independent parameter that does not depend on the confining potential $V({\bm i})$. Typically, in the numerical calculations $V({\bm i})$ is a hard--wall potential ( zero inside a finite region and infinite otherwise) that does not beak the  structural inversion symmetry.

\begin{figure}[tbp]
\begin{center}
\includegraphics[width=0.48\textwidth]{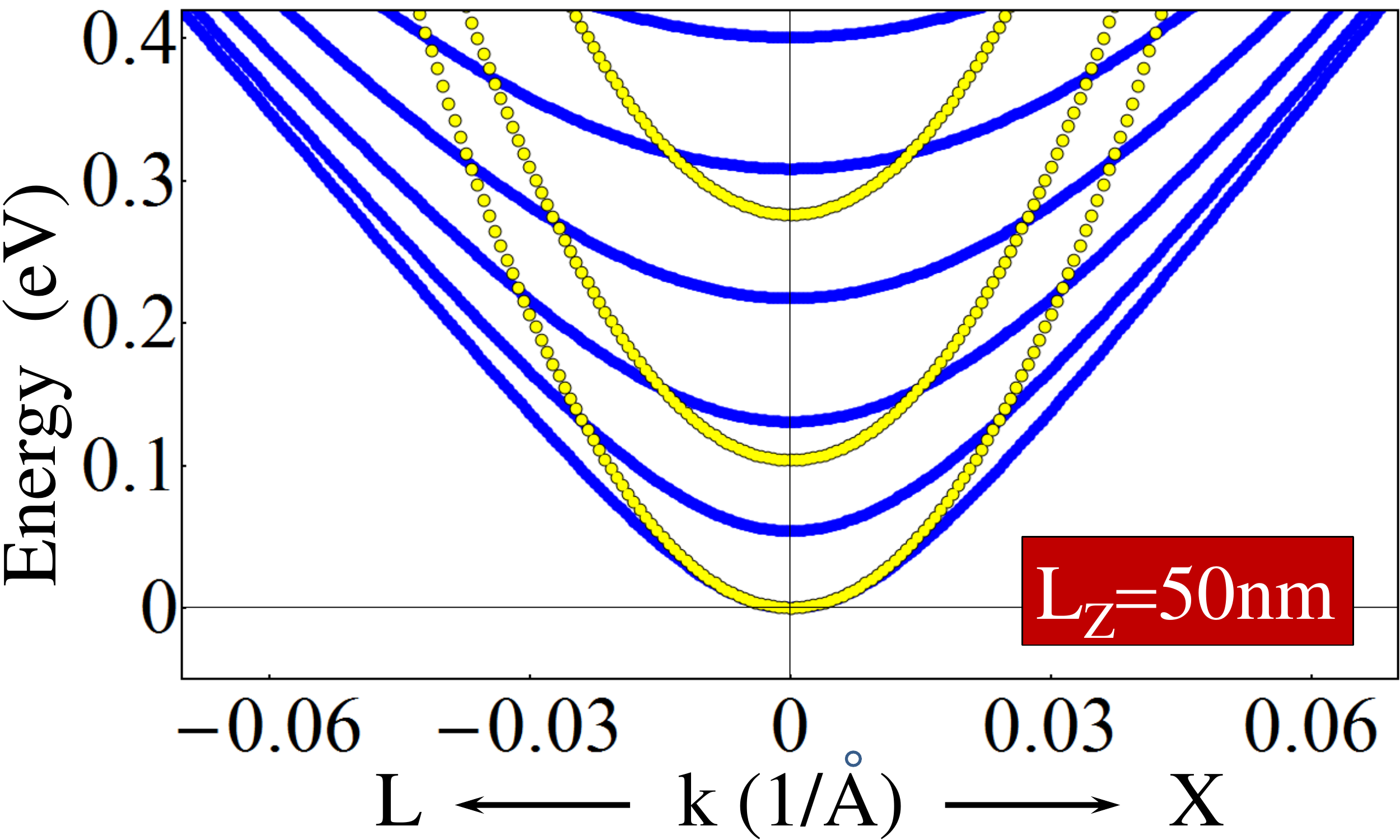}
\vspace{-7mm}
\end{center}
\caption{(Color online) Comparison between the conduction band spectra of a $50$nm InSb film obtained using the 8--band (blue) and the 2--band (yellow) models. Note that both models predict the same value of the effective mass for the lowest energy confinement--induced sub--band, but the effective masses of higher energy  sub--bands, as well as the values of the energy gaps between sub--bands, are significantly different.}
\vspace{-6mm}
\label{Fig3_2}
\end{figure}

Expanding the basis (\ref{Phim}) to include  the eigenstates of the total angular momentum momentum ${\bm J}$ with $j=1/2$ (corresponding to the split--off band), as well as s--type states (corresponding to the conduction band), allows to construct an 8--band tight--binding model for both electron--doped and hole--doped SMs that is equivalent in the long--wavelength limit to the 8--band Kane model~\cite{Kane1957}. There are two key differences between this model and the simpler models described above. First, the Rashba--type spin--orbit coupling in the 8--band model is determined by implicitly by the asymmetric confining potential $V({\bm i})$ and does not involve any additional independent parameter. Consequently, this model captures the correlation between the strength of the spin--orbit coupling and the transverse profile of the wave function. In turn, this profile plays a key role in the SC proximity effect, as we will show in section \ref{Sec3_2}. By contrast, these correlations are not captured by the simplified models. The second key difference becomes manifest when describing low--dimensional nanostructures. In the large length scale limit, the 2--band and 4--band  models described above generate conduction and valence band spectra, respectively,  that agree quantitatively with the results obtained using the 8--band model. This is illustrated in Fig. \ref{Fig3_1}, which shows a comparison between the 3D spectra obtained using these models in the absence of spin--orbit coupling. The large discrepancies at wave vectors away from the $\Gamma$ point suggest that for systems with reduced dimensionality theses models will generate significantly different results. This behavior is illustrated in Fig. \ref{Fig3_2}, which shows a comparison between the conduction band spectra of a SM film of thickness $L_z=50$nm obtained using the 8--band and the 2--band models. In essence, the spectrum of the 8--band model is characterized by strong non--parabolicity and by a sub--band--dependent effective mass. Also, the quasi--2D effective mass predicted by the 8--band model is strongly dependent on the film thickness. These features cannot be reproduced by the simplified 2--band model. Moreover, similar discrepancies characterize the valence bands obtained using the 8--band and 4--band models. All these differences   become even more pregnant in quasi--1D nanostructures. Nonetheless, since Majorana physics is mainly controlled by a relatively small number of low--energy states, the simplified models can provide a reasonably accurate description of the semiconductor system  with a proper choice of effective model parameters and within a limited range for the control parameters, i.e., chemical potential and Zeeman splitting.

An important aspect that has to be addressed when solving numerically a specific model for the Majorana wire is represented by the large number of degrees of freedom in the problem. We emphasize that  Majorana fermions are zero--energy bound states localized near the ends of a SM wire that is proximity--coupled to a superconductor, hence we are interested in modeling a finite quasi--1D system with no particular symmetries, in the presence of disorder and external potentials. For a typical electron--doped SM nanowires modeled using the 2--band model (\ref{HSM}) the number of degrees of freedom is of the order $10^7$--$10^9$. One possible solution is to use a coarse--grained lattice model with an effective lattice constant $a_{\rm eff}$ much larger the actual SM lattice constant. For example, choosing $a_{\rm eff}=40a$ reduces the number of degrees of freedom to about $10^3$--$10^4$. However, while this is numerically convenient, it becomes difficult to address the short--range properties of the system, for example the effects of certain types of disorder. More importantly,  the properties associated with Majorana physics are basically controlled by a reduced number of low--energy degrees of freedom, hence it is more natural to project the Hamiltonian onto the relevant  low--energy sub--space than to perform an overall energy--independent reduction of the Hilbert space. In addition, the simple models used in the calculations are expected to be highly inaccurate at high energy. The technical details associated with the projection onto a low--energy subspace are presented in Appendix \ref{App1} for the case of the 2--band tight--binding model.

\subsection{Proximity effect in semiconductor nanowire--superconductor hybrid structures}\label{Sec3_2}

A critical ingredient of any recipe for realizing Majorana fermions in a solid state hybrid structure is the proximity--induced superconductivity. In essence, the electrons from the SM nanowire acquire SC correlations by partly penetrating in the nearby bulk s--wave superconductor. In conjunction with spin--orbit coupling, these correlations generate effective p--type induced superconductivity in the SM nanowire. More specifically, each double degenerate confinement--induced band generates a combination of $p_x + i p_y$ and  $p_x - i p_y$ superconductivity. Further, the Zeeman splitting breaks time--reversal symmetry and selects one of these two combinations, making the SM nanowire a direct physical realization of Kitaev's toy model for a 1D
spinless p--wave superconductor \cite{Kitaev2001}. While induced pairing is the most prominent aspect of the proximity to the bulk superconductor, another effect is the renormalization of the energy scale for the nanowire. Qualitatively, this renormalization can be understood as resulting from the reduction of the quasiparticle weight of the low--energy SM states due to the partial penetration of the corresponding wave functions into the SC. As a consequence, the low--energy effective model for the nanowire will contain rescaled values of the original parameters (e.g., hopping parameters, spin--orbit couplings, Zeeman field, etc.), as we show below in section \ref{Sec3_3}.

To address quantitative aspects of the SC proximity effect, one has to consider specific models for the relevant terms in the total Hamiltonian (\ref{Htot}): $H_{\rm SC}$, which describes  s--wave bulk superconductor, and  $H_{\rm SM-SC}$, representing the SM nanowire--superconductor coupling. At this point, one possible approach is to treat the pairing problem self--consistently \cite{Black2011,Lababidi2011,Tewari2011} in order to account for spatial variations of the order parameter near the interface and effects due to interactions inside the nanowire \cite{Lutchyn2011,Gangadharaiah2011,Sela2011,Stoudenmire2011} and the presence of a magnetic field. In this work, we will not discuss this aspect of the proximity effect and will model the bulk semiconductor at the mean--field level using a simple tight--binding Hamiltonian characterized by a constant pairing amplitude $\Delta_0$. Explicitly, we have
\begin{equation}
H_{SC} = \sum_{{\bm i}, {\bm j}, \sigma} \left(t_{\bm i \bm j}^{sc} - \mu_{sc}\delta_{\bm i \bm j}\right) a_{{\bm i}\sigma}^\dagger a_{{\bm j}\sigma} + \Delta_0\sum_{\bm i} (a_{{\bm i}\uparrow}^\dagger a_{{\bm i}\downarrow}^\dagger + a_{{\bm i}\downarrow}a_{{\bm i}\uparrow}), \label{HSC}
\end{equation}
where $i$ and $j$ label SC lattice sites, $a_{i\sigma}^\dagger$ is the creation operator corresponding to a single--particle state with spin $\sigma$ localized near site $i$, and $\mu_{sc}$ is the chemical potential. Similarly, we can write the SM--SC coupling as
\begin{equation}
H_{\rm SM-SC}=\sum_{\bm i_0, \bm j_0}\sum_{m, \sigma}\left[\tilde{t}_{\bm i_0 \bm j_0}^{m\sigma} c_{\bm i_0 m}^\dagger a_{\bm j_0\sigma} + h.c.\right], \label{Hsmsc}
\end{equation}
where $\bm i_0=(i_x, i_y, i_{0z})$ and $\bm j_0=(j_x, j_y, j_{0z})$ label lattice sites near the interface in the SM and SC regions, respectively, $m$ is the quantum number that labels the SM states (e.g., if the SM is described by a 2--band model $m\equiv \sigma$), and $\tilde{t}_{\bm i_0 \bm j_0}^{m\sigma}$ are coupling matrix elements between SM and SC local states. Since we are interested in the low--energy physics of the SM nanowire, it is convenient to integrate out the SC degrees of freedom and define an effective action for the wire. In the Green function formalism, this amounts to including of a surface self--energy contribution in the SM Green function~\cite{Stanescu2010,Sau2010a,Potter2011}. The basic structure of this self--energy contribution is illustrated by the local  term
\begin{equation}
\Sigma_{\bm i_0 \bm i_0}(\omega)=-|\tilde{t}|^2\nu_F\left[ \frac{\omega + \Delta_0 \sigma_y\tau_y}{\sqrt{ \Delta_0^2-\omega^2}}+\zeta \tau_z\right], \label{Sigma_0}
\end{equation}
where $|\tilde{t}|$ represents a measure of the SM--SC coupling,  $\nu_F$ is the surface density of states of the SC metal at the Fermi energy, $\sigma_\lambda$ and $\tau_\lambda$ are Pauli matrices associated with the spin and Nambu spaces, respectively, and $\zeta$ is a proximity--induced shift of the chemical potential. Note that \ref{Sigma_0} contains an anomalous term proportional to $\tau_y$ that describes the proximity--induced pairing.
Also, the diagonal term that is linear in frequency in the limit $\omega \rightarrow 0$ is responsible for the reduced quasiparticle weight and the corresponding energy renormalization. Since the basic aspects have been addressed by several authors
\cite{Stanescu2010,Sau2010,Sau2010a,Potter2011,Khaymovich2011,Grein2011,Sau2012}, we focus here on two problems that are critical to understanding the proximity effect in real SM--SC hybrid structures: i) the role of multiband physics, and ii) the dependence on specific features of the SM--SC coupling matrix elements.

\vspace{-1mm}
\begin{center}
{\it Proximity effect in multiband nanowires}
\end{center}
\vspace{-1mm}

The key feature that differentiates the multiband case from its single band counterpart is the emergence of proximity--induced inter--band pairing. In addition, the proximity--induced renormalization of the SM energy scales has a matrix structure, rather than being described by an overall factor. To illustrate these features, we consider an electron--doped SM nanowire with rectangular cross section and dimensions  $L_x\gg L_y\sim L_Z$ in contact with an s--wave SC. The nanowire is modeled using the 2--band model (\ref{HSM}) written in the basis $\psi_{{\bm n} \sigma}({\bm i})$, as described by equations (\ref{phinl}--\ref{Hnnp}), and the SM--SC coupling is given by Eq. (\ref{Hsmsc}). For simplicity, we assume lattice matching between the SM and the SC and nearest neighbor hopping across the interface, $\tilde{t}_{{\bm i}_0 {\bm j}_0}^{m\sigma} = \tilde{t}(i_y)\delta_{{\bm i}_0\!+\!{\bm d}~ {\bm j}_0}$, where $\bm d = (0, 0, d)$ is a nearest--neighbor position vector.  The dependence of $\tilde{t}=\tilde{t}(i_y)$ reflects the possibility of nonuniform SM--SC coupling across the nanowire, as in the recent experiments by Mourik et al. \cite{Mourik2012}. After integrating out the
SC degrees of freedom, the effective surface self--energy can be written in the spinor basis  $\psi_{{\bm n}}=(\psi_{\bm n \uparrow}, \psi_{\bm n \downarrow}, \psi_{\bm n \uparrow}, \psi_{\bm n \downarrow})^T$ as
\begin{equation}
\Sigma_{\bm n \bm n^\prime} = \sum_{i_x, i_y}\sum_{i_x^\prime, i_y^\prime} \psi_{\bm n}^T(\bm i_0)\tilde{t}(i_y) G_{\rm SC}(\omega,\bm j_0, \bm j_0^\prime) \tilde{t}(i_y^\prime)\psi_{{\bm n}^\prime}(\bm i_0^\prime), \label{Sigmannp}
\end{equation}
where $\bm j_0=\bm i_0+\bm d$ and $\bm j_0^\prime=\bm i_0^\prime+\bm d$ and $G_{SC}$ is a matrix that contains both normal and anomalous terns that represents the Green function of the superconductor at the interface. We note that there are three sources of proximity--induced inter--band coupling in Eq. (\ref{Sigmannp}): i) the position dependence of $G_{SC}$, ii) the non--vanishing coupling between states with arbitrary values of $n_z$ and $n_z^\prime$, and iii) the position dependence of  $\tilde{t}$.

The first source has not yet been explored, but could have significant effects when the SC itself has small characteristic length scales. To account for these effects, the SC Green function has to be calculated explicitly by taking into account all relevant details, including the size and geometry of the system. On the other hand, assuming a large superconductor with a planar surface, the SC Green function becomes $G_{SC}=G_{SC}(\omega, i_x-i_x^\prime, i_y-i_y^\prime)$ and can be expressed~\cite{Stanescu2010,Potter2011,Stanescu2011} in terms of its Fourier transform $G_{SC}(\omega,\bm k_{||})\approx -\nu_F[(\omega + \Delta_0 \sigma_y\tau_y)/\sqrt{ \Delta_0^2-\omega^2}+\zeta \tau_z]$. In this case, since the dependence of  $G_{SC}(\omega,\bm k_{||})$ on the in--plane wave--vector $\bm k_{||}$ is very weak~\cite{Stanescu2010,Stanescu2011,Alicea2012}, the surface self--energy contribution is practically local and does not represent an additional source of  inter--band coupling.

The second mechanism, which couples different confinement--induced $n_z$ bands,  is due to the presence of the interface and and does require any in--plane spatial inhomogeneity. Qualitatively, this coupling can be understood in terms of virtual processes in which a electron occupying a state from the $n_z$ band  tunnels into the superconductor, then returns to the SM wire into a state from a different band, $n_z^\prime$.  Assuming uniform hopping $\tilde{t}$ across the interface , the effective coupling due to such processes depends on the values of the wave function at the interface and can be expressed as $\gamma_{n_z n_z^\prime} = \nu_F |\tilde{t}|^2\phi_{n_z}(i_{0z})\phi_{n_z^\prime}(i_{0z})$, where  $\phi_{n_z}$ is given by Eq. (\ref{phinl}). We note that the effect of this  inter--band coupling becomes negligible in the limit of strong confinement, when  $\gamma_{n_z n_z^\prime}$ is much smaller than the inter--band gap.

The third source of proximity--induced inter--band coupling is due to non--homogeneous interface hopping, $\tilde{t}=\tilde{t}(i_y)$. This mechanism is discussed in detail in Ref. \cite{Stanescu2011}. The effective SM--SC coupling of a SM nanowire with rectangular cross section and nonuniform interface tunneling can be written as
\begin{equation}
\gamma_{\bm n \bm n^\prime} =  \delta_{n_x n_x^\prime} \nu_F\phi_{n_z}(i_{0z})\phi_{n_z^\prime}(i_{0z})\sum_{i_y=1}^{N_y} |\tilde{t}(i_y)|^2\phi_{n_y}(i_{y})\phi_{n_y^\prime}(i_{y}).  \label{gammannp}
\end{equation}
Neglecting the spatial dependence of the SC Green function at the interface, the proximity--induced effective self--energy becomes
\begin{equation}
\Sigma_{\bm n \bm n^\prime} = -\gamma_{\bm n \bm n^\prime}\left[\frac{\omega + \Delta_0 \sigma_y\tau_y}{\sqrt{ \Delta_0^2-\omega^2}} + \zeta \tau_z\right], \label{Sigmannp1}
\end{equation}
where $\gamma_{\bm n \bm n^\prime}$ is given by Eq. (\ref{gammannp}) and $\zeta$ is a constant that depends on the details of the SC band structure.  Note that Eq. (\ref{Sigmannp1}) is valid for frequencies inside the SC gap, $|\omega|<\Delta_0$.
Also, we emphasize that the discussion leading to Eq. (\ref{Sigmannp1}) was based on the 2--band tight--binding model for the SM wire, which is constructed s--type localized orbitals  using as a basis, with spin being the only internal degree of freedom.  In more complex models, the local states are labeled by a quantum number $m\neq\sigma$, or by a set of quantum numbers. The hopping matrix elements across the interface, which parametrize the coupling Hamiltonian (\ref{Hsmsc}), depend explicitly on these quantum numbers and, in turn, the proximity--induced surface self--energy will depend on the details of this coupling. An example illustrating the the proximity effect for hole--doped SM nanowires is given in Appendix \ref{App2}.

\subsection{Effective low--energy Bogoliubov--de Gennes Hamiltonian for the Majorana wire}\label{Sec3_3}

In the presence of proximity--induced superconductivity, the low--energy physics of the SM nanowire is described by the Green function matrix $G_{\bm n \bm n^\prime}(\omega)$ that includes nonzero anomalous terms. Using the results of sections \ref{Sec3_1} and \ref{Sec3_2}, one can write the inverse of the Green function matrix as
\begin{equation}
[G^{-1}]_{\bm n \bm n^\prime}(\omega) = \omega - H_{\bm n \bm n^\prime} - \Sigma_{\bm n \bm n^\prime}(\omega),  \label{G1}
\end{equation}
where $H_{\bm n \bm n^\prime}$ is the effective low--energy Hamiltonian for the SM nanowire in the Nambu space and $\Sigma_{\bm n \bm n^\prime}(\omega)$ is the proximity--induced self--energy. For an electron--doped wire described using the 2--band tight--binding model, the self--energy is given by Eq. (\ref{Sigmannp1}), while the low--energy Hamiltonian can be obtained by expanding Eq. (\ref{Hnnp}) to include both particle and  hole sectors. Explicitly, we have
\begin{eqnarray}
&~&H_{{\bm n n}^\prime}=\left[\epsilon_{\bm n}+ \Gamma \sigma_x\right]\tau_z\delta_{{\bm n n}^\prime} \nonumber  \\
&+& i\alpha \delta_{n_z n_z^\prime}\left[q_{n_x n_x^\prime}{\sigma}_y\delta_{n_y n_y^\prime} - q_{n_y n_y^\prime}{\sigma}_x \tau_z\delta_{n_x n_x^\prime} \right],  \label{HnnpSC}
\end{eqnarray}
where the relevant quantities are the same as in Eq.  (\ref{Hnnp}), $\tau_z$ is a Pauli matrix in the Nambu space, and the identity matrices in the spin and Nambu spaces, $\sigma_0$ and $\tau_0$, respectively, have been omitted for simplicity. The eigenvalues of the low--energy states can be obtained by solving the Bogoliubov--de Gennes (BdG) equation
\begin{equation}
{\rm det}[G^{-1}(\omega)] = 0, \label{BdG1}
\end{equation}
where the frequency is restricted to values inside the bulk SC gap, $|\omega|<\Delta_0$, and the Green function is given by equations (\ref{G1}),  (\ref{Sigmannp1}), and (\ref{HnnpSC}).

Solving  Eq. (\ref{BdG1}) numerically can be rather demanding, but one can further simplify the problem by noticing that the relevant energy scale for majorana physics (e.g., the induced pair potential $\Delta$) is typically much smaller than the bulk SC gap  $\Delta_0$. Consequently,  one can focus on the the low--frequency limit $|\omega|\ll\Delta_0$ and consider the self--energy within the static approximation $\sqrt{\Delta_0^2-\omega^2}\approx\Delta_0$. In the static approximation, Eq. (\ref{G1}) becomes $[G^{-1}]_{\bm n \bm n^\prime} = \omega Q_{\bm n \bm n^\prime} - H_{\bm n \bm n^\prime} + \gamma_{\bm n \bm n^\prime}(\sigma_y\tau_y + \zeta/\Delta_0 \tau_z)$, where $Q_{\bm n \bm n^\prime} = \delta_{\bm n \bm n^\prime} + \gamma_{\bm n \bm n^\prime}/\Delta_0$ accounts for the proximity--induced normalization of the SM energy scales. Since $Q_{\bm n \bm n^\prime}$ is a positive definite matrix, the BdG equation (\ref{BdG1}) can be rewritten in the static approximation as ${\rm det}[\omega - H^{\rm eff}]=0$, where $H^{\rm eff}$ is a frequency--independent quantity that can be viewed as an effective BdG Hamiltonian for the SM nanowire with proximity--induced superconductivity. Explicitly, the effective Hamiltonian has the form
\begin{equation}
H_{\bm n \bm n^\prime}^{\rm eff} = \widetilde{Z}_{\bm n \bm m}H_{\bm m \bm m^\prime}\widetilde{Z}_{\bm m^\prime \bm n^\prime} - \Delta_{\bm n \bm n^\prime}\sigma_y\tau_y -  \delta\mu_{\bm n \bm n^\prime}\tau_z. \label{Heff}
 \end{equation}
 The first term in the right hand side of Eq. (\ref{Heff}) represents the SM Hamiltonian renormalized by the proximity effect. Note that, in general, this proximity--induced renormalization is described by a matrix, rather than an overall factor $Z$ representing the reduced quasiparticle weight. The renormalization matrix is the solution of the equation
\begin{equation}
 \widetilde{Z}_{\bm n \bm m}\left(\delta_{\bm m \bm m^\prime}+\frac{\gamma_{\bm m \bm m^\prime}}{\Delta_0}\right)\widetilde{Z}_{\bm m^\prime \bm n^\prime}=\delta_{\bm n \bm n^\prime},
\end{equation}
where $\gamma_{\bm m \bm m^\prime}$ is the effective SM--SC coupling matrix. In the single--band limit, we have $\widetilde{Z}^2 \equiv Z = (1+\gamma/\Delta_0)^{-1}$. Note that the renormalized Hamiltonian is written in the symmetric form $\widetilde{Z}H\widetilde{Z}$, rather than $Z H$, to ensure its hermiticity in the general, multi--band case.
The second term Eq. (\ref{Heff}) contains the proximity--induced pairing matrix
\begin{equation}
\Delta_{\bm n \bm n^\prime} =\widetilde{Z}_{\bm n \bm m}\gamma_{\bm m \bm m^\prime}\widetilde{Z}_{\bm m^\prime \bm n^\prime}. \label{Deltannp}
\end{equation}
In the single--band limit, this reduces to the induced SC pair--potential $\Delta = \gamma \Delta_0/(\gamma+\Delta_0)$. Finally, the last term in  (\ref{Heff}) describes proximity--induced energy shifts and inter--band couplings that have only a rather limited quantitative relevance, $\delta\mu_{\bm n \bm n^\prime}=\zeta\Delta_{\bm n \bm n^\prime}/\Delta_0$.

We emphasize here a key aspect of the proximity effect in multi--band SM-SC hybrid structure: the proximity--induced inter--band coupling and the induced inter--band pairing. These effects depend on both the details of the SM--SC coupling at the interface and the dimensions of the nanosystem. More specifically, considering a SM wire with rectangular cross section $L_y\times L_z$, the inter--band effects become critical whenever the gaps $\Delta E_{\bm n \bm n^\prime}$ separating different confinement--induced bands  ${\bm n}=(n_y, n_z)$ are comparable with the bulk SC gap $\Delta_0$. This situation occurs in wide wires and in the ``sweet spot'' regime~\cite{Lutchyn2011a,Stanescu2011}, when two spin sub--band become degenerate at finite Zeeman field.   By contrast, in the limit $\Delta E_{\bm n \bm n^\prime}\gg\Delta_0$ the inter--band effects are negligible and one can treat the proximity effect in the decoupled band approximation, i.e., $\gamma_{\bm n \bm n^\prime}\approx 0$ if $\bm n \neq \bm n^\prime$. In this approximation, the matrices describing the proximity effect become diagonal, e.g., $\gamma_{\bm n \bm n^\prime}=\gamma_{\bm n}\delta_{\bm n \bm n^\prime}$,  and we have $\widetilde{Z}_{\bm n} = 1/\sqrt{1+\gamma_{\bm n}/\Delta_0}$ and $\Delta_{\bm n} = \gamma_{\bm n}\Delta_0/(\gamma_{\bm n}+\Delta_0)$. Also, we note that there is a key difference between the proximity--induced inter--band effects associated with confinement in the directions normal and parallel to the interface, respectively. Specifically, assuming $n_z=n_z^\prime$, i.e., strong confinement in the normal direction (so that the low--energy physics is controlled by a single $n_z$ mode), and a  weaker confinement in the y--direction, along with with non--uniform SM--SC hopping $\tilde{t}(i_y)$, results in an effective SM--SC coupling $\gamma_{\bm n \bm n^\prime} = \gamma_{n_y n_y^\prime}$ that contains non vanishing off--diagonal terms. However, for typical SM--SC couplings, $\gamma_{n_y n_y^\prime}$ becomes negligible when the difference $|n_y-n_y^\prime|$ is large, as one can infer from Eq. (\ref{gammannp}), and only neighboring $n_y$ bands are significantly coupled. Consequently, when constructing the effective Hamiltonian (\ref{Heff}) it is enough to consider a reduced number of  $n_y$ bands to obtain an accurate description of the low--energy physics. This scenario was investigated in detail in Ref.~\cite{Stanescu2011}. On the other hand, when the confinement in the z--direction is weak, bands with arbitrary $n_z$ become coupled, as $\gamma_{n_z n_z^\prime} \propto \phi_{n_z}(i_{0z}) \phi_{n_z^\prime}(i_{0z})$.  In this case, it is critical to include the high--energy $n_z$ bands in the calculation, as they re--normalize the low--energy  physics via proximity--induced virtual processes.
The role of the proximity--induced inter--band coupling is studied in detail in Ref. \cite{Stanescu2013}. It is found that the proximity induced gap  is strongly suppressed in the intermediate and strong tunnel coupling regimes whenever the SM thickness exceeds a characteristic crossover value
determined by the band parameters of the SM~\cite{Stanescu2013}. Furthermore, the strong coupling regime is characterized by a small induced gap that decreases weakly with the SM--SC coupling strength~\cite{Stanescu2013}, in sharp contrast with expectations based on the decoupled band approximation, which predicts an induced gap of the order of the bulk SC gap $\Delta_0$.  
An example of low--energy spectrum for a SM thin film -- SC slab heterostructure with uniform coupling across a planar interface illustrating the theoretical scheme described above is presented in Appendix \ref{App3}.

\subsection{Low--energy physics in Majorana wires with disorder}\label{Sec4_3}

There are several types of disorder that may  play a role in the low--energy physics of semiconductor--supraconductor hybrid structures~\cite{Stanescu2011}: impurities inside the bulk s-wave SC, disorder in the SM nanowire, and disorder induced by random SM--SC coupling. Below, we mention a few representative sources that could generate these types of disorder and briefly discuss some of their main features. As a general remark, we point out that the theoretical treatment of disorder in semiconductor Majorana wires can be done along two different directions: calculations of disorder--averaged quantities and calculations involving specific disorder realizations. However,  the low--energy physics of the nanowire (on a scale of the order of the induced gap) is controlled by a small number of quantum states, typically less than 100 (see, for example, Figs. \ref{Fig4_4} and \ref{Fig4_5}, upper panels, in Appendix \ref{App4}) having most of their spectral weight inside the SM wire.  When disorder is located in the nanowire or at the interface, the energies and the wave functions characterizing these states depend significantly on the specific disorder realization and the relevance of the disorder--averaged quantities is questionable. The reason for this dependence can be understood qualitatively by noticing that features of the disorder potential with length scales smaller than $1/k_F$, where $k_F$ is the Fermi wave vector of the Majorana band, are irrelevant, since they are averaged over by the low--energy SM states. Typically,  $1/k_F$ is of the order $10^2$nm, hence the effective disorder potential in a nanowire of length $L_x\sim 1\mu$m is characterized by a small number of scattering centers and, consequently, the details of the disorder potential become relevant. Disorder--averaged quantities provide a good description of the low-energy physics (for example, the value of the induced SC gap) in long wires~\cite{Sau2012}, but do not capture the specific properties of a given small segment of that wire, or those of a short wire of similar length. 

Scattering off impurities inside a disordered bulk superconductor has a negligible effect on the topological SC phase of the nanowire~\cite{Lutchyn2012,Potter2011a}. In essence, this behavior is due the fact that the SM  effective impurity scattering rate  involves higher--order SM--SC  tunneling processes and is suppressed by the destructive quantum interference of quasi--particle and quasi--hole trajectories~\cite{Lutchyn2012}. Consequently, static disorder in the superconductor does not suppress the proximity induced topological superconductivity in the semiconductor.

By contrast, disorder in the SM nanowire or at the SM--SC interface can strongly affect the stability of  the topological SC phase. Some of the generic features of the low--energy states in the presence of disorder are illustrated in  Appendix \ref{App4}. Possible sources of disorder are the random variations of the width of the SM wire and, in general, surface roughness, and random potentials created by charged impurities located on or near the surface of the wire~\cite{Stanescu2011}. Another possible source of disorder is represented by the random coupling at the semiconductor--superconductor interface~\cite{Stanescu2011,Chevallier2013}. We note that this random coupling has a twofold manifestation: a fluctuating induced pairing potential and a random proximity--induced renormalization of the SM Hamiltonian (see Sec. \ref{Sec3_3}).  
In general, the suppression of topological superconductivity by disorder represents a serious challenge to the experimental realization of MFs. A recent proposal for optimizing the stability of the topological phase against disorder involves replacing the semiconductor wire by a chain of quantum dots connected by s--wave superconductors~\cite{Sau2012d}.
In addition to the adverse effect on the stability of the topological SC phase, the presence of disorder impacts the low--energy physics~\cite{Brouwer2011} of the SM nanowire--SC hybrid structures in a number of other ways, some of them being investigated theoretically in several recent studies. For example,  it was shown that the topological quantum phase transition is characterized by a quantized thermal conductance and electrical shot noise power that are independent of the degree of disorder~\cite{Akhmerov2011}. The robustness of the topological phase against disorder was shown to depend  non-monotonically on the Zeeman field applied to the wire~\cite{Brouwer2011a}. The interplay between disorder and interaction in one-dimensional topological superconductors was addressed in Ref.~\cite{Lobos2012}. Also, it was shown that in systems with disorder located at the end of the wire the weight of the characteristic Majorana--induced ZBCP in the differential tunneling conductance is strongly enhanced by mixing of sub--bands~\cite{Pientka2012}. On the other hand, the presence of disorder can generate a ZBCP even when the superconducting wire is topologically trivial due to the proliferation of disorder--induced low--energy states~\cite{Liu2012,Rainis2013}, or due to the weak antilocalization  resulting from random quantum interference by disorder~\cite{Pikulin2012}. A similar ZBCP can occur in the topologically trivial phase as a result of fermionic end states with exponentially small energy that emerge if the confinement potential at the end of the  wire is smooth~\cite{Kells2012}. In the light of the recent experiments on semiconductor nanowire--superconductor hybrid structures, ruling out the alternative mechanisms for the conductance peak represents a serious challenge.

\section{Recent experiments and theoretical interpretations}\label{Sec5}

In this section we briefly summarize several recent observations of experimental signatures consistent with the realization of Majorana bound states in semiconductor nanowire--superconductor structures. We also discuss a number of apparent discrepancies between the observed features and the theoretical predictions based on simple model calculations for the Majorana wire.

\subsection{Experimental signatures of Majorana fermions in hybrid superconductor--semiconductor nanowire devices}\label{Sec5_1}

The observation of a zero bias conductance peak in local tunneling conductance measurements on semiconductor nanowires coupled to an s-wave superconductor has been recently reported in Ref. ~[\onlinecite{Mourik2012}]. This observation, which is consistent with the theoretical predictions,  may represent the first experimental evidence of Majorana fermions in a condensed matter system. Soon after, observations of similar ZBCPs have been reported by two other groups~\cite{Deng2012,Das2012} and, recently, by two more groups~\cite{Churchill2013,Finck2013}. Since there are significant differences among the experimental setups,  establishing conclusively that the ZBCPs observed in different systems are due to the same mechanism remains a critical open question. Below, we briefly summarize the main results of Ref. ~[\onlinecite{Mourik2012}]. We mention that a measurement of the fractional a.c. Josephson effect has also been reported  in Ref.~[\onlinecite{Rokhinson2012}]. The observation suggests the presence of a Shapiro step with a height twice larger than the value expected for conventional superconductor junctions~\cite{Rokhinson2012}, which would be consistent with the presence of MFs.

\begin{figure}[tbp]
\begin{center}
\includegraphics[width=0.48\textwidth]{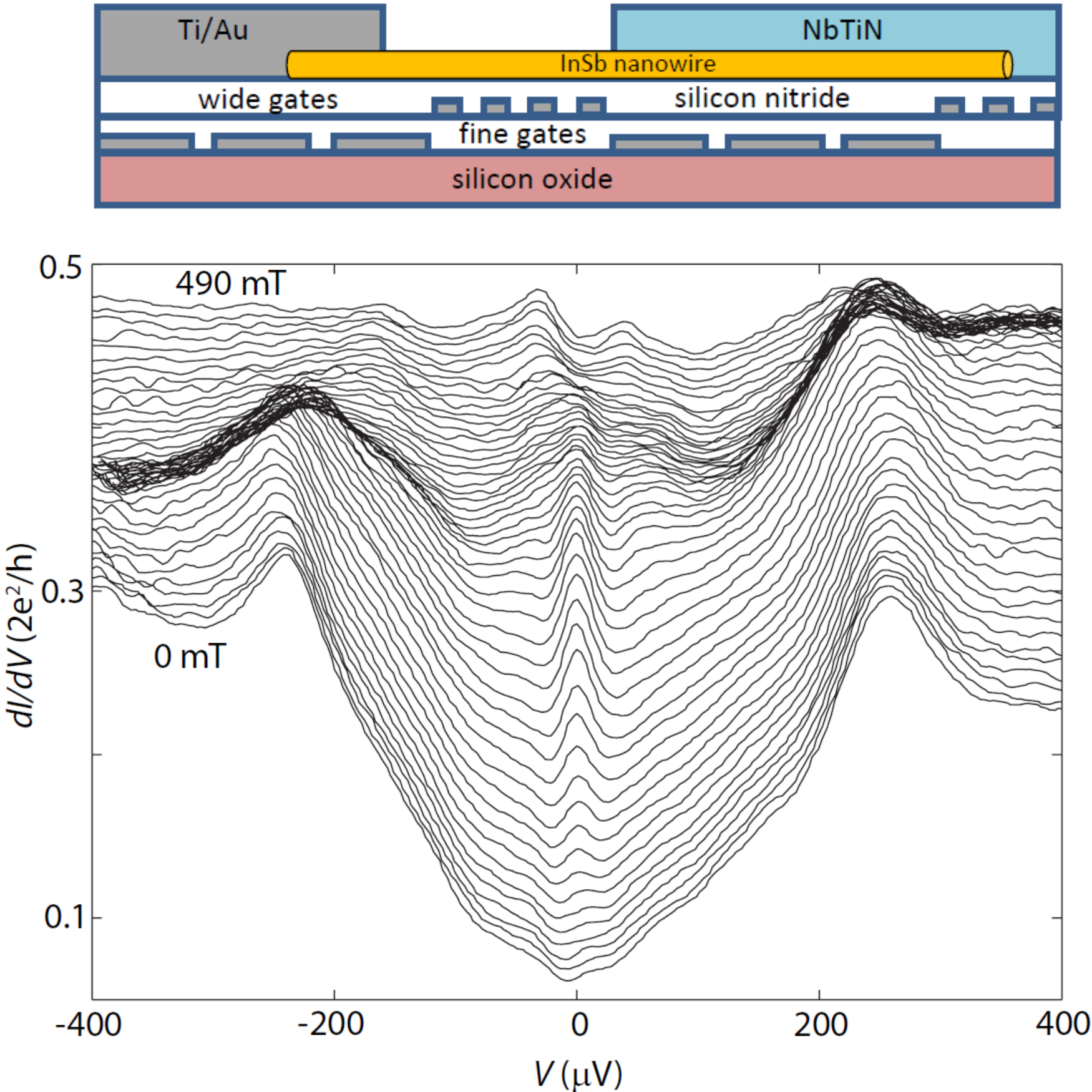}
\vspace{-7mm}
\end{center}
\caption{(Color online) Top panel:  Schematic representation of the experimental setup showing a cross--section through the device measured in Ref.~[\onlinecite{Mourik2012}]. Bottom panel: Differential conductance $dI/dV$ versus bias potential $V$ at a temperature $T=70$mK and for different values of the magnetic field from $0$ to $490$ mT in of $10$ mT. Traces are offset for clarity.  Figure adapted from Ref.~[\onlinecite{Mourik2012}].}
\vspace{-6mm}
\label{Fig5_1}
\end{figure}

As discussed in the previous sections, an optimal hybrid system capable of hosting Majorana bound states and of measuring the expected ZBCP should posses certain characteristics, such as i) a large SC gap, to protect the MF and to allow tuning the Zeeman field in a wide range without destroying the superconducting phase, ii) a long--enough wire length, to accommodate Majorana bound states at the wire ends without large overlap, iii) low disorder, iv) the ability to control the chemical potential, and v) the ability to tunnel from a normal lead, while minimizing the perturbation on the topological SC state.    
The basic experimental setup for the measurements reported in  Ref.~[\onlinecite{Mourik2012}] consists of a InSb nanowire  in contact with a superconductor (NbTiN) and a metallic (Au) lead, as shown schematically in the top panel of Fig. \ref{Fig5_1}.  To measure the differential conductance $dI/dV$, a bias voltage is applied between the normal lead  and the superconductor (SC) and a tunnel barrier is created in the region between the  metallic lead and the SC segment of the nanowire by applying a negative voltage to a narrow gate.
 The InSb nanowire is characterized by a high $g$-factor ($g\approx 50$), as well as a strong spin--orbit (SO) coupling (the Rashba parameter is $\alpha\approx 0.2$eV$\cdot$\AA), making it a good candidate for the realization of MFs in the presence of proximity--induced superconductivity.  The barrier potential  suppresses the background conductance due to Andreev precesses and helps to reveal the ZBCP generated by the possible presence of MFs that could emerge for values of the  magnetic field above a certain critical value. The dependence of the differential tunneling conductance on the bias voltage for different values of the magnetic field is shown in the bottom panel of  Fig. \ref{Fig5_1}. Note that a ZBCP emerges  when the applied magnetic field exceeds about $~ 100$ mT. The ZBCP is robust to further  increasing  the magnetic field until about $400$ mT, above which the single peak seems to split into a two--peak structure. These observations are consistent with the MF interpretation of the tunneling conductance, as shown, for example,  by the theoretical predictions illustrated in Fig. \ref{Fig2_2}.
 
\begin{figure}[tbp]
\begin{center}
\includegraphics[width=0.48\textwidth]{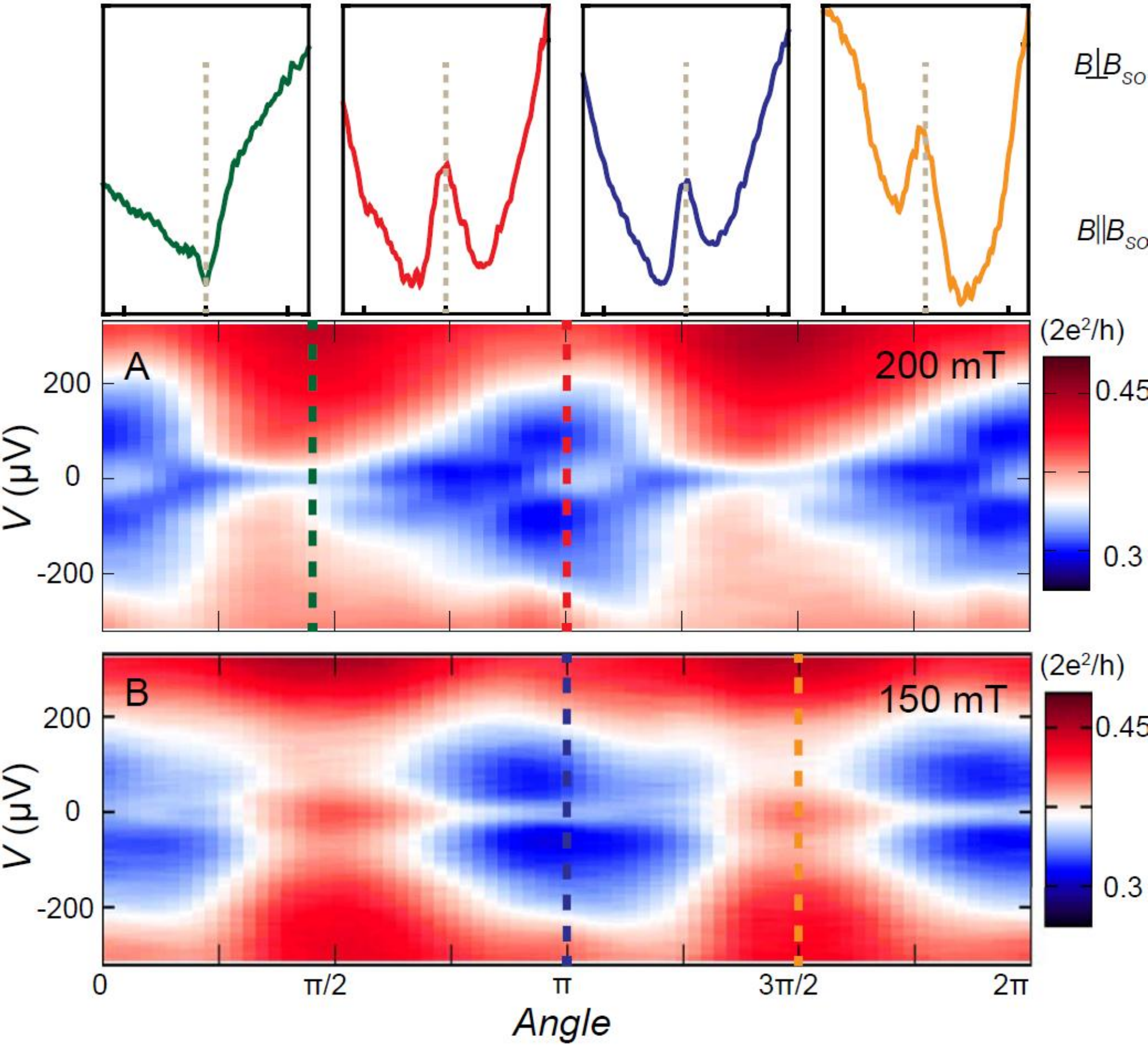}
\vspace{-7mm}
\end{center}
\caption{(Color online) Dependence of $dI/dV$ on the magnetic field orientation. (A) Rotation in the plane of the substrate for a magnetic field $|B| = 200$ mT. Note that the ZBCP vanishes when $B$ is perpendicular to the wire, i.e., parallel to the spin--orbit field. (B) Rotation in the plane perpendicular to the spin--orbit field for $|B| = 150$ mT. Note that a ZBCP is  present for all angles. The top panels  show line--cuts at angles marked with
corresponding colors in (A) and (B). Figure adapted from Ref.~[\onlinecite{Mourik2012}].}
\vspace{-6mm}
\label{Fig5_2}
\end{figure}

An important consistency check on the MF interpretation of the ZBCP is the requirement that the peak must disappear if the angle between the applied magnetic field and the direction of the effective spin--orbit field vanishes. The likely orientation of the spin--orbit field is along the direction perpendicular to the wire in a plane that is roughly parallel to the nanowire--SC and nanowire--substrate interfaces. The angle dependence of the ZBCP was measured in  Ref.~[\onlinecite{Mourik2012}] (see Fig. \ref{Fig5_2}) and found to be consistent with the Majorana scenario, thus  strengthening the identification of the observed ZBCP with the presence of a MF bound state at the normal lead--nanowire interface.

The main observations reported in Ref~[\onlinecite{Mourik2012}], in particular the emergence of a ZBCP at finite magnetic field,  have been subsequently confirmed by other groups~\cite{Deng2012,Das2012}. The similarities are rather surprising, considering that many of the relevant parameters that characterize the devices measured in these experiments are very different from those of  Ref~[\onlinecite{Mourik2012}]. For example, the experiments reported in Ref. ~\onlinecite{Das2012} use InAs nanowires in proximity to superconducting Al (instead of InSb and NbTiN). More importantly, the segment of the nanowire that is proximity coupled to the superconductor is very short (approximately $150-200$nm, which is comparable to the SC coherence length). Is it possible to realize Majorana bound states in such a short nanowire? A possible answer is provided in Ref. ~\onlinecite{Stanescu2012b}, which shows that, in a short wire, the lowest energy state represents a pseudo--Majorana mode that evolves continuously into a true zero--energy Majorana mode as the wire length increases.

In the recently reported experiments, there are several striking features, the most prominent being the absence of a signature associated with the closing of the quasiparticle gap at the topological quantum phase transition (TQPT) that separates the topologically trivial and nontrivial SC phases. Theoretically, this vanishing of the gap at the critical magnetic field {\em must} precede the emergence of the Majorana--induced ZBCP~\cite{Sau2010a,Read2000,Stanescu2011}. This can be clearly seen in the simulation shown in Fig. \ref{Fig2_2}, but is absent in the experimentally measured tunneling conductance (see Fig. \ref{Fig5_2}). Another striking feature is the ``soft'' nature of the induced SC gap, as consistently revealed by the differential conductance measured in the recent experiments~\cite{Mourik2012,Deng2012,Das2012}. Furthermore, the observed
ZBCP is more than an order of magnitude weaker than the quantized value of $2e^2/h$ predicted theoretically. Subsequent theoretical work~\cite{Lin2012} has attributed this discrepancy to a finite temperature effect in conjunction with a hybridization--induced splitting of the Majorana mode in finite wires.
 In addition to these unexpected experimental features, a number of alternative scenarios for the emergence of ZBCPs, which do not involve the presence of MFs but more conventional mechanisms involving strong disorder~\cite{Liu2012,Bagrets2012,Pikulin2012}, smooth confinement~\cite{Kells2012}, or Kondo physics~\cite{Lee2012}, have been recently proposed theoretically, making this problem an exciting but, at the same time, a rather fluid and confusing subject.  Below, we discuss several theoretical proposals that provide possible explanations for the experimentally observed features and suggest further tests that could better reveal the nature of the ZBCPs observed in charge transport measurements on semiconductor nanowire--superconductor structures.
We emphasize that the recent developments in this field reveal that key observable features are strongly dependent on various details of the system. Significant  understanding of the relevant physical mechanisms  responsible for these features, at least at a qualitative level,  can be achieved by realistically modeling the semiconductor--superconductor structures.

\subsection{Suppression of the gap--closing signature at the topological quantum phase transition}\label{Sec5_2}

\begin{figure}[tbp]
\begin{center}
\includegraphics[width=0.48\textwidth]{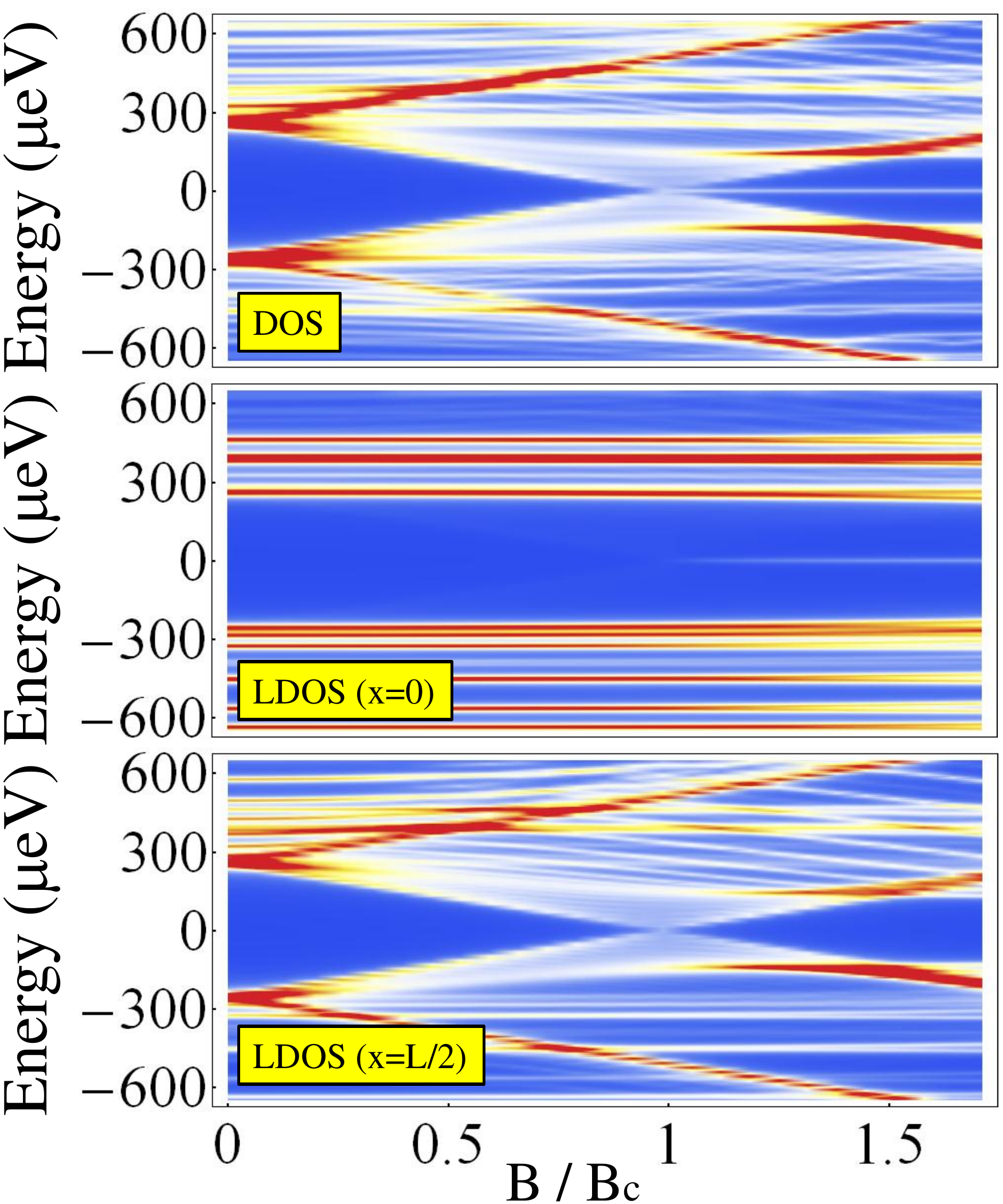}
\vspace{-7mm}
\end{center}
\caption{(Color online) {\em Top}: Density of states (DOS) as a function of $B$ for a wire with four partially occupied bands (seven spin sub--bands) and a chemical potential near the bottom of the fourth band (for $B=0$). At the TQPT characterized by $B_c\approx 0.2$T the bulk gap closes. {\em Middle}: Local density of states (LDOS) at the end of the wire as a function of magnetic field. The strong finite energy features have a weak dependence on $B$. For $B > B_c$ the Majorana  peak is clealy visible. Note that there is no visible signature of the bulk gap closing at the TQPT. {\em Bottom}:
LDOS at the middle of the wire. Note the closure of the gap at the TQPT and the absence of the zero--energy Majorana peak. Figure adapted from Ref.~[\onlinecite{Stanescu2012a}].}
\vspace{-6mm}
\label{Fig5_3}
\end{figure}

While the observations \cite{Mourik2012} of a ZBCP in conductance measurements on semiconductor nanowires coupled to superconductors may represent the first experimental evidence of MFs, the absence of any signature associated with the closing of the superconducting gap (see Fig. \ref{Fig5_1}) at the critical magnetic field asscociated with the topological quantum phase transition (TQPT) casts serious doubt on the MF interpretation of the ZBCP. The closing of the quasiparticle gap at the TQPT is a fundamental theoretical requirement. The question is whether or not this gap closure has a visible signature when the system is probed experimentally.  Ref.~[\onlinecite{Stanescu2012a}]  has offered a possible explanation for the observed non--closure of the gap  at the quantum phase transition between the trivial superconductor and the topological superconductor supporting the MFs in the end--of--wire tunneling experiments. By solving numerically an effective tight--binding model for multiband nanowires with realistic parameters, it has been shown~\cite{Stanescu2012a} that, in the vicinity of the topological transition,  the amplitude of the low--energy states at positions near the ends of the wire may be orders of magnitude smaller than the amplitudes of the localized MFs. Consequently, if the chemical potential of the system is located near the bottom of a confinement--induced semiconductor band, the contributions of these low--energy states to the end--of--wire local density of states (LDOS), and hence to the tunneling conductance, are essentially invisible (see Fig. \ref{Fig5_3}, middle panel).
This behavior results in an apparent non--closure of the gap, as reflected by these local quantities,  even though the magnetic field approaches the critical value and the system is driven through a TQPT. By contrast, the closing of the gap mandated by the topological transition  is clearly revealed by other quantities, such as the total density of states (DOS) (see Fig. \ref{Fig5_3}, top panel) and the LDOS near the middle of the wire (Fig. \ref{Fig5_3},  bottom panel).
A definite prediction of Ref.~[\onlinecite{Stanescu2012a}] is that a tunneling measurement near the middle of the wire should clearly reveal the closing of the gap at the critical field $\Gamma_c$, but the zero--bias peak associated with the presence of Majorana bound states should be absent. Correlated with an end--of--wire measurement characterized by a zero--bias peak for $\Gamma>\Gamma_c$, this  would constitute a powerful argument for the presence of zero--energy Majorana bound states. In addition, the model calculations have shown that the non--closure of the gap, as revealed by the end--of--wire LDOS,  is a non--universal phenomenon. For example, if a local measurement is performed in a regime characterized by a value of chemical potential that is not close to the bottom of a semiconductor band (i.e., $\Delta\mu \gg \Delta$), the closing of the gap should be visible. This regime is illustrated in Fig. \ref{Fig2_2}. Finally, we note that the dominant feature in the end--of--wire LDOS (see Fig. \ref{Fig5_3}, middle panel) are due to states associated with low--energy occupied bands that have significant amplitudes near the ends of the wire. In certain conditions (e.g., in the presence of disorder or in systems with smooth confinement) these states can have energies lower than the bulk quasiparticle gap and, consequently, can provide substantial contributions to the in--gap LDOS.

\subsection{Discriminating between Majorana bound states and garden variety low--energy states}\label{Sec5_3}

Despite their conceptual simplicity, the zero bias conductance peak experiments do not constitute a sufficient proof for the existence of Majorana bound states in semiconductor--superconductor hybrid structures. In a recent theoretical work~\cite{Kells2012}, it has been shown that a non--quantized  nearly zero bias peak, such as that observed in the recent experiments~\cite{Mourik2012,Deng2012,Das2012}, can arise even without end state MFs, provided the confinement potential at the wire end is smooth. A non--quantized nearly zero energy peak at the wire ends has also been shown to occur due to strong disorder effects \cite{Liu2012}, even when the nanowire is in the topologically trivial phase. In essence, these nearly--zero bias peaks occur at finite values of the Zeeman field due to the proliferation of low--energy states in the presence of disorder (see Sec. \ref{Sec4_3}) or in wires with smooth confinement (seeSec. \ref{Sec5_5}) .
To discriminate between different possible mechanisms responsible for the zero bias conductance occurring in a tunneling experiment,
a diagnostic signature for the Majorana--induced ZBCP has been recently proposed~\cite{Stanescu2012}. It was shown that, for smooth confinement at the ends of the wire, the emergence of the near ZBCPs is necessarily accompanied by a signature similar to the closing of a gap in the end--of--wire local density of state. This signature occurs even though there is no corresponding quantum phase transition, as the system stays in the topologically trivial phase, and traces the Zeeman field dependence of the nearly--zero energy states.
In the absence of such a gap closing signature, a ZBCP is unlikely to result from the soft confinement effect~\cite{Kells2012}.
Similarly, when the ZBCP appears at the wire ends from disorder effects (but without MFs), the emergence of the zero conductance is preceded by a signature similar to the closing of the gap~\cite{Liu2012,Stanescu2012}. So far, among all the scenarios that have been considered, the topological phase transition scenario involving the emergence of the MFs is the only one consistent with a ZBCP that  occurs beyond a certain critical magnetic field {\em and} the apparent non--closure of the quasiparticle gap before the emergence of the ZBCP. Since this is precisely what is observed in the experiments in Ref.~[\onlinecite{Mourik2012}], these theoretical results strengthen the identification of the observed ZBCP with topological Majorana bound states localized at the ends of the wire.

We note that the presence of low--energy sub--gap states in systems with disorder and smooth confinement may have another important consequence with dramatic experimentally--observable implications: the existence of a soft superconducting gap, i.e., a SC gap characterized by a non--vanishing density of states and a  v--shaped sub--gap tunneling conductance even at very low temperatures.  This feature is manifestly present in the experimental data~\cite{Mourik2012} (see, for example, Fig. \ref{Fig5_1}) and represents one of the most important open issues in this field. There are several critical aspects that need to be clarified, such as the origin of the in--gap spectral weight, why these in--gap contributions occur as a smooth background, rather than sharply defined peaks in the LDOS, and whether or not a Majorana--induced ZBCP can be well--defined in a system with soft gap. Recently, disorder induced by interface fluctuations was  identified~\cite{Takei2012} as the likely source of the in--gap states responsible for the soft gap.  Another theoretical work~\cite{Stanescu2012} has shown that there are two key ingredients that may explain the emergence of the soft gap in weakly confined wires: i) A finite potential barrier allows states with large spectral weight near the end to hybridize with metallic states from the leads. ii) States from lower--energy bands can penetrate through the barrier, hybridize strongly with the metallic states, and generate broad contribution to the LDOS. By contrast, the Majorana mode, which is associated with the top occupied band, couples weakly to the lead, hence it is weakly broadened and still generates a well defined ZBCP on top of the smooth background.

\begin{figure}[tbp]
\begin{center}
\includegraphics[width=0.48\textwidth]{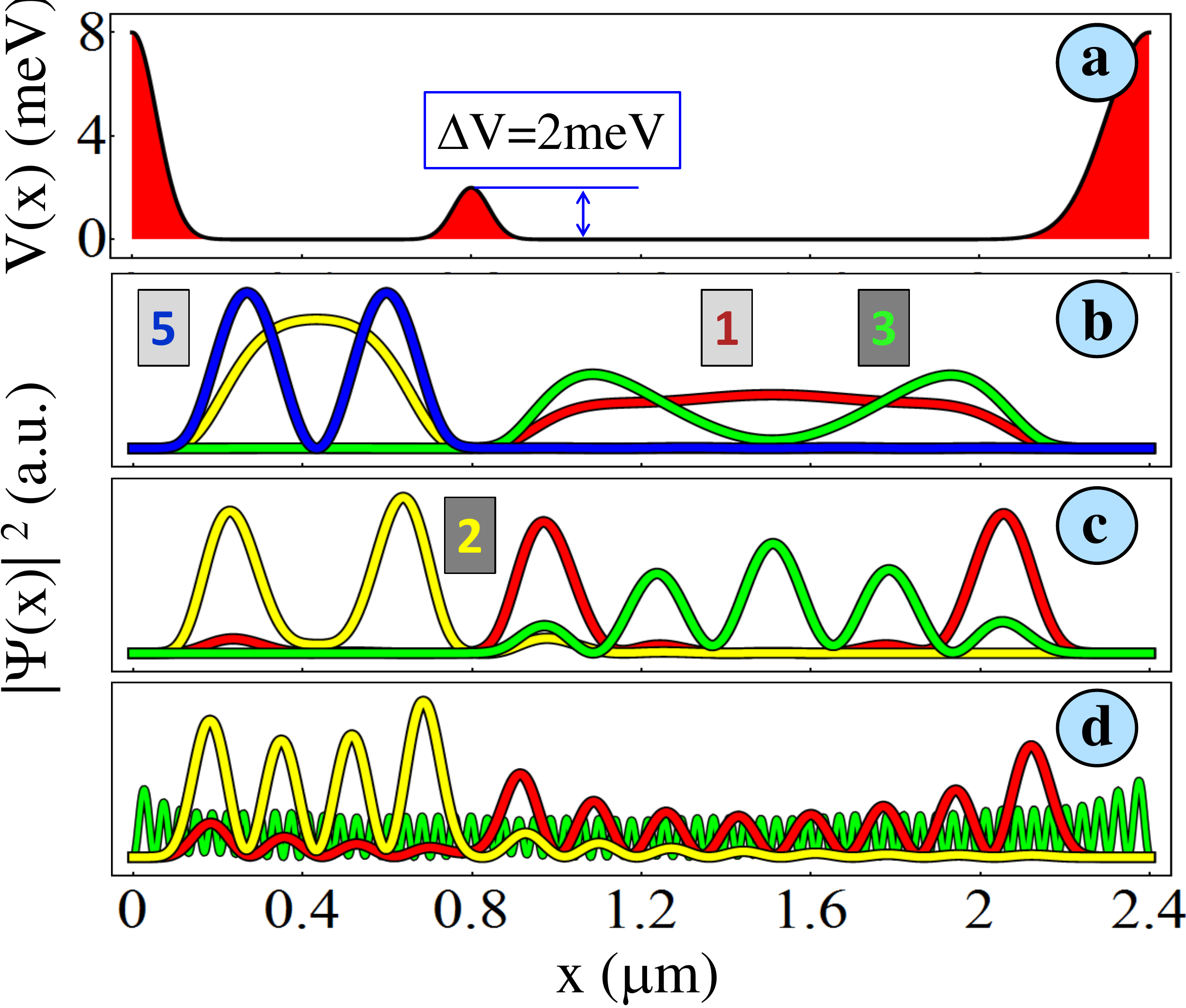}
\vspace{-7mm}
\end{center}
\caption{(Color online)  Semiconductor nanowire with four partially occupied bands in the presence of a nonuniform potential $V(x)$. The position-dependence of $V$ is shown in panel (a). The profiles of several low--energy states $\Psi_n$ with $|E_1|\leq|E_2|\leq\dots$ corresponding to different values of the Zeeman field are shown in panels (b) $\Gamma=0.25$meV  (trivial SC phase) , (c) $\Gamma=0.4$meV (topological SC phase), and (d) $\Gamma=0.8$meV. Note that the wire is effectively separated into two segments for states with a small characteristic wave vector. States corresponding to large $k_F$ values penetrate through the potential barrier $\Delta V$. With increasing Zeeman field, low--energy states associater with the low--energy bands emerge because of the smooth confinement,  e.g., the state $n=3$ (green line) in panel (d).}
\vspace{-6mm}
\label{Fig5_4}
\end{figure}

\subsection{Majorana physics in finite--size wires: A ``smoking gun" for the existence of the Majorana mode}\label{Sec5_5}

It has recently been proposed~\cite{DSarma2012} that direct observation of the splitting of the zero bias conductance peak could serve as a ``smoking gun'' evidence for the existence of the Majorana mode. In essence, the Majorana bound states come always in pairs~\cite{Kitaev2001} that are localized near the ends of the wire, if the system is clean enough. In any finite wire,  the wave functions of the two Majoranas overlap, leading to a splitting~\cite{Cheng2009,Prada2012,Lin2012,Stanescu2012b} of the Majorana mode, which is  a pure zero--energy mode only in the infinite wire limit. At fixed chemical potential, this hybridization--induced energy splitting is characterized by an oscillatory behavior that depends on the Fermi wave vector of the top occupied band and the length of the wire. By contrast, when the particle density is constant, the oscillations can be suppressed~\cite{DSarma2012}, but the splitting of the Majorana mode is still a generic feature. Furthermore, regardless of conditions, two independent tunneling measurements at the opposite ends of a wire should observe exactly the same splitting of the ZBCP~\cite{DSarma2012}, as long as: i) the peak is due to Majorana splitting, and ii) a single pair of Majorana bound states exists in the system. We emphasize that features other than the splitting itself observed in local measurements at the two ends of the wire may be completely different.  If confirmed experimentally, this would constitute strong evidence for the existence of the elusive Majorana mode in semiconductor--superconductor structures.

\begin{figure}[tbp]
\begin{center}
\includegraphics[width=0.48\textwidth]{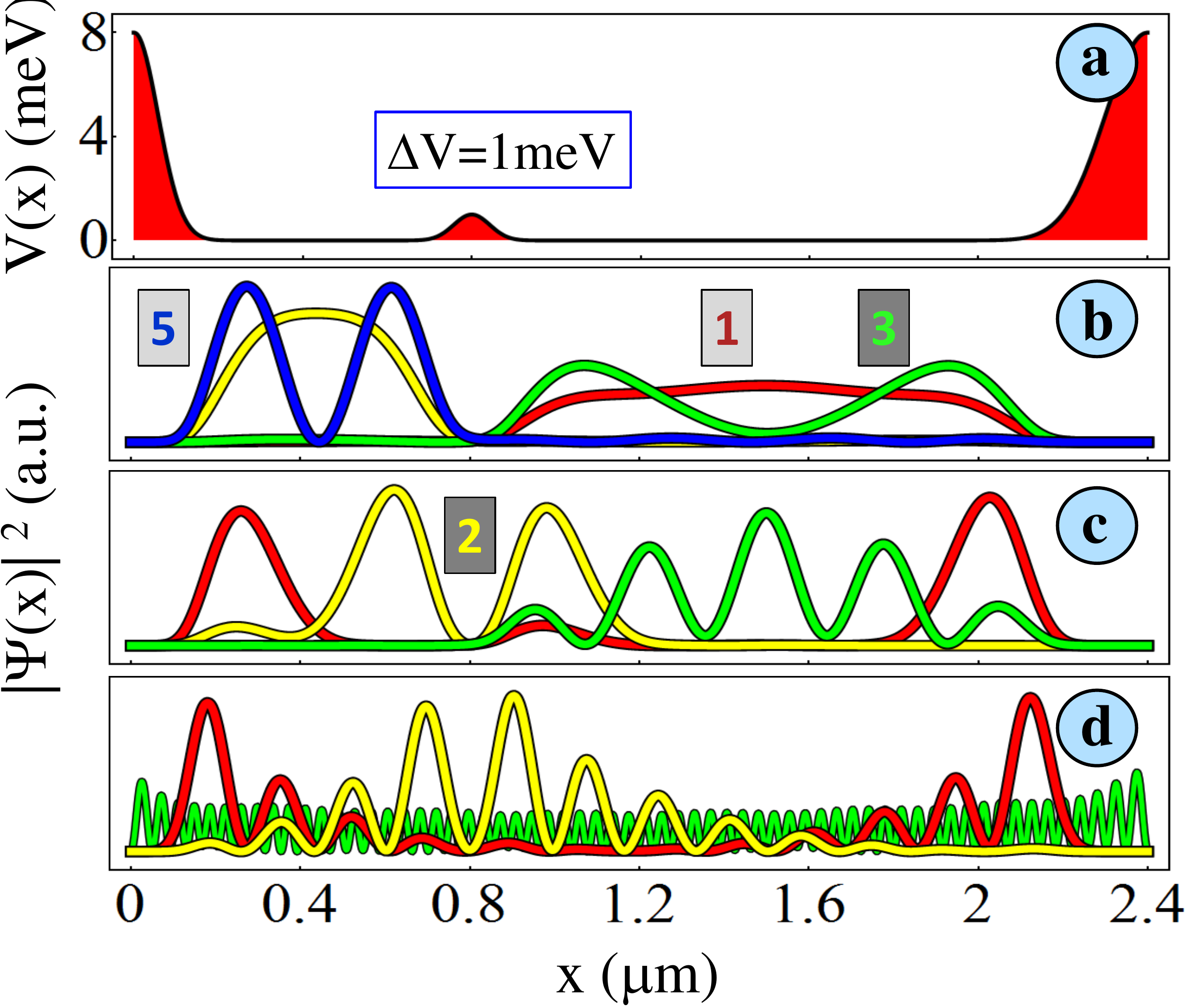}
\vspace{-7mm}
\end{center}
\caption{(Color online) Same as in Fig. \ref{Fig5_4}, but for a lower value of the potential barrier $\Delta V$. Note that, unlike Fig. \ref{Fig5_4}, the lowest energy state ($n=1$, red line), corresponding to two overlapping Majorana bound states, is peaked near the ends of the wire for Zeeman fields associated with the topological SC phase. }
\vspace{-6mm}
\label{Fig5_5}
\end{figure}

In the presence of disorder, several states with energies lower than the induced SC gap will emerge (see Sec. \ref{Sec4_3}). The energies of these sub--gap states depend on the strength of the disorder and the values of the Zeeman field. However, the Majorana mode is still protected, as long as the disorder strength does not exceed a certain critical value, and, consequently,  the hallmark signature described above should still be observable. Nonetheless, strong disorder can effectively cut the nanowire into two or more segments that can each host a pair of Majorana bound states. A natural question concerns the fate of the splitting oscillations generated by these multiple Majorana bound states that emerge in the presence of strong scattering centers. To address this question, we consider we consider a wire with rectangular cross section $L_y\times L_z=100\times 40$ nm, four partially occupied bands, and smooth confinement in the presence of a strong scattering center consisting in a potential barrier of hight $\Delta V$ located at distances $L_{x1}\approx 0.6 \mu$m and $L_{x2}\approx 1.2 \mu$m from the two ends of the wire, respectively. The confinement and the barrier are provided by the position--dependent potential $V(x)$ shown in the upper panels of Fig. \ref{Fig5_4} and Fig. \ref{Fig5_5}. When $\Gamma > \Gamma_c=0.35$ meV and $\Delta V=0$, two Majorana bound states are localized near the ends of the wire (i.e., $x\approx 0.2 \mu$m and  $x\approx 2.2 \mu$m, respectively), while for large $\Delta V$ the wire is split into two disconnected segments, each of them hosting a pair of Majoranas.  The behavior of the low--energy states in the intermediate regime is illustrated in figures \ref{Fig5_4} and \ref{Fig5_5}. As a general feature, we note that states corresponding to small characteristic wave vectors (i.e., long wavelength oscillations) tend to be contained in one of the two segments of the wire separated by $\Delta V$. By contrast, states with large characteristic wave vectors (rapid oscillations) can more easily penetrate through the finite barrier. Furthermore, we note that Fermi $k$--vector $k_F$ associated with the top occupied band increases with the Zeeman field, while the Fermi wave vectors corresponding to the lower energy bands are always larger than $k_F$. Consequently, in the presence of a large barrier, the two pairs of Majoranas are characterized by a strong intra--pair hybridization and a weak inter--pair hybridization, which leads to two low--energy modes mostly localized inside the  $L_{x1}$ and $L_{x2}$ segments, respectively. The corresponding wave functions (states $n=1$ - red line - and $n=2$ - yellow) are shown in Fig. \ref{Fig5_4}. Reducing $\Delta V$ (or increasing the Zeeman field) increases the inter--pair hybridization, which results in a lowest energy state with maxima  near the ends of the wire (see  Fig. \ref{Fig5_5}).

\begin{figure}[tbp]
\begin{center}
\includegraphics[width=0.48\textwidth]{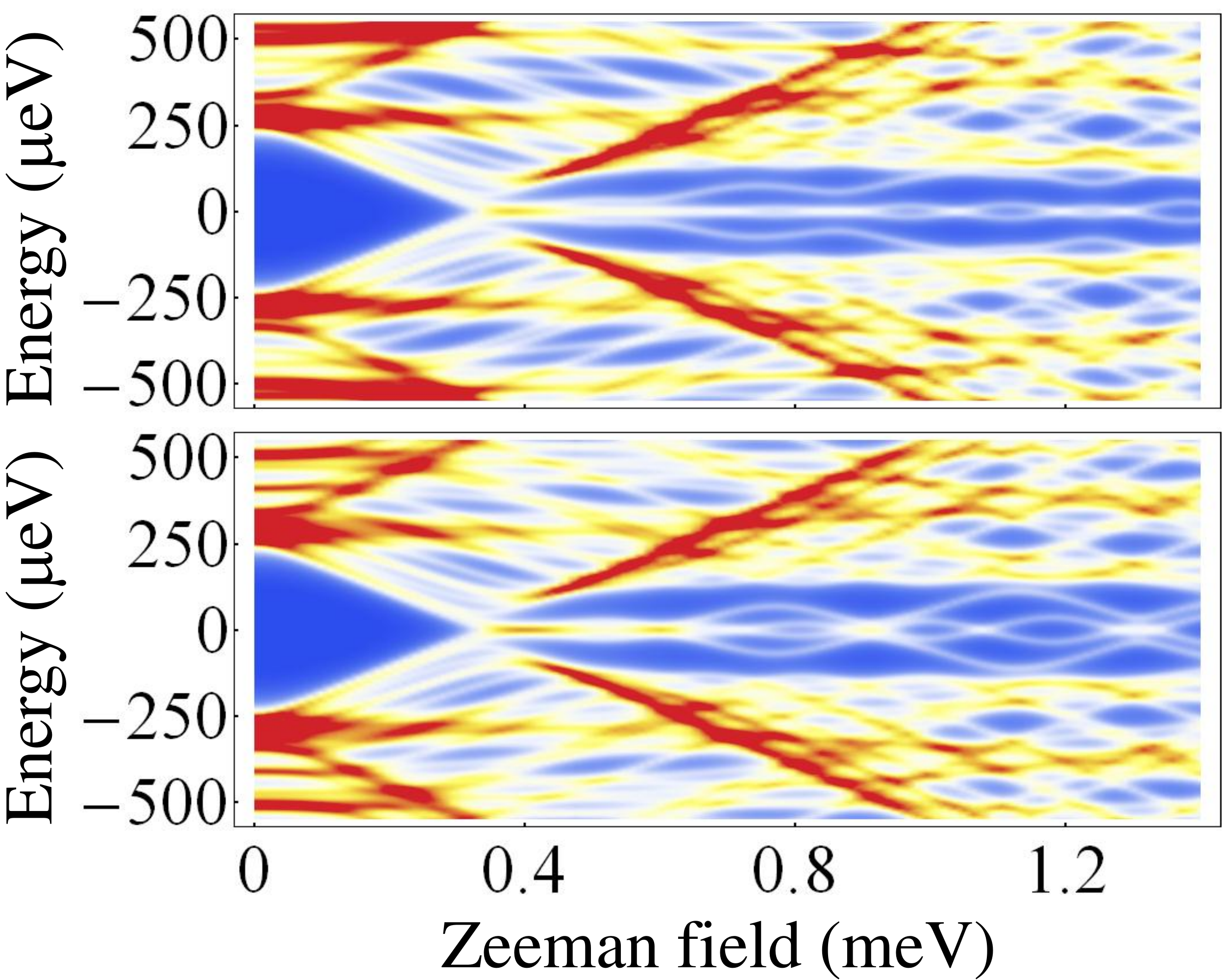}
\vspace{-7mm}
\end{center}
\caption{(Color online)  Density of states as a function of the Zeeman field in the presence of a nonuniform potential  with $\Delta V=1$meV (top) and  $\Delta V=2$meV (bottom). Note the closing of the bulk gap at $\Gamma_c\approx 0.35$meV  and the oscillations of the lowest energy modes for $\Gamma > \Gamma_c$. The corresponding LDOS at the opposite ends of the wire is shown in Fig. \ref{Fig5_7}.}
\vspace{-6mm}
\label{Fig5_6}
\end{figure}

The hybridization--induced splitting oscillations of the Majorana mode can be clearly seen in the field dependence of the density of states (DOS), as shown in Fig. \ref{Fig5_6}.  We want to emphasize two features. First, regardless of the number of coupled Majorana pairs, only one mode exhibits zero--energy crossings at discrete values of the Zeeman field, which is a characteristic signature of Majorana physics in finite nanowires\cite{Stanescu2012b}. By contrast, the other low--energy mode is characterized by minima that vanish only in the limit $\Delta V\rightarrow \infty$, i.e., vanishing inter--pair tunneling. Second, the period and the amplitude of the oscillations induce by intra--pair hybridization increase with decreasing wire length approximately as $1/L_x$~\cite{DSarma2012}. Since $L_{x2}\approx 2L_{x1}$, the ratio between the corresponding periods and amplitudes is approximately two, as evident from the lower panel of  Fig. \ref{Fig5_6}. We note that, in very short wires (quantum dots), the period of the oscillations may be large, so that practically only one zero--energy crossing is accessible~\cite{Stanescu2012b}. Also, in such quantum dot--superconductor structures the Coulomb interaction is expected to play an important role~\cite{Lee2013}.  

\begin{figure}[tbp]
\begin{center}
\includegraphics[width=0.48\textwidth]{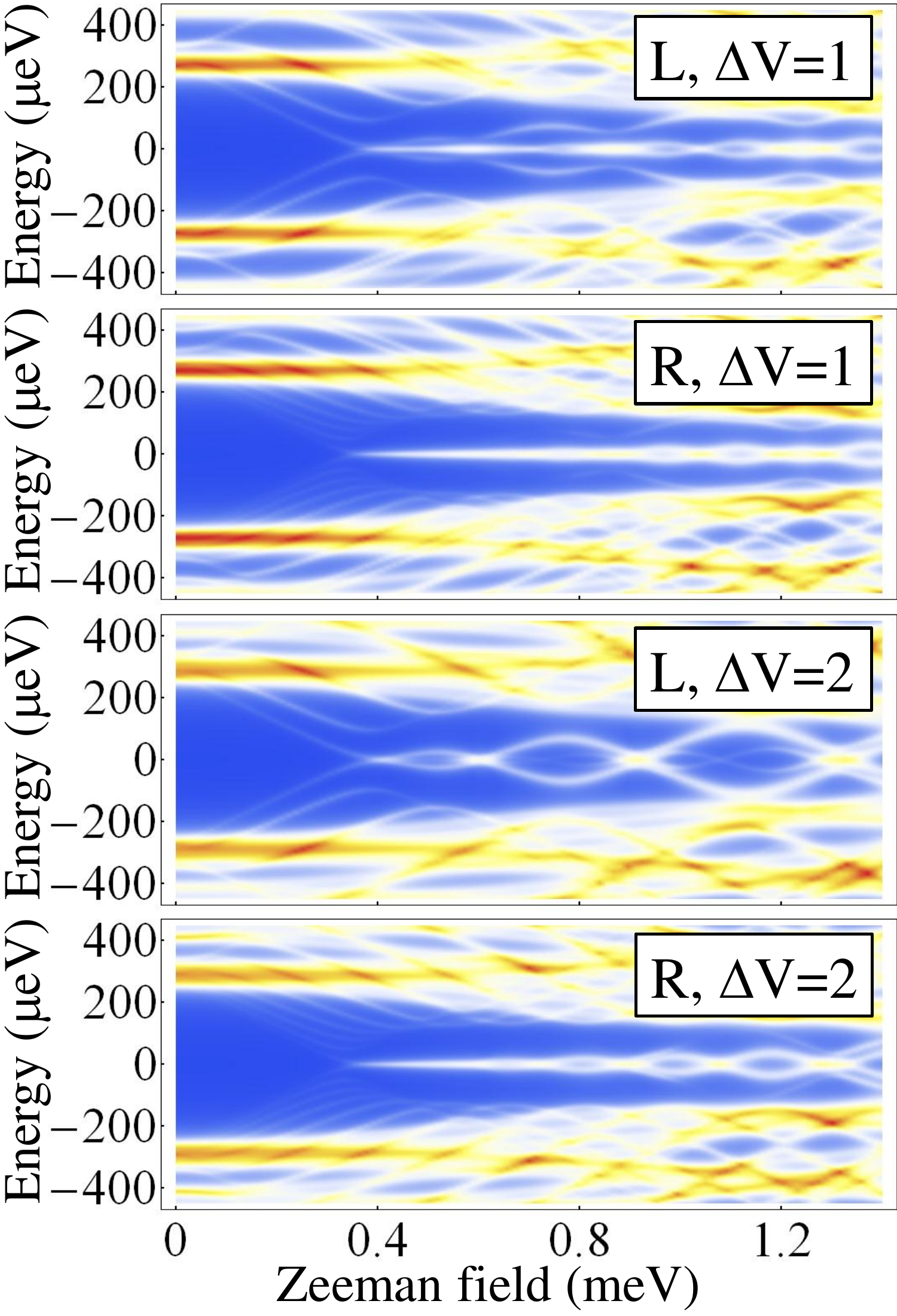}
\vspace{-7mm}
\end{center}
\caption{(Color online) Splitting oscillations of the ZBCP as reflected by the LDOS at the left (L) and right (R) ends of the wire. The top two panels correspond to the parameters of Fig. \ref{Fig5_5}, while the lower two panel are for the system illustrated in Fig. \ref{Fig5_4}. }
\vspace{-6mm}
\label{Fig5_7}
\end{figure}

The characteristic signatures of the splitting oscillations in a set of two independent local measurements at the opposite ends of the wire, as reflected by the corresponding local density of states (LDOS), are shown in Fig. \ref{Fig5_7}. Several features need to be emphasized. As discussed in Sec. \ref{Sec5_2}, the suppression of the gap closing signature stems from the spatial properties of various low--energy states. In short wires, there is no qualitative difference between extended states and states localized near the wire ends, and consequently, a gap closing signature should be visible. This signature should be stronger at the left end of the wire, since the segment $L_{x1}$ is shorter. However, the corresponding features are rather weak (see  Fig. \ref{Fig5_7})  and could be very hard to resolve when on top of an incoherent in--gap background, such as that responsible for the soft gap observed experimentally~\cite{Mourik2012,Deng2012,Das2012}. In an actual experiment, the low--energy states  couple to metallic states from the leads, while the states associated with low--energy bands are expected to have significant weight outside the superconducting region of the wire~\cite{Stanescu2012}. This leads to strong broadening and could explain the the soft gap observed in the experiments while hiding all the weak features that are present in Fig. \ref{Fig5_7}. Consequently, we expect the splitting oscillations measured at the two ends of a wire containing a strong scattering center to be uncorrelated, as shown by the dominant low--energy features corresponding to $\Delta V=2$meV in Fig. \ref{Fig5_7}. By contrast, in the presence of a weak scattering center ($\Delta V=1$meV in Fig. \ref{Fig5_7}), the dominant low--energy features reveal the same ZBCP slitting at both ends of the wire. However, the amplitude of the splitting is strongly reduced because of the larger $L_x$ and the oscillations may not be visible at finite energy resolution.  Therefore, we conclude that the optimal experimental setup for observing correlated Majorana--induced splitting oscillations should use high quality wires with $L_x < 1\mu$m.

\section{Conclusions}\label{Sec6}

The emergence of a zero--bias conductance peak above a certain critical value of the Zeeman field represents the necessary condition for the existence of zero--energy Majorana bound states in spin-orbit-coupled semiconductor-superconductor hybrid structures. Strong experimental evidence for signatures consistent with Majorana physics were recently reported in charge transport measurements on such nanowire heterostructures. These encouraging experimental developments have also raised a number of rather unexpected questions that need to be clarified before claiming victory. Nonetheless, corroborating these observations by performing a ``smoking gun'' measurement, such as the proposed observation of the correlated splitting of the zero--bias peak, could be a step towards unambiguously establishing the existence of Majorana bound states in semiconductor nanowires. This goal could be achieved in the near future. In addition,
a fractional AC Josephson effect, $2e^2/h$ quantized conductance through a MF state, or a signature of a Zeeman-tuned TQPT, say, by tunneling to a region away from the wire ends, can decisively establish the existence of MFs in semiconductor nanowire heterostructures.

However, the search for the non-Abelian Majorana fermion zero modes in semiconductor heterostructures is far from over. Future experiments have to address the sufficient conditions for the existence of the Majorana mode, which may consist a tunneling based interference measurement such as that discussed in Sec.~IIH or the direct observation of non--Abelian braiding statistics in some form. This direction will involve challenging experimental problems regarding the controlled engineering of complex devices, as well as basic aspects related to quantum decoherence and the manipulation of Majorana bound states.  An important lesson provided by the recent developments is that, in real systems, key observable features are determined by various details of the structure, despite the topological nature of the superconducting state predicted to host the Majorana bound states. Therefore, to be able to discriminate between alternative scenarios and to clarify the mechanisms responsible for various features observed in the experiments, it is critical to develop realistic models of the heterostructures as discussed in Sec.~III.

\vspace{2mm}

{\em Acknowledgment:} We would like to thank our collaborators J. D. Sau, R. M. Lutchyn, C. Zhang, and S. Das Sarma. This work was supported in part by WV HEPC/dsr.12.29, DARPA-MTO (FA9550-10-1-0497), and NSF (PHY-1104527).


\appendix

\section{Inter--band pairing and the ``sweet spot''}\label{App0}

\begin{figure}[tbp]
\begin{center}
\includegraphics[width=0.48\textwidth]{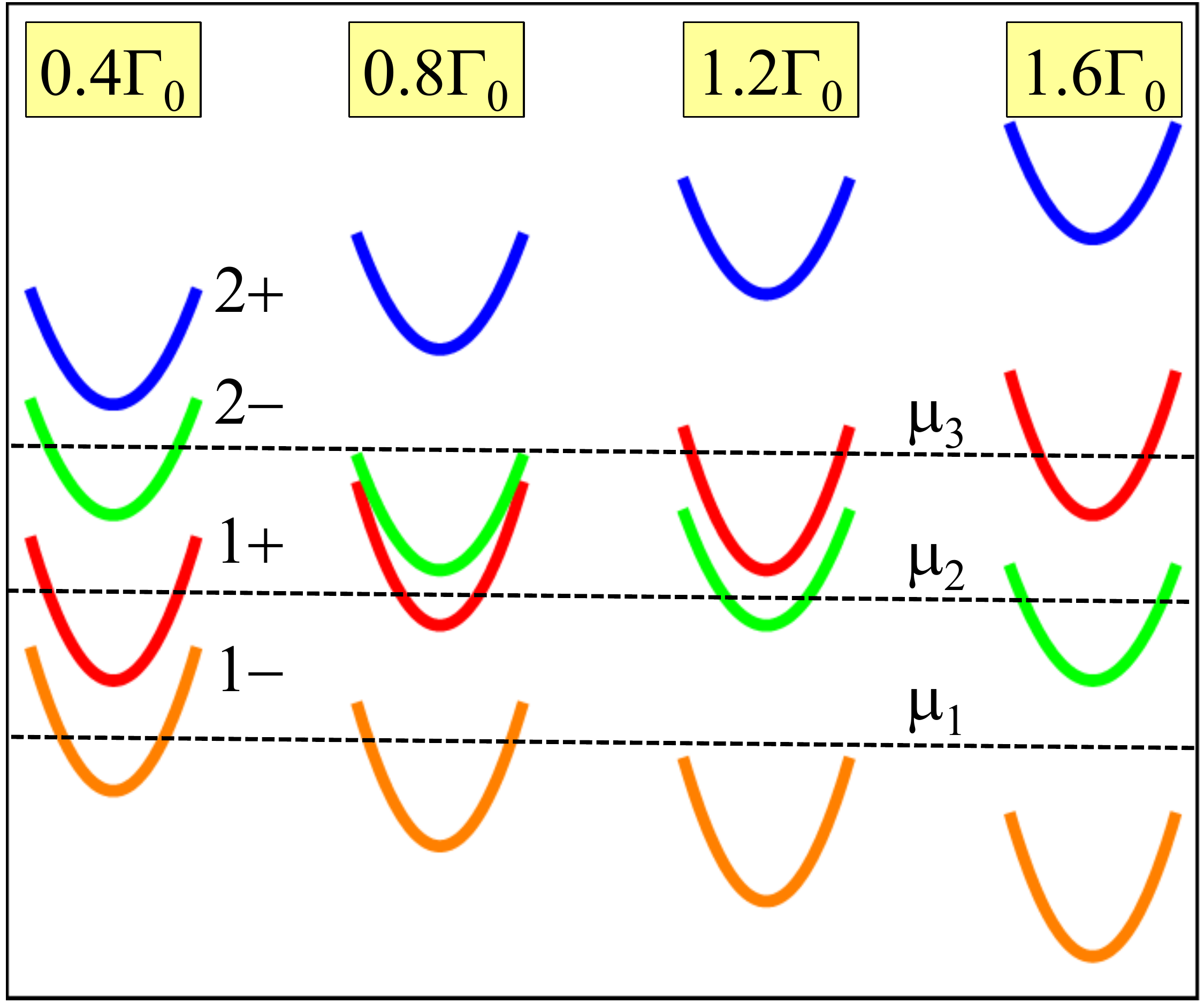}
\vspace{-7mm}
\end{center}
\caption{(Color online) Schematic representation of the evolution of the SM nanowire spectrum with the applied Zeeman field. The curves represent the energy as function of the wave vector $k_x$ for four different values of the Zeeman field $\Gamma$. For $\Gamma=\Gamma_0$ (not shown) the sub--bands $1+$ and $2-$ are degenerate at $k_x=0$. In the presence of induced superconductivity, a system with a value of the chemical potential $\mu=\mu_1$ or $\mu=\mu_3$ is in a topological SC state, while for $\mu=\mu_2$ the SC state is topologically trivial.  However, in the presence of inter--band paring, $\Delta_{12}\neq 0$, the SC state corresponding to $\mu=\mu_2$ becomes  topologically nontrivial in the vicinity of $\Gamma=\Gamma_0$ (the sweet spot).}
\vspace{-6mm}
\label{Fig4_1}
\end{figure}

In the 'standard model' of the Majorana nanowire, which involves  a semiconductor (SM) wire with spin--orbit coupling, proximity--induced superconductivity, and Zeeman spin splitting, the emergence of zero energy Majorana bound states requires values of the chemical potential consistent with an odd number of partially occupied sub--bands. As an example, let us consider the case illustrated in Fig. \ref{Fig4_1} involving two bands, $n=1$ and $n=2$, and four values of the Zeeman field $\Gamma$. The splitting between the pairs of sub--bands with opposite helicity (i.e., $n-$ and $n+$) is proportional to the applied field and, for $\Gamma=\Gamma_0$, the sub--bands $1+$ and $2-$ are degenerate at $k_x=0$, $\epsilon_{1+}(0)=\epsilon_{2-}(0)=\mu_2$. The condition for the existence of a topological superconducting (SC) phase is satisfied for all four values of $\Gamma$ when the chemical potential is $\mu=\mu_1$ (one partially occupied sub--band) or $\mu=\mu_3$ (three partially occupied   sub--bands), but the system is topologically trivial when $\mu=\mu_2$ (two occupied sub--bands). This picture holds as long as the inter--band paring is zero. However, when $\Delta_{12}\neq 0$, the number of partially occupied sub--bands does not represent a good criterion for establishing the topological nature of a given SC phase. More specifically,  a quasi-1D nanowire with inter--band pairing and control parameters in the vicinity of the ``sweet spot'' $\Gamma-\Gamma_0$ and $\mu=\mu_2$ is in the topologically nontrivial SC phase provided the inter sub--band spacing $|\epsilon_{1+}-\epsilon_{2-}|$ is smaller that the off-diagonal coupling $\Delta_{12}$~\cite{Lutchyn2011a}. In the limit of strong inter--band mixing, the sweet spot regime  is characterized by a topological state that is robust against chemical potential fluctuations, such as those created by disorder. This stronger immunity can be understood in terms of the range of chemical potentials consistent with a topological SC state for a given value of the Zeeman field. Considering  example shown in Fig. \ref{Fig4_1}, for $\Gamma=0.4\Gamma_0$ there are two narrow topological regions (for chemical potentials in the vicinity of $\mu_1$ and $\mu_3$, respectively), separated by a topologically trivial phase. By contrast, for $\Gamma$ sufficiently close to $\Gamma_0$ the range of $\mu$ consistent with a topological SC state extends from the bottom of sub--band $1-$ to the bottom of $2+$, without passing through any topological quantum phase transition.

How can inter--band mixing occur in SM nanowires with proximity--induced superconductivity? In essence, non--uniform interface tunneling generically induces inter--band coupling. Consider, for example, a hybrid structure containing a SC and a long SM nanowire with rectangular cross section with $L_y\gg L_z$. The low--energy physics is controlled by the lowest energy transverse mode, $n_z=1$, but , in general, involves several $n_y$ bands. Now let us assume that only half of the nanowire is covered by the superconductor, so that $\tilde{t}(y)=\tilde{t}$ for $y<L_y/2$ and  $\tilde{t}(y)=0$ for $y>L_y/2$. Using Eq. (\ref{gammannp}), the effective SM--SC coupling becomes
\begin{equation}
\gamma_{n_y n_y^\prime} = \gamma_0 \sum_{i_y=1}^{N_y/2} \phi_{n_y}(i_y) \phi_{n_y^\prime}(i_y),
\end{equation}
where $\gamma_0=\nu_F |\tilde{t}|^2[\phi_1(i_{0z})]^2$. The diagonal coupling is $\gamma_{n_y n_y}=\gamma_0/2$, but, in addition, each band couples to all other bands of opposite parity. The inter--band couplings for neighboring bands, $n_y^\prime=n_y\pm 1$, are comparable to the diagonal value, e.g., $\gamma_{12}\approx 0.85 \gamma_{11}$. As described in Section \ref{Sec3_3}, the off--diagonal SM--SC coupling generates proximity--induced inter--band pairing, $\Delta_{n_y n_y^\prime}$, which leads to the sweet spot physics described above. Finally, we note that the experimental realization of non--uniform SM--SC coupling by partially covering the SM nanowire  with superconductor~\cite{Mourik2012} can provide additional benefits, such as reducing SC--generated screening and allowing gate--voltage control of the chemical potential.

\section{Projection of the effective tight--binding Hamiltonian onto a low--energy subspace}\label{App1}

To realize the projection of the effective tight--binding Hamiltonian onto a low--energy subspace, one needs to identify a convenient low--energy basis. While, in general, this problem has to be addressed using a combination of analytical and numerical tools, the 2--band lattice model described above has a simple analytical solution.  Specifically, for an electron--doped nanowire with rectangular cross section and dimensions $L_x \gg L_y\sim L_z$, the single particle quantum problem corresponding to $H_0$ from Eq. (\ref{HSM}) has eigenstates $\psi_{{\bm n} \sigma}({\bm i}) = \prod_{\lambda=1}^3 \phi_{n_\lambda}(i_\lambda) \chi_\sigma$, where ${\bm n}=(n_x, n_y, n_z)$ with $1\leq n_\lambda \leq N_\lambda$, $\chi_\sigma$ is an eigenstate of the ${\sigma}_z$ spin operator, and
\begin{equation}
\phi_{n_\lambda}(i_\lambda) = \sqrt{\frac{2}{N_\lambda+1}}\sin\frac{\pi n_\lambda i_\lambda}{N_\lambda+1}. \label{phinl}
\end{equation}
In Eq. (\ref{phinl}) $\lambda = x, y, z$ and $L_\lambda = a N_\lambda$, where $a$ is the lattice constant.  The eigenvalues corresponding  to $\psi_{{\bm n} \sigma}$ are
 \begin{equation}
 \epsilon_{\bm n} \!=\! -2 t_0 \left( \cos\frac{\pi n_x}{N_x\!+\!1} \!+\!  \cos\frac{\pi n_y}{N_y\!+\!1} \!+\!  \cos\frac{\pi n_z}{N_z\!+\!1}-3\right)\!-\!\mu, \label{epsn}
\end{equation}
where $\mu$ is  the chemical potential.

Next, we assume that only a few bands are occupied, and that the low--energy subspace is defined by the eigenstates satisfying the condition $\epsilon_{\bf n} <\epsilon_{\rm max}$, where the cutoff energy  $\epsilon_{\rm max}$ is typically of the order $100$meV. Using this low-energy basis, the matrix elements of the total Hamiltonian can be written explicitly, as described in Ref. \cite{Stanescu2011}. The matrix elements of the SOI Hamiltonian from Eq. (\ref{HSM}) are
\begin{eqnarray}
\langle\psi_{{\bm n}\sigma}|H_{\rm SOI}|\psi_{{\bm n^\prime}\sigma^\prime}\rangle &=& \alpha \delta_{n_z n_z^\prime}\left\{ q_{n_x n_x^\prime}(i{\sigma}_y)_{\sigma \sigma^\prime} \delta_{n_y n_y^\prime}\right. \nonumber \\
&-& \left.  q_{n_y n_y^\prime}(i{\sigma}_x)_{\sigma \sigma^\prime} \delta_{n_x n_x^\prime} \right\}, \label{HSOI}
\end{eqnarray}
where
\begin{equation}
q_{{n}_\lambda {n}_\lambda^\prime} = \frac{1-(-1)^{n_\lambda+n_\lambda^\prime}}{N_\lambda+1}\frac{\sin\frac{\pi n_\lambda}{N_\lambda+1}\sin\frac{\pi n_\lambda^\prime}{N_\lambda+1}}{\cos\frac{\pi n_\lambda}{N_\lambda+1}-\cos\frac{\pi n_\lambda^\prime}{N_\lambda+1}}.  \label{qnn}
\end{equation}
The first term in Eq. (\ref{HSOI}) represents the intra--band Rashba spin--orbit interaction, while the second term couples different confinement--induced bands. Similarly, assuming that the Zeeman splitting $\Gamma$ is generated by a magnetic field oriented along the wire (i.e., along the $x$-axis),  $\Gamma = g^* \mu_B B_x /2$, where $g^*$ is the effective g--factor for the SM nanowire,  the matrix elements for the corresponding term in Eq. (\ref{Htot}) are
\begin{equation}
\langle\psi_{{\bm n}\sigma}|H_{\rm Z}|\psi_{{\bm n^\prime}\sigma^\prime}\rangle =\Gamma \delta_{{\bm n}{\bm n}^\prime} \delta_{\bar{\sigma} \sigma^\prime}, \label{HZeemannn}
\end{equation}
where $\bar{\sigma}=-\sigma$. Adding together these contributions, the effective Hamiltonian describing the low--energy physics of the semiconductor nanowire in the presence of a Zeeman field becomes $H_{{\bm n n}^\prime} = \langle \psi_{\bm n} |H_{\rm SM} + H_{\rm Z}| \psi_{\bm n}^\prime\rangle$, with $\psi_{\bm n}$ representing the spinor $(\psi_{{\bm n}\uparrow}, \psi_{{\bm n}\downarrow})$. Explicitly, we have
\begin{eqnarray}
&~&H_{{\bm n n}^\prime}=\left[\epsilon_{\bm n}+ \Gamma \sigma_x\right]\delta_{{\bm n n}^\prime} \nonumber  \\
&+& i\alpha \delta_{n_z n_z^\prime}\left[q_{n_x n_x^\prime}{\sigma}_y\delta_{n_y n_y^\prime} - q_{n_y n_y^\prime}{\sigma}_x \delta_{n_x n_x^\prime} \right],  \label{Hnnp}
\end{eqnarray}
where ${\bm n} =(n_x,n_y,n_z)$, $\epsilon_{\bm n}$ is given by Eq. (\ref{epsn}) and $q_{n_\lambda n_\lambda^\prime}$ by Eq.  (\ref{qnn}). Note that a similar effective low--energy Hamiltonian can be written for an infinite quasi--1D wire. In the limit $L_x\rightarrow \infty$, the wave vector $k_x$ becomes a good quantum number and we have
\begin{equation}
H_{{\bm n n}^\prime}(k_x)=[\epsilon_{\bm n}(k_x) +\alpha_R k_x \sigma_y + \Gamma\sigma_x]\delta_{{\bm n n}^\prime} - i \alpha q_{n_y n_y^\prime} \sigma_x \delta_{n_z n_z^\prime},
\end{equation}
where $\alpha_R=\alpha a$ and ${\bm n} =(n_y,n_z)$ labels the confinement--induced bands with energy
 \begin{equation}
 \epsilon_{\bm n}(k_x) \!=\! \frac{\hbar^2 k_x^2}{2 m^*} -2 t_0 \left(\cos\frac{\pi n_y}{N_y\!+\!1} \!+\!  \cos\frac{\pi n_z}{N_z\!+\!1}-2\right)\!-\!\mu. \label{epsk}
\end{equation}

\section{Proximity effect in hole--doped semiconductors}\label{App2}

 To illustrate this the dependence of the SC proximity effect on the details of the SM--SC coupling, we consider the case of hole--doped SM nanowires. We use the 4--band Luttinger--type model defined by equations (\ref{Phim}) and (\ref{H0L}) to describe a hole--doped SM nanowire with rectangular cross section proximity coupled an s-wave SC. We assume that the $z$ axis is perpendicular to the interface and corresponds to the $(0,0,1)$ crystal axis of the underlying $fcc$ lattice. The specific form of the SM--SC coupling Hamiltonian (\ref{Hsmsc}) depends on symmetry of the localized states that define the effective SC Hamiltonian (\ref{HSC}). Assuming that $\mu_{sc}$ lies within a band with s-type character, we notice that the orbitals that are responsible for the coupling across the interface are the $|Z\rangle$ orbitals. Consequently, only the $\Phi_{\pm\frac{1}{2}}$ eigenstates couple to the SC. The corresponding matrix elements in equation (\ref{Hsmsc}) can be written as
\begin{equation}
\tilde{t}_{\bm i_0 \bm j_0}^{~s\!\frac{1}{2} \sigma}= \sqrt{\frac{2}{3}}\tilde{t} ~\delta_{\bm i_0\!+\!\bm d~\bm j_0} \delta_{s \sigma},
\end{equation}
where $s\frac{1}{2}=\pm\frac{1}{2}$ and $\tilde{t}$ is a constant,  while the coupling of the $3/2$ states vanishes, $\tilde{t}_{\bm i_0 \bm j_0}^{~s\!\frac{3}{2} \sigma}=0$. We address the following question: What is the consequence of this selective coupling on the strength of the proximity effect in hole--doped nanowires with rectangular cross section? Specifically, we focus on the effective SM--SC coupling $\gamma_v$ for the top valence band.

\begin{figure}[tbp]
\begin{center}
\includegraphics[width=0.48\textwidth]{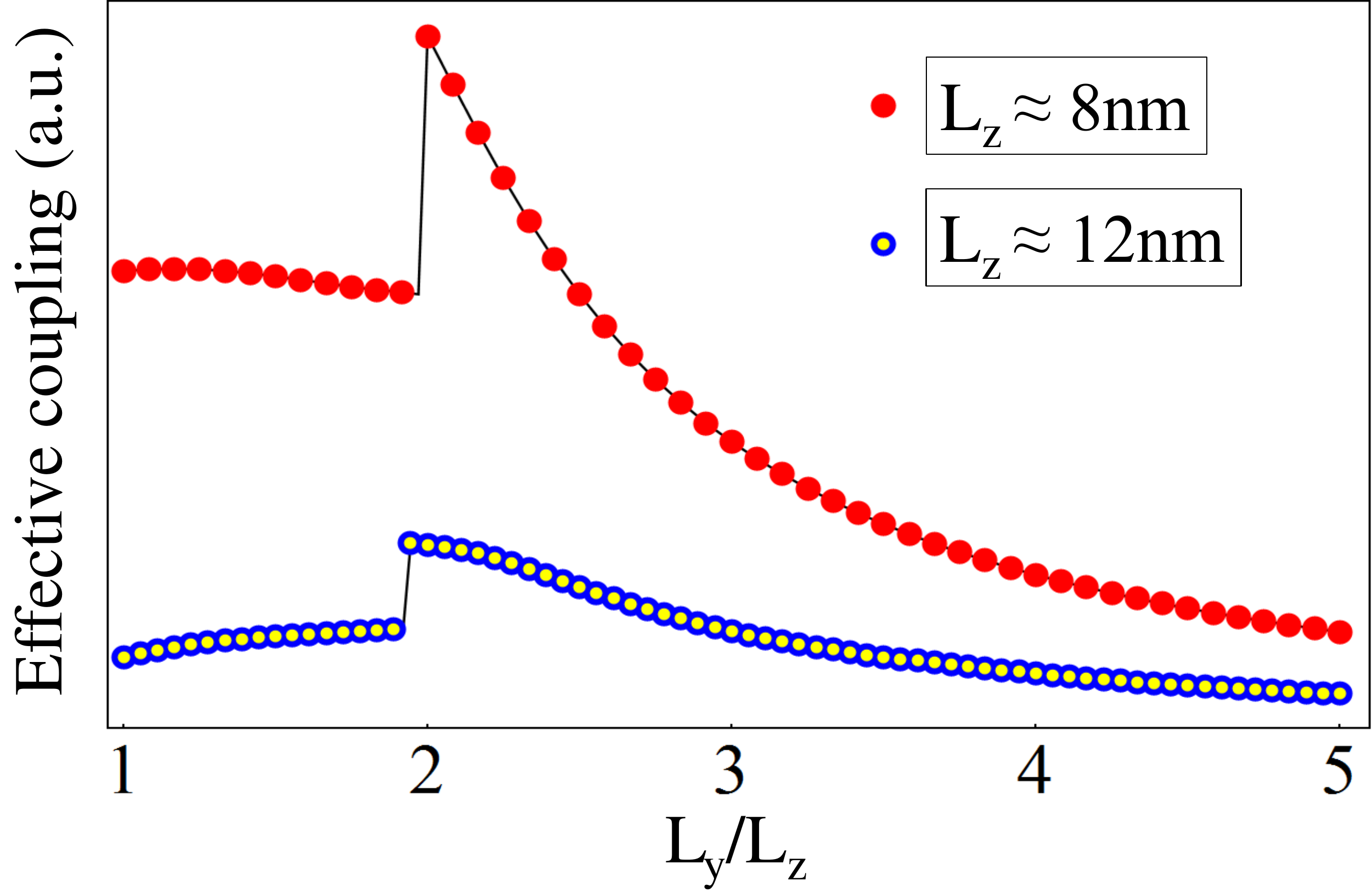}
\vspace{-7mm}
\end{center}
\caption{(Color online) Dependence of the effective SM--SC coupling of the top valence band in a hole--doped SM wire with rectangular cross section $L_y \times L_z$ on the $L_y/L_z$ ratio. The strength of the coupling decreases with increasing the wire thickness $L_z$ but, in contrast with electron--doped wires, it shows a strong dependence on the wire width $L_y$. In the limit $L_y\rightarrow\infty$ the coupling vanishes because the top valence band has purely heavy--hole character, i.e.,   $\psi_{\pm\frac{1}{2}}^{(s)}=0$. }
\vspace{-6mm}
\label{Fig3_3}
\end{figure}

To determine the effective coupling, we calculate numerically the wave functions $\Psi_{\pm}$ corresponding to the double degenerate top valence band at $k_x=0$.  These states can be expressed as four--component spinors, $\Psi_{s}(i_y, i_z) = [\psi_{\frac{3}{2}}^{(s)},  \psi_{\frac{1}{2}}^{(s)},  \psi_{-\frac{1}{2}}^{(s)},  \psi_{-\frac{3}{2}}^{(s)}]^T$, and we have
\begin{equation}
\gamma_c = \frac{2}{3}\nu_F|\tilde{t}|^2\sum_{i_y=1}^{N_y}|\psi_{\pm\frac{1}{2}}^{(s)}(i_y, i_{0z})|^2.  \label{gammac}
\end{equation}
For comparison, in the case of an electron--doped wire with uniform tunneling across the SM--SC interface, the effective coupling of the lowest conduction band can be obtained from Eq. (\ref{gammannp})  for $\tilde{t}(i_y)=\tilde{t}$ and $\bm n = \bm n^\prime = (1, 1, 1)$. We have $\gamma_c =  \nu_F |\tilde{t}|^2 \phi_1^2(i_{0z})\approx  2\pi^2\nu_F |\tilde{t}|^2/(N_z+1)^3$. Note that $\gamma_c$ is independent of the wire width $L_y$ and decreases approximately as $1/L_z^3$ with increasing wire thickness $L_z$, due to the decrease of the wave function amplitude  at the interface.   By contrast, the effective coupling of the top valence band is characterized by a strong dependence on $L_y$ for a fixed wire thickness $L_z$. The numerical results are shown in Fig. \ref{Fig3_3}.  Note that in the quasi--2D limit, $L_y\rightarrow\infty$,  the effective coupling of the top valence band to the bulk SC vanishes. This can be understood by noticing that the top valence band of a SM slab is a purely heavy--hole band and,  for a $(0, 0, 1)$ surface orientation, the corresponding states have vanishing $\psi_{\pm\frac{1}{2}}^{(\pm)}$ components. Consequently, we conclude that no proximity--induced superconductivity can occur in the top valence band of planar SM--SC heterostructures with a $(0, 0, 1)$ interface. This result holds for different interface orientations. Nonetheless, in quasi--1D nanowires, the top valence band is a superposition of heavy--hole and light--hole states and, consequently, the effective coupling to the SC is nonzero, as shown in Fig. \ref{Fig3_3}. Note that the amplitude of the of the light--hole component varies non--monotonically with $L_y$ and has a sharp discontinuity for a width--to--thickness ration $L_y/L_z \approx 2$. A quantitative description of the SC proximity effect in hole--doped nanowires, requires a more detailed modeling of the wire (e.g., using the 8--band model).

\section{Proximity effect in multi--band systems and the collapse of the induced gap}\label{App3}

\begin{figure}[tbp]
\begin{center}
\includegraphics[width=0.48\textwidth]{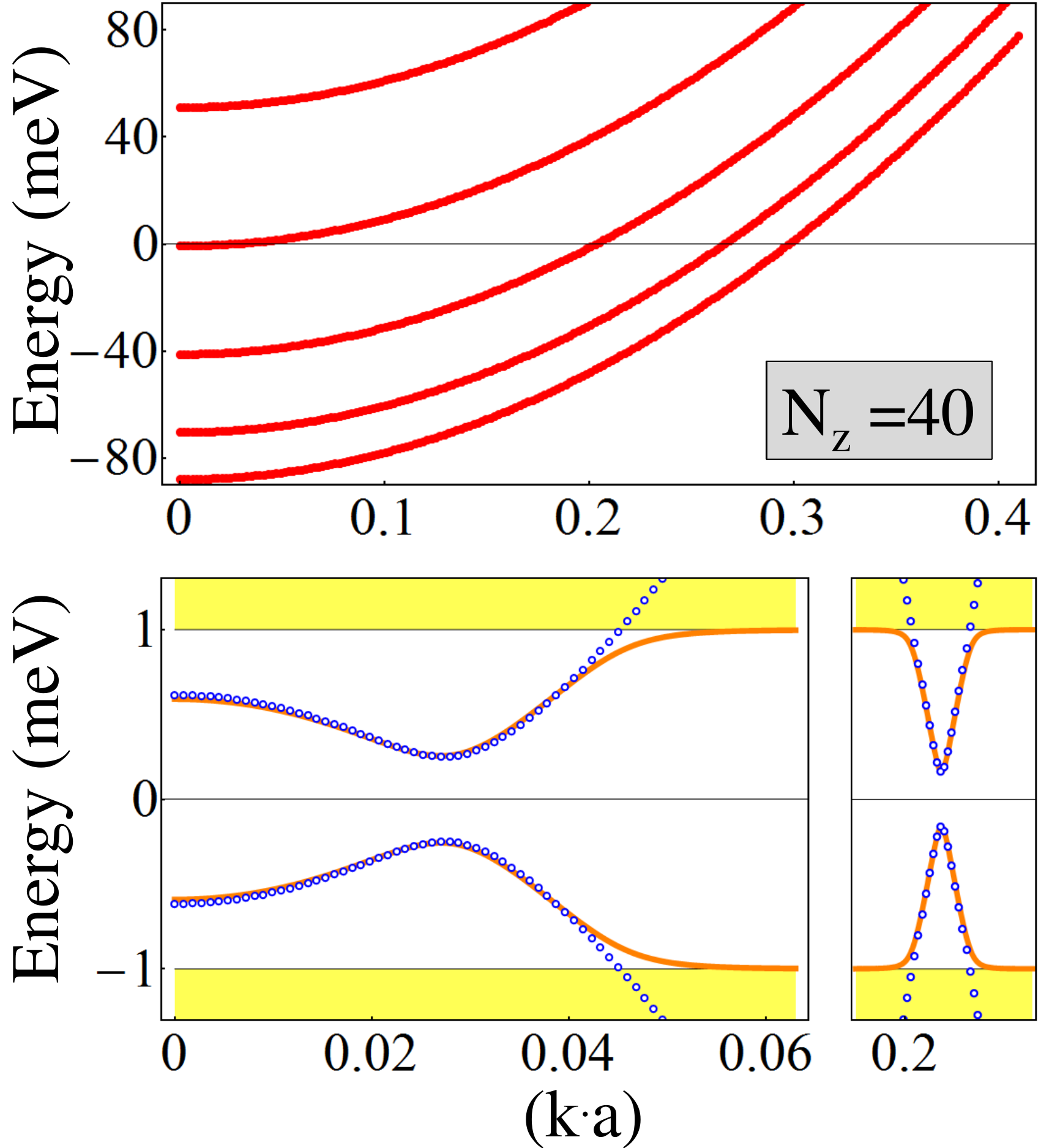}
\vspace{-7mm}
\end{center}
\caption{(Color online) Top panel: Spectrum of a SM slab with $N_z=40$ layers along an arbitrary direction in the $(k_x, k_y)$ plane.  The SM slab is described by a tight--binding model with nearest--neighbor hopping $t=1$eV on a simple cubic lattice with lattice constant $a$. The energy is measured with respect to the chemical potential. Bottom panel:  Comparison between the BdG spectrum obtained using Eq. (\ref{BdG1}) (orange lines) and the effective Hamiltonian (\ref{Heff}) (blue circles). The bulk SC gap is $\Delta_0=1$meV and the effective SM--SC coupling corresponds to $\gamma_4=\Delta_0/3$ for the top occupied band.  The minima of the BdG spectrum occur in the vicinity of the Fermi k--vectors corresponding to different $n_z$ bands (only the minima corresponding to $n_z=4$ and $n_z=3$ are shown) and are determined by the corresponding induced SC gap ($\Delta_4=0.25$meV and $\Delta_3\approx0.14$meV). }
\vspace{-6mm}
\label{Fig3_4}
\end{figure}

\begin{figure}[tbp]
\begin{center}
\includegraphics[width=0.48\textwidth]{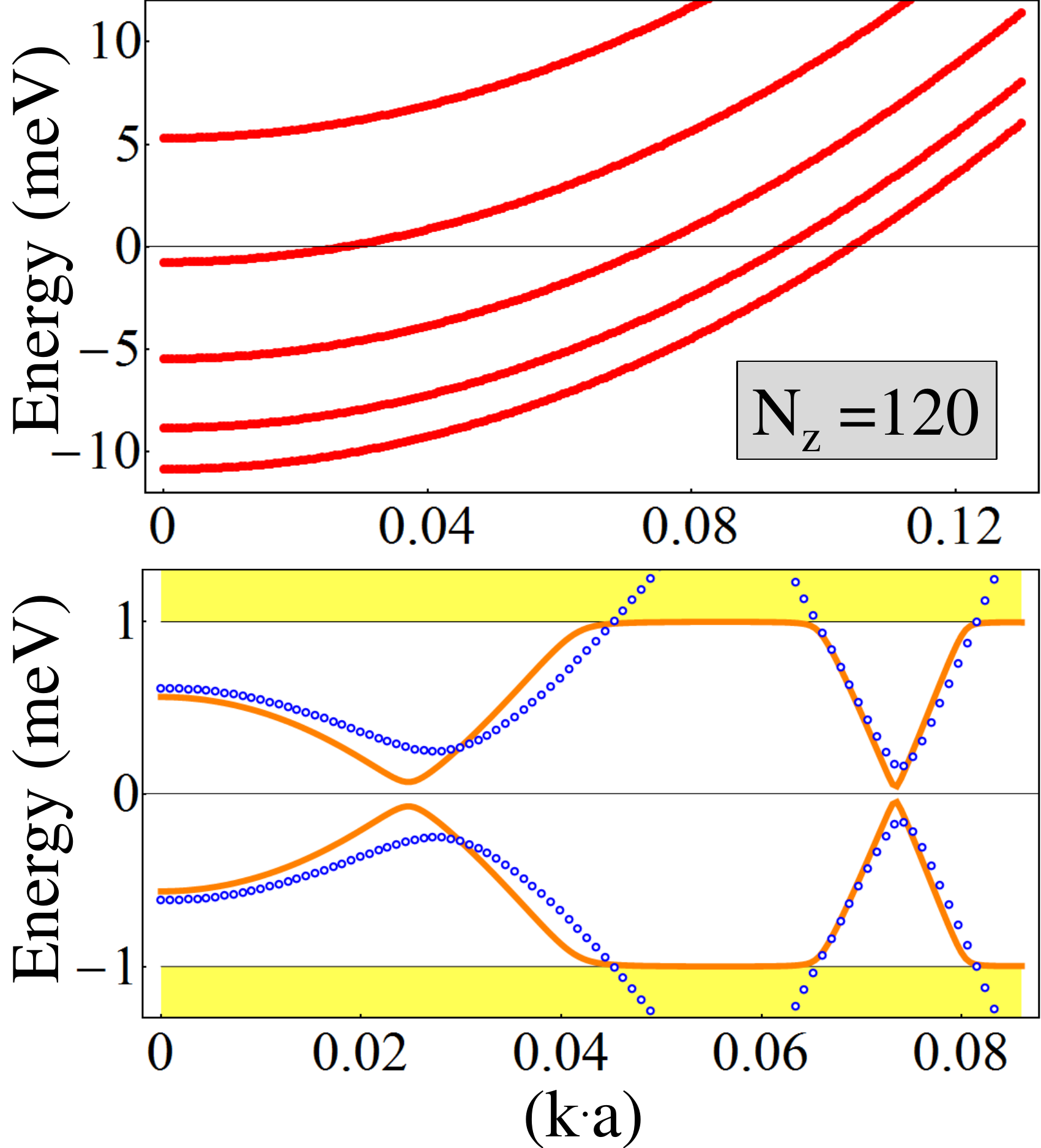}
\vspace{-7mm}
\end{center}
\caption{(Color online) Top panel: Normal spectrum of a SM slab with $N_z=120$. The other parameters are the same as in Fig. \ref{Fig3_4}. Bottom panel: Collapse of the induced SC gap due to proximity--generated inter--band coupling. All parameters (except $N_z$) are the same as in  Fig. \ref{Fig3_4}.  The solution obtained using Eq. (\ref{BdG1}) (orange lines) includes inter--band  coupling, while the effective Hamiltonian (\ref{Heff}) (blue circles) is constructed in the decoupled band approximation, $\gamma_{n_z n_z^\prime} \propto \delta_{n_z n_z^\prime}$.}
\vspace{-6mm}
\label{Fig3_5}
\end{figure}

We illustrate the role of multi--band physics in proximity--coupled finite size systems by calculating the low--energy spectrum of a SM thin film -- SC slab heterostructure with uniform coupling across the planar interface. This allows us to focus on the proximity--induced inter--band coupling involving bands with different $n_z$ quantum numbers, a problem that was not previously investigated in the literature. First, we consider the limit $\Delta E_{\bm n \bm n^\prime}\gg\Delta_0$ and we test the accuracy of the static approximation that allows us to define the effective Hamiltonian (\ref{Heff}). For simplicity, the SM film is described using the 2--band model with nearest neighbor hopping $t=1$eV,  no Rashba spin--orbit coupling ($\alpha=0$), and no Zeeman field ($\Gamma=0$). The normal state spectrum of a slab containing $N_z=40$ layers is shown in the upper panel of Fig. \ref{Fig3_4}. The chemical potential is near the bottom the fourth band, $n_z=4$, and intersects the bands with $n_z\leq4$ at different Fermi wave vectors $k_{F n_z}$. The SM slab is coupled to a SC with a bulk gap $\Delta_0=1$meV.  The effective SM--SC coupling has the form $\gamma_{n_z n_z^\prime} = \gamma_0\phi_{n_z}(i_{0z}) \phi_{n_z^\prime}(i_{0z})$, where $\phi_{n_z}(i_{0z})$ is given by Eq. (\ref{phinl}),  and the constant $\gamma_0$ is chosen so that the effective coupling of the fourth band be $\gamma_{4}\equiv \gamma_{4 4} = \Delta_0/3$. Since $\Delta E_{\bm n \bm n^\prime}\gg\Delta_0$, the effective Hamiltonian is constructed using the decoupled band approximation and the corresponding low--energy spectrum is compared with the solution of the full BdG equation (\ref{G1}). The results are shown in the lower panel of Fig. \ref{Fig3_4}. Note that the effective Hamiltonian description is highly accurate at energies within the SC gap, except near the gap edge where the static approximation $\sqrt{\Delta_0^2-\omega^2}\approx\Delta_0$ manifestly breaks down.

The role of proximity--induced inter--band coupling is illustrated in Fig. \ref{Fig3_5}, which shows the spectra corresponding to a SM film with $N_z=120$ layers. The other parameters are identical with those in Fig. \ref{Fig3_4}, including the effective  SM--SC coupling $\gamma_{n_z n_z^\prime}$.  Note that the decoupled band approximation fails, as the inter--band gaps are now comparable with $\Delta_0$ (see the upper panel of  Fig. \ref{Fig3_5}). In particular, inter--band coupling results in a collapse of the induced SC gap. A detailed analysis of this effect is presented in Ref. \cite{Stanescu2013}. We emphasize  that this effect is not due to the decrease of the wave function amplitude at the interface with increasing $N_z$, as this is compensated for by increasing the transparency of the interface, i.e., $\gamma_0$, but stems from the off--diagonal elements of the effective coupling matrix $\gamma_{n_z n_z^\prime}$. Also, we note that in this regime the construction of the effective Hamiltonian (\ref{Heff}) has to involve the high--energy $n_z$ bands, as they are intrinsically coupled to the low--energy bands and renormalize them strongly, which ultimately leads to the collapse of the induced SC gap.

\section{Low--energy states in the presence of disorder}\label{App4}

\begin{figure}[tbp]
\begin{center}
\includegraphics[width=0.48\textwidth]{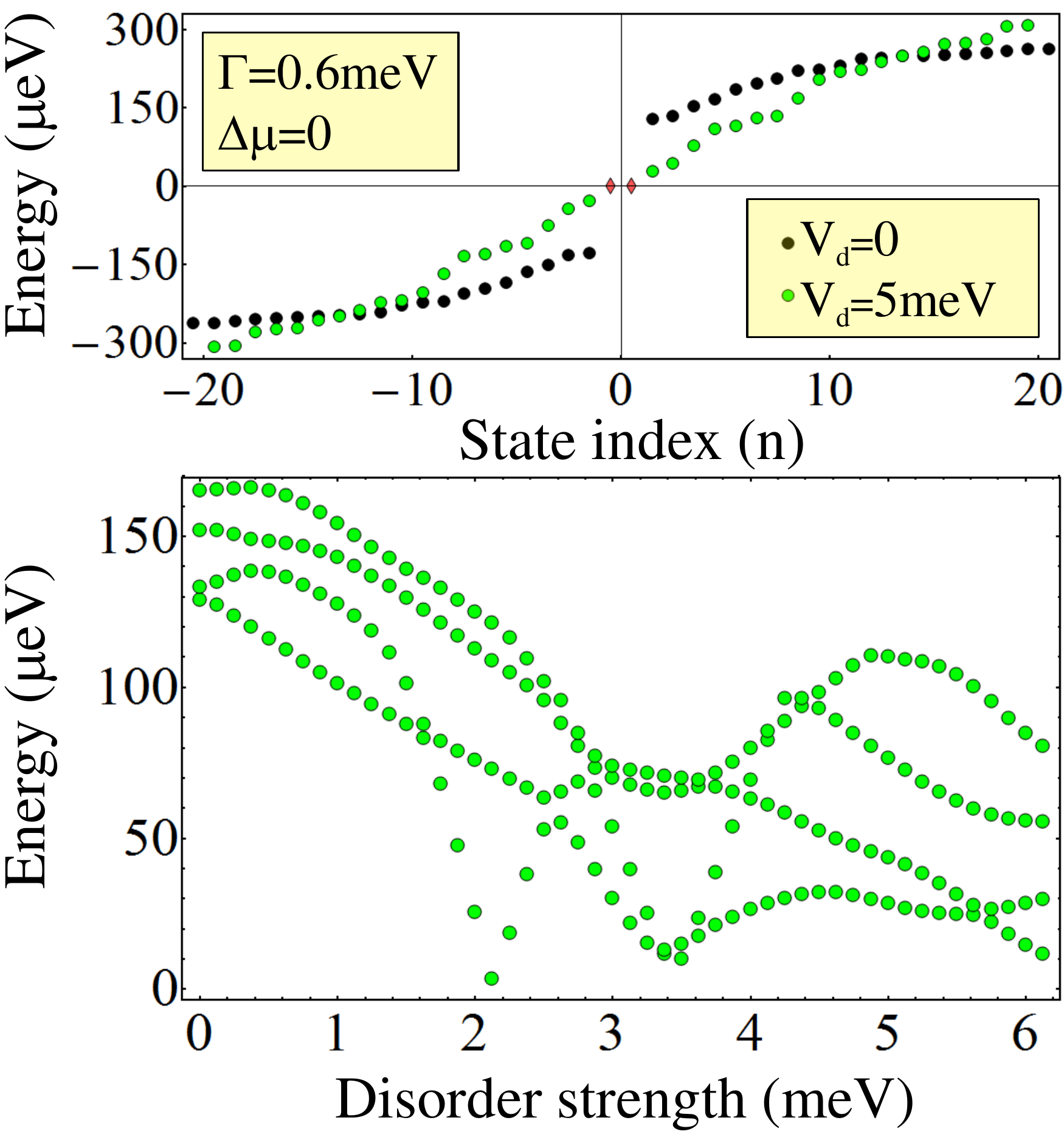}
\vspace{-7mm}
\end{center}
\caption{(Color online)
{\em Top}: BdG spectrum of a SM wire with proximity--induced superconductivity and Zeeman splitting $\Gamma=0.6$meV. The chemical potential relative to the bottom of the third band is $\Delta\mu=0$. In a clean wire ($V_d=0$) the Majorana bound states (red diamonds) are protected by a gap of the order $130\mu$eV. This gap collapses in the presence of strong disorder. {\em Bottom}: Dependence of the non--zero lowest--energy modes on the strength $V_d$ of the disorder potential. }
\vspace{-6mm}
\label{Fig4_4}
\end{figure}

\begin{figure}[tbp]
\begin{center}
\includegraphics[width=0.48\textwidth]{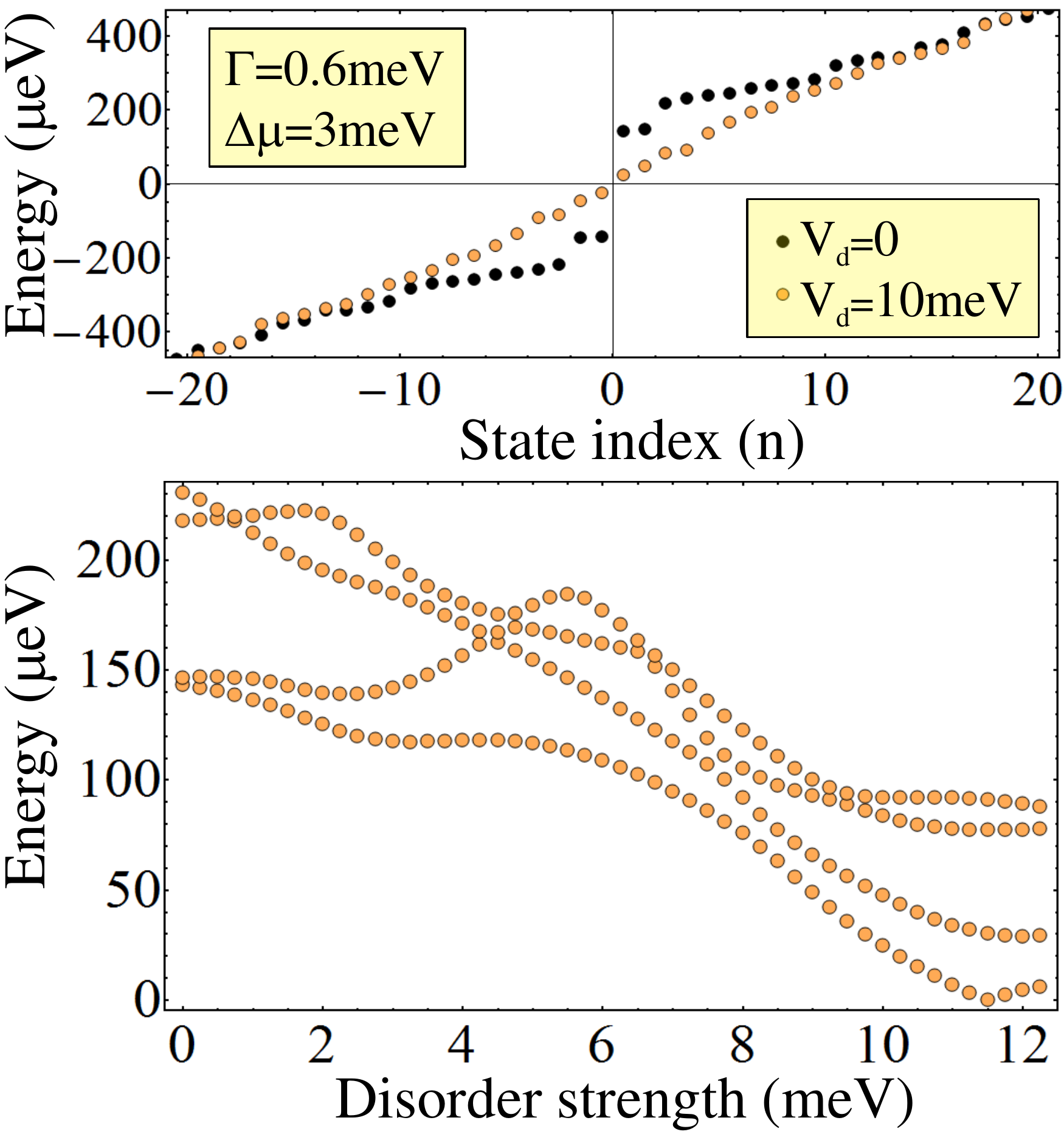}
\vspace{-7mm}
\end{center}
\caption{(Color online) {\em Top}: BdG spectrum of a SM wire with parameters corresponding to the topologically trivial phase. Note that there are no zero--energy states. The quasiparticle gap collapses in the presence of strong disorder. {\em Bottom}: Dependence of the  lowest--energy modes on the strength $V_d$ of the disorder potential. Note that the proliferation of disorder--induced low--energy states requires a stronger disorder potential that in the topological SC phase (see Fig. \ref{Fig4_4}). }
\vspace{-6mm}
\label{Fig4_5}
\end{figure}

We illustrate some of the generic features of low--energy BdG spectrum of a disordered SM--SC hybrid system by considering  a SM nanowire of length $l_x=3\mu$m and rectangular cross section with $L_y=80$nm and $L_z=40$nm in the presence of a disorder potential $V_(\bm r) = V_d f_V(\bm r)$, where the disorder profile is described by the random function $f_V$ with $|f_V(\bm r)|\leq 1$ and $V_d$  represents the amplitude of the disorder potential. The low--energy spectrum is calculated using a 2--band model, as described in Sec. \ref{Sec3_1}. All the calculation are done for a specific disorder realization, i.e., a fixed profile $f_V$, but for variable disorder strength $V_d$. First, we consider a system with three partially occupied bands (five spin sub--bands) in the presence of a Zeeman splitting $\Gamma=0.6$meV and a fixed chemical potential $\Delta \mu=0$, as measured relative to the bottom of the third band in the absence of Zeeman splitting. The clean wire ($V_d=0$) is in the topological SC phase that supports zero--energy Majorana bound states. The corresponding BdG spectrum is shown in the top panel of Fig. \ref{Fig4_4} (black circles and red diamonds for the Majorana states). In the presence of disorder ($V_d\neq 0$), the mini--gap that protects the Majorana bound state becomes smaller and, eventually, collapses.  The dependence of the non--zero lowest--energy modes on the strength of the disorder potential is shown in the bottom panel. We note that, in the presence of disorder,  multiple zero--energy modes are possible as disorder effectively cuts the wire into disconnected topologically nontrivial segments supporting Majorana bounds states at their ends. Typically, these segments are relatively short and the states localized at the ends of each segment (which can be viewed as representing a Majorana chain) overlap significantly  and are characterized by nearly--zero energies  that oscillate with the chemical potential and the Zeeman splitting.

\begin{figure}[tbp]
\begin{center}
\includegraphics[width=0.48\textwidth]{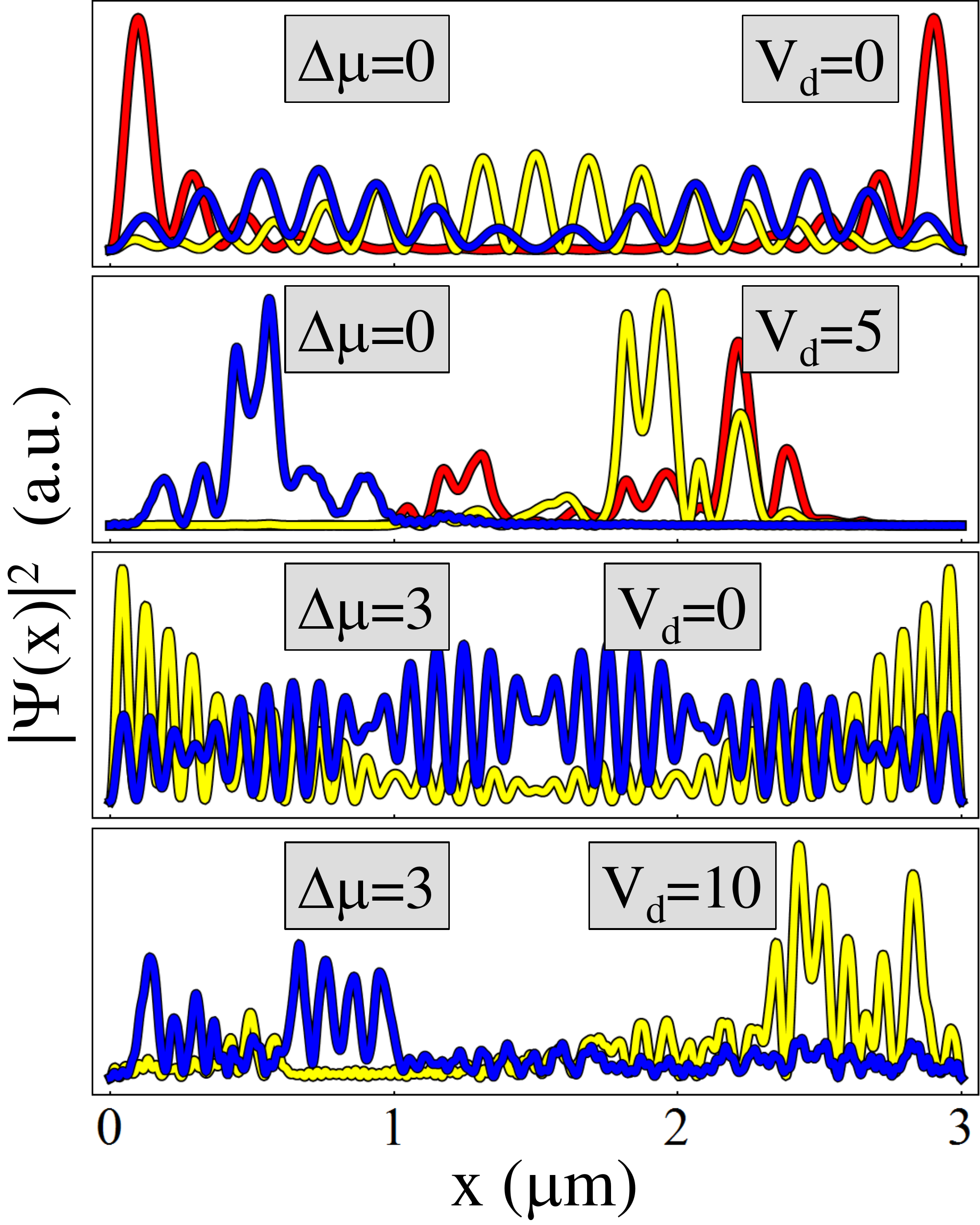}
\vspace{-7mm}
\end{center}
\caption{(Color online) Position dependence of the amplitude of the lowest--energy states for a wire with $L_x=3\mu$m and $\Gamma=0.6$meV. The top panel corresponds to a clean wire in the topological SC phase. The red line with maxima near the ends of the wire represents the Majorana bound states. In the presence of disorder, all the states become localized in various regions of the wire. The wave function of the lowest energy state of clean wire ($V_d=0$) in the topological trivial state is characterized by an envelope with maxima near the ends of the wire and oscillations corresponding to a certain value of $k_F$. The lowest--energy states associated with the low--energy occupied bands (not shown) have similar characteristics, but higher values of $k_F$. For given values of the disorder strength and Zeeman splitting, states with larger characteristic Fermi wave vectors (shorter oscillation period) exhibit weaker localization.}
\vspace{-6mm}
\label{Fig4_6}
\end{figure}

A similar proliferation of low--energy states with increasing the strength of the disorder can be seen in a system with parameters corresponding to the trivial SC phase, as shown in Fig. \ref{Fig4_5}. Note that in this case the decrease of the quasiparticle gap with $V_d$ is slower than that corresponding to $\Delta\mu=0$. This behavior can be understood qualitatively by noting that disorder tends to localize the low--energy states and that this effect depends of the characteristic Fermi wave vector associated with those states. Increasing $\Delta\mu$ corresponds to larger values of $k_F$ and to a weaker localization. This behavior is illustrated by the profiles of the low--energy states shown in Fig. \ref{Fig4_6}.  The low--energy states corresponding to $\Delta\mu=0$ exhibit stronger localization that those for $\Delta\mu=3$meV, in spite of the smaller amplitude of the disorder potential. We note that the low--energy states associated with the low--energy bands are characterized by large values of $k_F$ and, consequently, are harder to localize.  Finally, we note that the other key parameter that controls the effect of disorder on the low--energy states is the Zeeman splitting. Increasing $\Gamma$, which breaks time--reversal symmetry,  facilitates the collapse of the SC gap. Again, this effect is stronger for states with low values of the characteristic $k_F$, which typically corresponds to the top occupied band, although significant band mixing is possible for certain types of disorder. Consequently, in a wire with hard confinement the nearly--zero energy states  with most of the spectral weight coming from low--energy bands require higher values of the Zeeman splitting than the disorder--induced  nearly--zero energy states associated with the top band. On the other hand, the energy of these states can become exponentially small even in the absence of disorder if the confinement potential is smooth~\cite{Kells2012}. Furthermore, the states associated with low--energy bands can penetrate through finite barrier potentials and extend into the normal section of the wire~\cite{Stanescu2012}. 

\bibliography{RevPaperREF}
\bibliographystyle{unsrt}
\end{document}